\documentclass[amsmath,amssymb,nofootinbib,prd]{revtex4}
\pdfoutput=1
\usepackage{graphicx}
\usepackage{amsmath}
\usepackage{amssymb}
\usepackage{amsfonts}
\usepackage{color}
\usepackage{soul}

\newcommand{\ud}{\mathrm{d}}
\newcommand{\p}{\partial}

\newcommand{\Q}{\mathcal{Q}}

\def\be{\begin{equation}}
\def\ee{\end{equation}}
\def\bea{\begin{eqnarray}}
\def\eea{\end{eqnarray}}

\begin{document}

\title{Generalisation of the Kaiser Rocket effect in general relativity in the wide-angle galaxy 2-point correlation function}

\author{Daniele Bertacca}

\affiliation{Dipartimento di Fisica e Astronomia ``G. Galilei", Universit\'a degli Studi di Padova, via Marzolo 8, I-35131, Padova, Italy\\
INFN Sezione di Padova,  I-35131 Padova, Italy }

\begin{abstract}
We study wide-angle correlations in the galaxy power spectrum in redshift space, including  all general relativistic effects and  the Kaiser Rocket effect in general relativity. 
We find that the Kaiser Rocket effect becomes important on large scales and at high redshifts, and leads to new contributions in wide-angle correlations.
We believe this effect might be very important for future large volume surveys.

\end{abstract}

\date{\today}

\maketitle

\section{Introduction}

  This Local Group (LG) of galaxies contains 14 members within $\sim~1.4$ Mpc from the LG barycenter (not including satellites of M31 and MW), e.g. see \cite{Kogut:1993ag}. 
The LG forms a bound object and resides in a mildly over-dense region characterised by a  small velocity shear  and moves with a non-vanishing velocity relative to the general expanding background. 
 This motion of LG galaxies carries an imprint  as a  dipole moment in the galaxy distribution and can be measured  using a variety of galaxy catalogs in the full-sky redshift surveys.
 
  A straightforward way to measure the dipole is based on temperature maps of the Cosmic Microwave Background (CMB) radiation, identifying the motion of the LG equal to the measure of the dipole anisotropy form the CMB radiation.
 The velocity of LG in the CMB frame from this analysis is $622\pm33$ km\,s$^{-1}$ in the $l=277\pm3^{\circ}$ and $b=33\pm3^{\circ}$ direction (Galactic coordinates)
 i.e. towards the constellation of Hydra (e.g. see \cite{Yahil:1977zz,Kogut:1993ag,Fixsen:1996nj,Hinshaw:2008kr}). (For completeness, see also \cite{Gibelyou:2012ri, Nusser:2014sha, Maartens:2017qoa, Pant:2018smd} and see \cite{Baleisis:1997wx, Blake:2002gx, Singal:2011dy, Rubart:2013tx, Kothari:2013gya, Schwarz:2015pqa, Tiwari:2015tba, Colin:2017juj, Bengaly:2017slg}  for the radio dipole.)
  Of course, a comparison of this motion with the dipole moment of the galaxy distribution can be a direct measure of the growth. 
 Following the standard cosmological paradigm (e.\ g.\ see \cite{Peebles1980}), the LG acceleration should be the result of the cumulative gravitational pull of the surrounding distribution matter in the Universe.  A recent analysis of this issue
 by \cite{Davis:2010sw}  shows a  good agreement between the local velocity and gravitational fields.   
Now, it is important to  ask whether, for the observed large scale structure which are traced by the galaxy distribution, we should take into account the LG motion.
From the above considerations, it is clear that that the observed galaxy overdensity in {\it redshift-space} has to be measured in the LG frame and not  in the CMB frame.

In a generic redshift survey, we can compute  the dipole moment from  (e.g. see \cite{Strauss:1995fz})
\[{H_0 \beta_0 \over 4 \pi} \sum_i W_{\rm g}(r_i) {\delta_{\rm g}}_i ~ { {\bf r}_i \over r_i^3}\;,\]
where the summation is over the grid points, ${\bf r}_i$ is the distance of the grid cell $i$ from the LG position, ${\delta_{\rm g}}_i$ is the overdensity contrast at a given cell $i$ and  the window function $W_{\rm g}(r_i)$ specifies the finite survey volume at $r_i$. In particular, we should consider a window  that has a cutoff both at the largest possible radius and a small distance in order to minimise the shot noise (see also \cite{Juszkiewicz1990,Lahav-Kaiser-Hoffman1990, Peacock1992} where they pointed out  that the structure outside the window could be decisive in the measuring the dipole). Here, in linear theory, $\beta_0$ is related to linear galaxy bias and the rate of the growth. An interesting analysis was recently done in \cite{Nusser:2014sha} where they concluded that  the CMB frame can be  gradually reached and they showed that the LG motion cannot be recovered to better than $150-200$ km s$^{-1}$ in amplitude and within an error of  $\simeq 10^{\circ}$ in direction, which is inevitable whether the analysis is done both the redshift and in real space.

At this point, an important effect that we have to take into account is the impact of the {\it rocket  effect} (also called {\it Kaiser rocket}), see \cite{Kaiser:1987qv}.
Indeed, when we try to correct the redshift with the LG peculiar velocity  without considering the following quantity
\[\frac{W_r( r) }{r}\left(2 + \frac{\partial \ln \bar n_{\rm g}(r)}{\partial \ln r}\right) \;, \]
where $W_r( r)\propto \bar n_{\rm g}(r)$ is the (normalised) radial selection function (i.e. $\int r^2 W_r dr=1$), we have a spurious contribution. Finally, the signature of the {\it  rocket effect} cannot be neglected if we consider the reconstructed LG motion at radii  larger than 100$h^{-1}$ Mpc, for example see \cite{Nusser:2014sha}. 
Clearly, the {\it rocket effect} can be  corrected if the selection function is well constrained by observations \cite{Strauss:1995fz}. 
Therefore, it is crucial to evaluate the Kaiser rocket effect well. In fact it is  useful to understand if it is only (if ignored) a possible source of systematic effects or, if isolated and measured, it allows us to estimate cosmological parameters and break degenerations next-generation galaxy surveys .
Indeed, with next-generation galaxy surveys [such as Euclid\footnote{http://www.euclid-ec.org} and measurements of HI from the Square Kilometre Array survey\footnote{http://www.skatelescope.org} (SKA)], covering large volumes with dramatically improved statistics, we are about to enter the era of precision cosmology in galaxy surveys.

Observations are performed along the past light-cone, which brings in a series of  local and non-local (i.e. integrated along the line of sight) corrections, usually called GR projection effects (hereafter they will be abbreviated as {\it GR effects} or {\it corrections}), which are not included in the ``standard"  treatment  (e.g. see~\cite{Kaiser:1987qv, Hamilton:1997zq}).
GR corrections arise because we observe galaxies on the past light-cone and not a constant time hypersurface. Indeed the fact that the volume element constructed by using observables differs from the physical volume occupied by the observed galaxies, the observed galaxy density map is affected by these distortions. 
The study of these GR effects on first-order statistics of large scale structure, for example to compute the power-spectrum, the two-point correlation function  or angular two-point correlation (both for the galaxy and continuum radio sources), has received significant attention in recent years, see e.g. \cite{Yoo:2008tj,Yoo:2009au,Yoo:2010ni,Bonvin:2011bg,Challinor:2011bk,Bruni:2011ta,Baldauf:2011bh,Jeong:2011as,Yoo:2011zc,Bertacca:2012tp,Yoo:2012se,Yoo:2013tc,Raccanelli:2013multipoli,DiDio:2013bqa,Yoo:2013zga,DiDio:2013sea,Bonvin:2013ogt,Raccanelli:2013gja,Bacon:2014uja,Chen:2014bba,Raccanelli:2015vla,Alonso:2015uua,Montanari:2015rga,Chen:2015wga,Alonso:2015sfa,Fonseca:2015laa,Bonvin:2015kuc,Gaztanaga:2015jrs, Cardona:2016qxn,Raccanelli:2016avd,DiDio:2016ykq,Borzyszkowski:2017ayl,Scaccabarozzi:2017ncm,Abramo:2017xnp,Scaccabarozzi:2018vux,Tansella:2017rpi,Lepori:2017twd,Villa:2017yfg, Tansella:2018hdm,Schoneberg:2018fis,Tansella:2018sld,Breton:2018wzk}.

Recently, using a GR analysis, \cite{Scaccabarozzi:2018vux} has correctly  pointed out that,  in the galaxy two-point correlation function,  the dipole at the observer position is often ignored in literature (even though this contribution could be larger than the other relativistic and projection contributions at large redshift). Taking into account this claim, in this work we want compute the dipole contribution and its all possible correlation with local and non-local GR corrections. Finally, we will extend the work made by \cite{Bertacca:2012tp}, where the authors developed the fully general relativistic wide-angle formalism with GR and wide-angle GR corrections (see  also \cite{Raccanelli:2013multipoli,Raccanelli:2013gja}), adding the dipole effect in their analysis.

In general, the Kaiser Rocket effect is associated with the dipole moment of the density field of galaxies (or velocity field). Instead in our paper we want to focus mainly on the impact of the Kaiser Rockect effect in the 2-point statistics.

The paper is organised as follows: in Section \ref{sec:dipole_velocity} we introduce how, in literature,  the dipole term to the galaxy correlation function has been analysed. Instead, in Section \ref{Dipole-Sec}, we write the observed overdensity and the list of all terms that we observe on the past light-cone.
In Section \ref{RSDcorrelation}
 we briefly review the results in \cite{Bertacca:2012tp,Raccanelli:2013multipoli, Raccanelli:2013gja}
and Section \ref{Analysis-dipole}  is devoted to dipole contribution in GR. In Sections \ref{dipole-corr} and \ref{localtermswithdipole} we present a formalism to compute the dipole correlation terms and we describe the various effects in more detail.
In Section \ref{sec:numresults} we investigate different configurations formed by the observer and the pair of galaxies and we try to figure out if these effects could be important for future large volume surveys.
Finally, in Section \ref{sec:Conlusions} we draw our conclusions and discuss results and future prospects.

\section{The Dipole velocity field  and the Rocket  Effect}\label{sec:dipole_velocity}

Let us discuss more about how we obtain the {\it rocket  effect}. In the classic/standard prescription (e.g. see \cite{Peebles1980, Juszkiewicz1990,Lahav-Kaiser-Hoffman1990, Strauss:1995fz}), using  the continuity equation and assuming the linear theory, it guarantees that the peculiar velocities of the galaxies in the LG frame are small with respect to the distances $r$, we can write the velocity field in the following way:
\begin{equation}\label{v(r)}
{\bf v}({\bf r})=\frac{f{\mathcal H}}{4 \pi}  \int_{V^{\mathcal R}} {\rm d}^3{\bf r'} \; \frac{ ({\bf r'}-{\bf r})}{{\left| {\bf r'}-{\bf r}\right|}^3 } \delta^{\mathcal R}_{\rm m}{(\bf r')}\;.
\end{equation}
Let us point out again that, for simplicity, we take ${\bf v}_{\rm g}={\bf v}$, i.e. there is no the velocity bias. (In general, in literature, instead of $f$ and $\delta^{\mathcal R}_{\rm m}$, it is written $\beta$ and\footnote{If we define Eq. (\ref{v(r)}) with $\beta$ we could make a possible mistake because $b$ depends on both the space and time. Consequently it is not correct that $1/b$ can be out of the 3D integral.} $\delta^{\mathcal R}_{\rm g}$.) From the above relation, we assume that this velocity field is mainly determined by all matter that is clustering (in particular the CDM). 
Of course, in order to apply the linear relation, one should smooth the density field on small scales. It also removes the issue of the large velocity dispersion (which cannot be described by linear theory). With this smoothing, we suppress completely the behaviour of the cluster of galaxies which typically collapse to nearby galaxies associated with the prominent cluster (in this case we are also removing the {\it fingers-of-God} distortions). In addition, using this approach we may prevent an important issue/problem related to the fact that the redshift-distance relation along the line of sight could not be necessarily monotonic in the vicinity of the cluster of galaxies, e.g. see \cite{Yahil-Strauss-Davis-Huchra1991,Davis-Strauss-Yahil1991,Strauss:1995fz}. 

In order to extract ${\bf v}({\bf 0})$, from Eq. (\ref{v(r)}), we set ${\bf r}={\bf 0}$, i.e 
\begin{equation}\label{v(0)}
{\bf v}({\bf 0})=\frac{f_0 H_0}{4 \pi}  \int_{V^{\mathcal R}} {\rm d}^3{\bf r'} \; \frac{ \hat{\bf r'}}{{r'}^2 } \delta^{\mathcal R}_{\rm m}{(\bf r')}\;,
\end{equation}
where $f_0=f(\eta_0)=f(z=0)$. Then,  it yields
\begin{equation}\label{v_r(0)}
v_r({\bf 0})=\frac{1}{H_0}\hat{\bf r} \cdot {\bf v}({\bf 0})
=\frac{f_0}{4 \pi}  
\int_{V^{\mathcal R}} {\rm d}^3{\bf r'} \; \frac{ \hat{\bf r} \cdot \hat{\bf r'}}{{r'}^2 } \delta^{\mathcal R}_{\rm m}{(\bf r')}\;.
\end{equation}
Using the identity
\begin{equation}\label{Y1}
\left(\hat{\bf r} \cdot \hat{\bf r'}\right) = {4 \pi \over 3} \sum_{{\bar m}=-1}^{1} Y_{1 {\bar m}}( \hat{\bf r})  Y_{1 {\bar m}}^*(\hat{\bf r'})
\end{equation}
we can rewrite the dipole contribution (e.g. see \cite{Fisher:1994cm}) in the following way
\begin{equation}
 \delta_{\rm g}^{\rm Dipole}({\bf r})= \frac{f_0}{3  r} \left(2 + \frac{\partial \ln \bar n_{\rm g}(r)}{\partial \ln r}\right)  \sum_{{\bar m}=-1}^{1} Y_{1 {\bar m}}(\hat{\bf r})   \int  {\rm d}\tilde r  ~ {\delta^{\mathcal R}_{\rm m}}_{1 \bar m}(\tilde r)
\end{equation}
and, projecting directly this physical quantity only on a sphere,  it turns out
\begin{eqnarray}
{\delta_{\rm g}^{\rm Dipole}}_{\ell m}(r)&=& \frac{i^\ell f_0}{6 \pi^2  r}  \left(2 + \frac{\partial \ln \bar n_{\rm g}(r)}{\partial \ln r}\right)  \delta^K_{\ell 1} \int \frac{{\rm d}^3{\bf k}}{k} ~ Y^*_{\ell m}(\hat{\bf k}) \, \delta_{\rm m}^{\mathcal{R}}({\bf k}, \eta)\;.
\end{eqnarray}

Let us conclude this part mentioning a different approach used in \cite{Nusser:1993sx}  (see also \cite{Nusser:2011vz,Nusser:2014sha}), where, starting from the Zel'dovich approximation, conservation of galaxies and assuming the velocity is irrotational, they studied the displacement of the galaxies at $z \to 0$ between redshift to real space. Through this method they reconstruct the smooth peculiar velocity field from the observed distribution of galaxies. In particular,  this technique involves the expansion in spherical harmonics and correcting eventually each mode with the peculiar velocities of the galaxies in redshift space. 
In sections  \ref{Dipole-Sec} we will study the effect of the dipole on the large scale structure.

\subsection{Observed galaxy density perturbation in General Relativity}\label{}

Let us discuss briefly the observed number density of tracers contained within a volume defined in terms of the observed coordinates.
The spatial volume seen by a source with  (comoving) 4-velocity $u^\mu$ is $\ud V_{\rm R}=\sqrt{-g}\,\epsilon_{\alpha\beta\gamma\delta}
\, u^\alpha 
\,a_{\rm R}^3\,
\ud x_{\rm R}^\beta \ud x_{\rm R}^\gamma \ud x_{\rm R}^\delta$ where $\epsilon_{\alpha\beta\gamma\delta}$ is the fully antisymmetric Levi-Civita symbol.
We start from writing down the total number of galaxies  contained within a volume $V_{\rm R}$ (defined in terms of the $x_{\rm R}^\mu$ coordinates)
\begin{equation}
 \label{N}
N=
\int_{V_{\rm R}} \sqrt{- g_{\rm R}(x_{\rm R}^\alpha)}\: a_{\rm R}^3(x_{\rm R}^0) \: n^{\rm R}_{\rm g}(x_{\rm R}^\alpha)\: \ud V_{\rm R}\;,
\end{equation}
where 
 $n^{\rm R}_{\rm g}(x_{\rm R}^\alpha)$ denotes the actual number density of galaxies as a function of the comoving coordinates $x_{\rm R}^\alpha$ and $g_{\rm R}(x_{\rm R})$ is the determinant of the comoving metric.
\begin{figure*}[htb!]
\includegraphics[width=0.8\columnwidth]{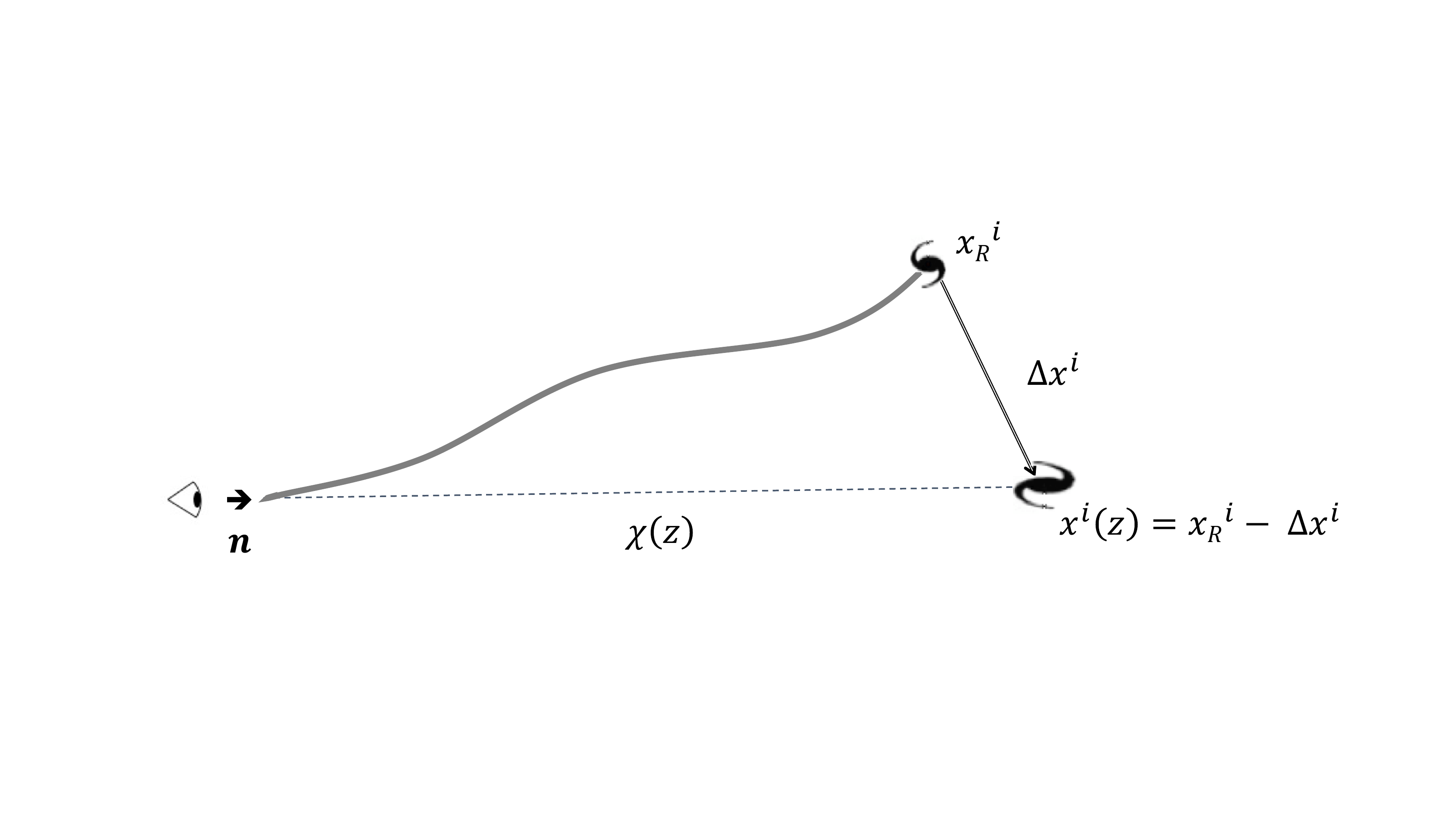}
\caption{Positions a galaxy both on the light-cone and in real-space.}\label{lcone}
\end{figure*}
The observed galaxy overdensity is a function of the observed direction ${\bf n}$ (or, equivalenlty, $n^i$) and redshift $z$ (see Fig.\ \ref{lcone}) and the equivalent relation in redshift space is
\begin{equation}
\label{N2}
N =
\int_{ {V}} { a^3( x^0) \,  n_{\rm g}\left( x^0, {\bf x}\right)} \, \ud^3{\bf x} \;.
\end{equation}
where the observed comoving volume is $\ud^3 {\bf x} = \ud V$ and $n_{\rm g}( x^0, {\bf x})$ is the observed galaxy density.
Then,  by equating the total number of galaxies $N$  in Eqs.\ (\ref{N}) and (\ref{N2}) and expanding to linear order in the perturbations,  we can write  the matter density contrast in redshift space as  
\begin{eqnarray}\label{eq_delta_rsd} 
 \Delta_{\rm g}=\delta_{\rm g}+\Delta_{\rm RSD}\;, 
\end{eqnarray} 
where 
\[\delta_{\rm g}(x^\alpha(z)) = {n_{\rm g}(x^\alpha(z)) - \bar n_{\rm g}(z)\over \bar n_{\rm g}(z)}\;,\]
where $\bar n_{\rm g}(z)$ is the average density of galaxies at the observed redshift $z$. We have conveniently collected the corrections
due to the metric distortions into the term $\Delta_{\rm RSD}$,  e.g. see  \citep{Yoo:2008tj, Yoo:2009au, Bonvin:2011bg, Challinor:2011bk, Jeong:2011as}.
It is important to notice that $\Delta_{\rm RSD}$ receives contributions from three terms:
the determinant $\sqrt{-g_{\rm R}(x_{\rm R})}$, the spatial Jacobian determinant of the mapping from real to redshift space and  $[a_{\rm R}(x_{\rm R})]^3\,n^{\rm R}_{\rm g}(x_{\rm R})$.

In order to write explicitly the above terms we need to choose a gauge. To correctly incorporate galaxy bias in the expression for the overdensity  we should treat the galaxy bias using the synchronous-comoving gauge.
This gauge is entirely appropriate to describe the matter overdensity. Indeed, in this gauge, the bias is defined in the rest frame of CDM which is assumed to coincide with the rest frame of galaxies on large scales. 
Finally, in $\Lambda$CDM, the CDM rest frame is defined in the synchronous-comoving gauge, in which the galaxy and matter overdensities are gauge invariant, e.g. see \cite{Jeong:2011as}.
The synchronous-comoving gauge is defined by $g_{00}=-1$,    $g_{0i}=0$ and $v^{i}=0$ (where $v^i$  is the galaxy peculiar velocity) and, at linear order, it turns out
\begin{equation}\label{sync}
ds^2=a_{\rm R}^2(\eta_{\rm R})\Big\{-\ud\eta_{\rm R}^2+\Big[\big(1-2{\cal R}
\big)\delta_{ij}+2\partial_i \partial_j E\Big] \ud x_{\rm R}^i \ud x_{\rm R}^j \Big\}.
\end{equation}
where, in $\Lambda$CDM, we have ${\cal R}'=0$
\cite{Ma:1995ey,Matarrese:1997ay} (a prime denotes
$\partial_\eta$)\footnote{For example, see Appendix A of Ref.  \cite{Jeong:2011as}, 
where the authors describes clearly the relation among the common parametrizations of synchronous-comoving gauge and relating all metric perturbations to the matter density perturbation in the adiabatic case.}.

We can write the observed overdensity at observed redshift $z$ and in the unit direction ${\bf n}$ as
\begin{equation}
\label{delta_g} 
\Delta_{\rm g} ( {\bf n},z) =  {\Delta}_{\rm loc} ( {\bf n},z)+
{\Delta}_\kappa( {\bf n},z)+ {\Delta}_{\rm I} ( {\bf n},z) + \Delta_o ( {\bf n},z) \;.
 \end{equation}
Here $ {\Delta}_{\rm loc} $ is a local term  evaluated at the source,
which includes the galaxy density perturbation, the redshift
distortion and the change in volume entailed by the redshift
perturbation. $ {\Delta}_\kappa$ is the weak lensing convergence
integral along the line of sight, $ {\Delta}_{\rm I}$ is a time
delay integral along the line sight and $\Delta_o$  incorporates all the terms  that are evaluated at the observer. In the gauge, i.e. using Eq. (\ref{sync}), we
have \cite{Jeong:2011as}
\bea
\label{delta-loc}{\Delta}_{\rm loc} &=&b \delta + \left[b_e -
\left(1+2\mathcal{Q}\right)+\frac{(1+ {z})}{H}\frac{dH}{dz}
-\frac{2}{ {\chi}}\left(1-\mathcal{Q}\right)\frac{(1+ {z})}
{H}\right]\left( \p_\| E' + E'' \right) \nonumber\\
&&- \frac{(1+ {z})}{H} \p^2_\| E' -\frac{2}{
{\chi}}\left(1-\mathcal{Q}\right)\left(\chi {\cal R}+E'\right)\;, \\
\label{delk}
 {\Delta}_\kappa &=& \left(1-\mathcal{Q}\right)\nabla^2_\perp
 \int_0^{ {\chi}} d\tilde\chi  \left( {\chi}-\tilde\chi\right)
 \frac{ {\chi}}{\tilde\chi}\left(E''-{\cal R} \right)\;,\\
  \label{deli}
 {\Delta}_{\rm I} &=& -\frac{2}{ {\chi}}\left(1-\mathcal{Q}\right)
 \int_0^{ {\chi}} d\tilde\chi\left(E''-{\cal R} \right) \nonumber \\
&&+  \left[b_e - \left(1+2\mathcal{Q}\right)+\frac{(1+
{z})}{H}\frac{dH}{dz}-\frac{2}{{\chi}}\left(1-\mathcal{Q}\right)\frac{(1+ {z})}{H}\right]
\int_0^{ {\chi}} d\tilde\chi E'''\;,\\
\label{del_o}
{\Delta}_o  &=&\left[3-b_e - \frac{(1+ {z})}{H}\frac{dH}{dz}
+ \frac{2}{ {\chi}}\left(1-\mathcal{Q}\right)\frac{(1+ {z})}
{H}\right] (\p_\| E')_o \\
&& + \left[-b_e +
\left(1+2\mathcal{Q}\right)-\frac{(1+ {z})}{H}\frac{dH}{dz}
+\frac{2}{ {\chi}}\left(1-\mathcal{Q}\right)\frac{(1+ {z})}
{H}\right] (E'')_o + \frac{2}{ {\chi}}\left(1-\mathcal{Q}\right) (E')_o\;,
\eea
where $\chi(z)$ is the comoving distance, $b( {z})$ is the bias,
\begin{equation}\label{Q}
\mathcal{Q}( {z}) = {\p \ln N_{\rm g} \over \p \ln \mathcal{L}}
\Big|_{\mathcal{ L}=\mathcal{L}_{\rm lim}}, 
\end{equation} 
 is the magnification bias \cite{Jeong:2011as}
 and
\begin{equation}
b_e ( {z})=-(1+ {z}) {\p \ln  N_{\rm g}\over \p z}
\end{equation}
is the evolution bias.
Here  $N_{\rm g}=a^3n_{\rm g}$ denotes the comoving number density of galaxies with luminosity larger than $L$ 
and the derivative is evaluated at the (redshift-dependent) limiting luminosity of the survey.\footnote{For 
simplicity, we are assuming that the list of targets for spectroscopic observations is flux limited. In case also a size cut 
is applied, another redshift-dependent function should be added to ${\mathcal Q}$ since gravitational lensing also alters the size of galaxy images \citep{Schmidt:2009rh}.} 
Finally, the directional
derivatives are defined as
\bea\label{nabla^2_perp}
\p_\|= n^j \p_j,~~~\p_\|^i =
n^i \p_\|\;, ~~~\p_\|^2 =
\p_{\| i}\p_\|^i =\p_\| \p_\|, ~~ \p_{\perp}^i = (\delta^{ij} -  n^i n^j )\p_j \;,~~~
 \nabla^2_\perp = \p_{\perp i}\p_\perp^i =\nabla^2 - \p_\|^2 - 2 {\chi}^{-1}\p_\|.
\eea
The local term $\Delta_{\rm loc}$ contains the Newtonian local terms, and in addition some GR corrections. The line of sight term $\Delta_{\rm I}$ is a pure GR correction. The lensing term $\Delta_\kappa$ is the same as in  the Newtonian analysis. 
It is useful to relate the metric perturbations to the matter density
contrast in synchronous gauge.  Removing the residual gauge ambiguity and consequently and using \be E''+aHE'-4\pi G \rho_m E =0\,,\ee
we obtain
\bea \label{EqE}
E'     &=&-\frac{H}{(1+z)}f\nabla^{-2}\delta, \\
E''    &=&-\frac{H^2}{(1+z)^2} \Big(\frac{3}{2}
\Omega_{\rm m}-f\Big)\nabla^{-2}\delta, \\
E'''   &=&-3 \frac{H^3}{(1+z)^3}\Omega_{\rm m}
\left(f-1\right)\nabla^{-2}\delta ,\\
{\cal R} &=&  \frac{H^2}{(1+z)^2} \Big(\frac{3}{2}
\Omega_{\rm m}+f\Big) \nabla^{-2}\delta.
\eea
Here $\Omega_{\rm m}(z)$ is the matter density and  $f(z)$ is the growth rate,
\begin{equation}
f={d\ln D \over d\ln a},~~~ \delta({\bf x},z)= \delta({\bf x},0){D(z) \over D(0)},
\end{equation}
where $D $ is the growing mode of $\delta$.
For intensity mapping surveys of the H{\sc I} 21cm emission (e.g. \cite{Abdalla:2004ah}), \cite{Hall:2012wd, Alonso:2015uua} pointed out that we can use the above ration assuming $\mathcal{Q} = 1$ and hence $\Delta_\kappa=0$.
In other words,
\be
\Delta_{\rm IM}( {\bf n},z) =\Delta_{\rm g}( {\bf n},z, \Q=1)\;.
\ee

\section{Dipole}\label{Dipole-Sec}

Via \cite{Scaccabarozzi:2018vux} we know that the most important  contribution in ${\Delta}_o$ is the dipole and all the other terms are negligible (at least for $z<5-10$) for scales less than $1/H_0$. [For a further discussion about this point see the comment below Eq. (\ref{eq:psis}).] In other words, we are able to simplify this quantity as
\be \label{Dv_||o}
{\Delta}_o \simeq {\Delta}_{v_\| o}=\left[3-b_e - \frac{(1+ {z})}{H}\frac{dH}{dz}
+ \frac{2}{ {\chi}}\left(1-\mathcal{Q}\right)\frac{(1+ {z})}
{H}\right] (\p_\| E')_o \;.
\ee
Then from now on we will consider only the following quantity
\be\label{Delta_g}
\Delta_{\rm g}= \Delta+  {\Delta}_{v_\| o}  \quad {\rm where} \quad  \Delta= {\Delta}_{\rm loc}+  {\Delta}_\kappa +  {\Delta}_{\rm I} 
\ee

In this case we have to generalise the results computed in \cite{Bertacca:2012tp} and we need to understand if this local effect is really relevant and/or the same order of GR and wide-angle contributions.
Here below we  shortly review the results obtained in Refs.\  \cite{Bertacca:2012tp, Raccanelli:2013multipoli} (see also \cite{Raccanelli:2013gja} where they analyzed in details the integrated effects),  and in the section \ref{Analysis-dipole} we study the dipole effect within the two point correlation function.

\section{Redshift-space correlation function using only $\Delta$.}
\label{RSDcorrelation}

\begin{center}
\begin{figure*}[htb!]
\includegraphics[width=1.0\columnwidth]{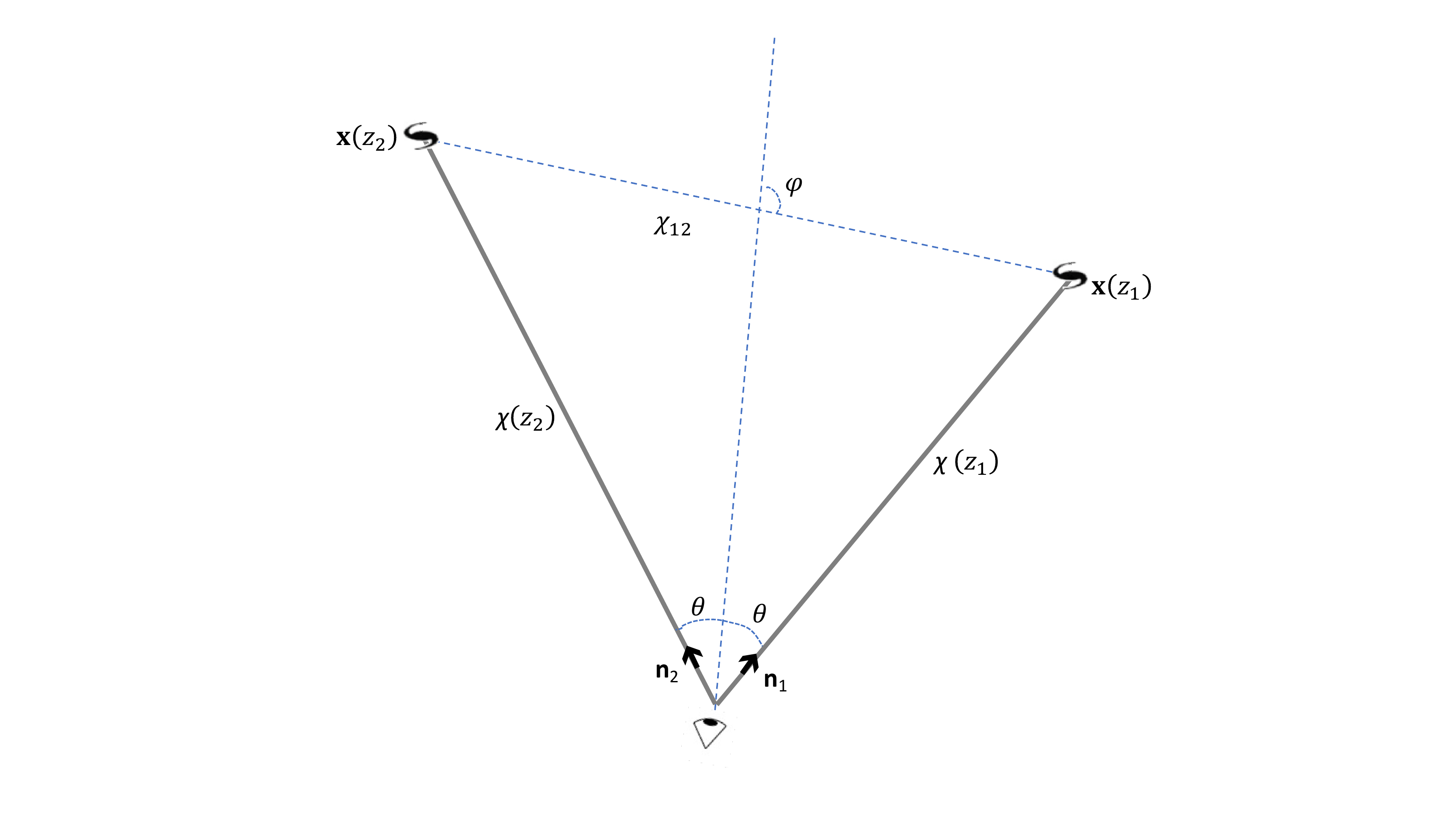}
\caption{ Geometry of the problem: the triangle formed by the observer and the pair of galaxies on the lightcone.
}\label{lcone2}
\end{figure*}
\end{center}

First of all, let us start to compute the observed galaxy correlation function  (see Fig.~\ref{lcone2}):
\be \label{xi}
\xi_\Delta( {\bf x}_1, {\bf x}_2) = \xi_\Delta( {\bf n}_1, {\bf n}_2, z_1,z_2) = \langle \Delta( {\bf n}_1,z_1) \Delta( {\bf n}_2,z_2)  \rangle\,.
\ee
where ${\bf x}$ is related to the comoving distance $\chi$ by
\begin{equation}
{\bf x}=\chi(z) {\bf n}, ~~\chi(z)=\int_0^z{d  z' \over H(z')}.
\end{equation}
In   \cite{Bertacca:2012tp} the authors applied the decomposition used in previous analyses based on  \cite{Szalay:1997cc,Bharadwaj:1998bq, Matsubara:1999du, Szapudi:2004gh, Papai:2008bd}
where they expanded the redshift space correlation function using tripolar spherical harmonics, with the basis  following functions
\bea\label{tripo}
&& S_{\ell_1\ell_2 L}( { {{\bf n}}}_1,  { {{\bf n}}}_2, { {{\bf
n}}}_{12}) =\left[ {(4\pi)^3 \over (2\ell_1+1) (2\ell_2+1)
(2L+1)} \right]^{1/2}\! \sum_{m_1,m_2,M} \left(
\begin{array} {ccc} \ell_1 & \ell_2 & L \\ m_1 & m_2 & M
\end{array} \right)
Y_{\ell_1 m_1}( { {{\bf n}}}_1) Y_{\ell_2 m_2}( { {{\bf n}}}_2) Y_{L
M}( { {{\bf n}}}_{12}),  \nonumber\\
\eea
where
$$ -\ell_1 \leq m_1 \leq \ell_1\;, \quad -\ell_2 \leq m_2 \leq \ell_2\;, \quad -L \leq M \leq L, \quad {\rm and } \quad
\left(
\begin{array} {ccc} \ell_1 & \ell_2 & \ell_3 \\ m_1 & m_2 & m_3
\end{array} \right)
$$
is the Wigner 3$j$ symbol (see also \cite{Raccanelli:2010hk, Samushia:2011cs, Raccanelli:2013multipoli, Raccanelli:2013gja, Raccanelli:2016avd}). 
Then  the  correlation functions in redshift space can be written in the following way:
 \bea
\label{xi} \xi_{AB}( {\bf x}_1, {\bf x}_2) = \langle
\Delta_A( {\bf x}_1) \Delta_B( {\bf x}_2)  \rangle = \xi_{BA}({\bf x}_2, {\bf x}_1), \quad {\rm where}  \quad A,B= {\rm loc,
\kappa,I}\,. \label{xiab}
\eea

First of all, let us define the tensor $\mathcal A$ which spherical transforms the matter overdensity as \cite{Szalay:1997cc}
\be
\label{A-tensor} \mathcal{A}^n_\ell ({\bf x},z)= \int
\frac{d^3k}{(2\pi)^3} \, {\mathcal{P}_\ell ( {{\bf n}}
\cdot  {\hat{\bf k}}) \over (ik)^n}  ~ {\rm e}^{i{\bf k} \cdot {\bf x}}\;
\delta({\bf k},z)\,,
\ee
where ${\bf x}=\chi {\bf n}$, $\mathcal{P}_\ell$ is a Legendre polynomial. 
For simplicity, we start with considering only local terms. Using the decomposition relation by Eq. (\ref{A-tensor}), Eq. (\ref{delta-loc}) turns out
\bea
\label{delta_loc-calA}
\Delta_{\rm loc} = b\left[ \left(1+\frac{1}{3}\beta\right)
\mathcal{A}^0_0+ \gamma \mathcal{A}^2_0  + \frac{\beta\alpha}{\chi}
\mathcal{A}^1_1  +\frac{2}{3}\beta \mathcal{A}^0_2,\label{deltas}\right]
 \eea
 where
\bea
\label{alpha}
\alpha(z)&=& \alpha_{\rm Nwt} (z) -  \frac{\chi(z) H(z)}{(1+z)} \left[\frac{3}{2}\Omega_{\rm m} (z) -
1-2\mathcal{Q}(z)\right] \; , \\
\label{beta}
\beta(z)  &=& \frac{f(z)}{b(z)},  ~~~~ f=-{d \ln D(z) \over d \ln (1+z)}\;, \\
\label{gamma}
\gamma(z) &=& \frac{H(z)}{(1+z)} \left\{\frac{H(z)}{(1+z)}
\left[\beta(z) -\frac{3}{2}\frac{\Omega_{\rm m} (z)}{b(z)}
\right]b_e(z) + \frac{3}{2}\frac{H(z)}{(1+z)} \beta(z) \big[
\Omega_{\rm m} (z) - 2 \big]
\right.  \nonumber \\
& & \left. -\frac{3}{2} \frac{H(z)}{(1+z)}  \frac{\Omega_{\rm m}
(z)}{b(z)} \left[1-4\mathcal{Q}(z) + \frac{3}{2}\Omega_{\rm m}
(z)\right] +\frac{3}{\chi(z)}
\big[1-\mathcal{Q}(z)\big]\frac{\Omega_{\rm m} (z)}{b(z)} \right\} \,. 
\label{eq:gammaz}
\eea
Here
\be
\label{alpha_nwt}
{\alpha_{\rm Nwt}(z) \over  \chi(z)}= - \frac{H(z)}{(1+z)}\left\{b_e(z)-\frac{2}{\chi(z)}\big[1-\mathcal{Q}(z)\big]\frac{(1+z)}{H(z)}\right\}=  \frac{d \ln{N_{\rm g}}}{d \chi}+ \big[1-\mathcal{Q}(z)\big]\frac{2}{\chi}\;.
\ee
is the Newtonian usual part of $\alpha$ considered in \cite{Hamilton:1997zq}. Note that considering a $\Lambda$CDM
 background, we have used in the above relations the following identity $(1+z)/H \, (\p  H/\p z) = 3 \Omega_{\rm m}(z)/2$. Correlating the tensor $\mathcal A$ [defined in Eq. (\ref{A-tensor}) ], we have
\bea \label{calS}
{\mathcal S}_{\ell_1\ell_2}^{n_1+n_2}(z_1,z_2;{\bf n}_1,{\bf n}_2)&\equiv&
\langle \mathcal{A}^{n_1}_{\ell_1} ( {\bf x}_1,
{z}_1)\mathcal{A}^{n_2}_{\ell_2} ( {\bf x}_2, {z}_2) \rangle \nonumber\\
&=&
(-1)^{\ell_2} \int  \frac{d^3k}{(2\pi)^3} (ik)^{-(n_1+n_2)}
\mathcal{P}_{\ell_1}( {\hat{\bf k}} \cdot   {\bf n}_1)
\mathcal{P}_{\ell_2}( {\hat{\bf k}} \cdot   {\bf n}_2)
\exp{\left(i{\bf k} \cdot  {\bf x}_{12}\right)} \; P_\delta(k;
{z}_1, {z}_2)\;,
\eea
where ${\bf x}_{12}= {\bf x}_1- {\bf x}_2 \equiv \chi_{12} {\bf n}_{12}$ and
 $P_\delta(k; {z}_1, {z}_2)$ it the power spectrum. Here we have used the identity $\mathcal{P}_{\ell}( -{\hat{\bf k}} \cdot   {\bf n})=(-1)^\ell \mathcal{P}_{\ell}( {\hat{\bf k}} \cdot   {\bf n})$.
Expanding  in spherical harmonics 
\be  \label{LegandrePol}
\mathcal{P}_\ell ( {{\bf n}} \cdot  \hat{\bf k}) =\frac{4\pi}{2\ell+1}\sum_{m=-\ell}^\ell  Y^*_{\ell m} (\hat {\bf k}) Y_{\ell m} ( {\bf n})
 \ee
and 
\be \label{expkx}
{\rm e}^{i{\bf k} \cdot
{\bf x}}=4\pi \sum_{\ell=0}^{\infty} \sum_{m=-\ell}^\ell  i^\ell j_\ell (\chi k)Y^*_{\ell m} (\hat {\bf k}) Y_{\ell m} ( {\bf n}) \,,
\ee
  and applying the Gaunt
integral 
\begin{eqnarray}
\label{gaunt}
 \int {\rm d}^2 \hat {\bf k} ~ Y_{\ell_1 m_1} (\hat {\bf k})  Y_{\ell_2 m_2} (\hat {\bf k}) Y_{\ell_3 m_3} (\hat {\bf k})
 &=&\sqrt{\frac{(2\ell_1+1)(2\ell_2+1)(2\ell_3+1)}{4\pi}} 
 \left( \begin{array}{ccc} \ell_1 & \ell_2 & \ell_3 \\ 0 & 0 & 0 \end{array} \right)
\left( \begin{array}{ccc} \ell_1 &\ell_2& \ell_3 \\ m_1 & m_2 &m_3 \end{array} \right) \;,
\end{eqnarray}
 we find
\bea
{\mathcal S}_{\ell_1\ell_2}^{n_1+n_2}(z_1,z_2;{\bf n}_1,{\bf n}_2) &=&\sum_{L} (-1)^{\ell_2} \; i^{L-n_1-n_2} 
\left( \begin{array} {ccc} \ell_1 & \ell_2 & L \\ 0 & 0 & 0 \end{array} \right) 
S_{\ell_1\ell_2L}(
{\bf n}_1,  {\bf n}_2, {\bf n}_{12})\,
\xi_L^{n_1+n_2}(\chi_{12};z_1, z_2) , \nonumber\\ \\ 
&& |\ell_1-\ell_2|
\leq L \leq \ell_1+\ell_2, \label{aa}\nonumber
\eea
where we used the tripolar basis defined in \eqref{tripo}.
Here, we have defined
\begin{equation}\label{xiLn}
\xi_L^{n}(\chi; z_1, z_2) = \int \frac{dk}{2\pi^2} k^{2-n}
j_L(\chi k) \, P_\delta(k; z_1, z_2)
\end{equation}
which describes the spherical Bessel transformation of the matter power spectrum $P_\delta(k; z_1, z_2)$~\cite{Szalay:1997cc}.
Finally, using the tripolar decomposition of $\xi_{loc}$, we find 
\begin{equation}
\label{eq:xi_loc}
{\xi}_{\rm loc}( {\bf x}_1, z_1, {\bf x}_2, z_2) = b(z_1) b(z_2)
\sum_{\ell_1,\ell_2,L,n}
{\left(B_{\rm loc\,loc}\right)}_{n}^{\, \ell_1\ell_2L}( {\chi}_1, {\chi}_2)\, 
S_{\ell_1\ell_2L}( {{{\bf n}}}_1, { {{\bf n}}}_2,  { {{\bf n}}}_{12}) \,
\xi_L^{\, n}(\chi_{12}; z_1, z_2) \, .
\end{equation}
  The ${\left(B_{\rm loc\,loc}\right)}_{n}^{\, \ell_1\ell_2L}$ coefficients contain the local corrections due to the functions $\alpha$, $\beta$ and\footnote{Precisely, in \cite{Bertacca:2012tp}, it used a different notation. Indeed, ${\left(B_{\rm loc-loc}\right)}_{n}^{\, \ell_1\ell_2L}$ here is ${B_{\rm ss}}_{n}^{\, \ell_1\ell_2L}$ there.} $\gamma$ \cite{Bertacca:2012tp}.

Furthermore another very interesting expression of the local correlation function can be achieved if we rewrite the  tripolar spherical harmonics basis $S_{\ell_1\ell_2L}$  as the combination of two Legendre polynomials
which depend on the angular dependence  $\varphi$ and  $\theta$, see Fig.\ \ref{lcone2}. This method appears to be more natural for looking at the behaviour of the  multipoles. 
Then we can obtain the following decomposition 
\begin{equation}\label{xisum}
\xi_{\rm loc} (z_2, \theta, \varphi) = b(z_1) b(z_2) \sum_{\tilde{L} , \tilde{\ell}} \; C_{\tilde{L} \tilde{\ell}_3} (z_2, \theta, \varphi) \; \mathcal{P}_{\tilde{L} }(\cos\varphi) \; \mathcal{P}_{ \tilde{\ell_3}}(\cos\theta) 
\end{equation}
and the coefficients $C_{\tilde{L} \tilde{\ell}_3}$ are given \cite{Raccanelli:2013multipoli}
\bea
\label{eq:psis}
C_{00}&=&\left(1+\frac{\beta_1}{3}+\frac{\beta_2}{3} + \frac{29}{225} \beta_1 \beta_2 \right) \xi_0^0 -\left(\gamma_1+\gamma_2+\frac{1}{3}\beta_1\gamma_2+\frac{1}{3}\gamma_1\beta_2+\frac{1}{9}\beta_1\beta_2\frac{\alpha_1}{\chi_1}\frac{\alpha_2}{\chi_2}\right) \xi^2_0 + \nonumber \\
&&+ \gamma_1 \gamma_2 \xi^4_0+\sin\varphi\sin \theta \left[\left(1+\frac{1}{3}\beta_1\right)\beta_2\frac{\alpha_2}{\chi_2}+\left(1+\frac{1}{3}\beta_2\right)\beta_1\frac{\alpha_1}{\chi_1}\right] \xi_1^1 + \nonumber \\ 
&&-\sin\varphi\sin \theta\left(\gamma_1 \beta_2\frac{\alpha_2}{\chi_2} +\beta_1\frac{\alpha_1}{\chi_1} \gamma_2 \right)\xi^3_1 
-\left(\frac{2}{9}\beta_1+\frac{2}{9}\beta_2+\frac{44}{315}\beta_1\beta_2\right)\xi^0_2 + 
\nonumber \\ 
&&+\frac{2}{9}\left(\beta_1\beta_2\frac{\alpha_1}{\chi_1}\frac{\alpha_2}{\chi_2}+\gamma_1\beta_2+\beta_1 \gamma_2\right)\xi^2_2 + \frac{32}{1575}\beta_1 \beta_2 \xi^0_4 \;, \nonumber \\ \nonumber\\
 C_{11}&=& 
\left[-\left(1+\frac{7}{25}\beta_1\right)\beta_2\frac{\alpha_2}{\chi_2}+\left(1+\frac{7}{25}\beta_2\right)\beta_1\frac{\alpha_1}{\chi_1}\right] \xi_1^1 + \left( \gamma_1 \beta_2\frac{\alpha_2}{\chi_2} - \beta_1\frac{\alpha_1}{\chi_1}  \gamma_2 \right) \xi^3_1  + \nonumber\\
&&+2 \sin\varphi\sin \theta\left(\beta_2-\beta_1\right)\xi^0_2 + 2 \sin\varphi\sin \theta\left(\beta_1\gamma_2-\gamma_1\beta_2\right)\xi^2_2 + \frac{2}{25}\left( \beta_1\frac{\alpha_1}{\chi_1}\beta_2-\beta_1 \beta_2\frac{\alpha_2}{\chi_2} \right) \xi^1_3\;, \nonumber \\
   C_{02}&=& - \frac{16}{315} \beta_1 \beta_2 \xi_0^0+ \frac{4}{9}\beta_1\beta_2\frac{\alpha_1}{\chi_1}\frac{\alpha_2}{\chi_2} \xi^2_0 - \frac{8}{15} \sin\varphi\sin \theta\beta_1\beta_2 \left(\frac{\alpha_1}{\chi_1}+\frac{\alpha_2}{\chi_2}\right) \xi_1^1 + \nonumber \\ 
 &&+ \left(\frac{2}{9}\beta_1+\frac{2}{9}\beta_2+\frac{100}{441}\beta_1\beta_2\right)\xi^0_2  -\frac{2}{9}\left(\beta_1\beta_2\frac{\alpha_1}{\chi_1}\frac{\alpha_2}{\chi_2}+\gamma_1\beta_2+\beta_1 \gamma_2\right)\xi^2_2\; +\nonumber\\
&&  + \frac{2}{15} \sin\varphi\sin \theta\beta_1\beta_2 \left(\frac{\alpha_1}{\chi_1}+\frac{\alpha_2}{\chi_2}\right) \xi_3^1- \frac{88}{2205}\beta_1 \beta_2 \xi^0_4 \;, \nonumber \\ 
 C_{20} &=& \left(\frac{2}{9}\beta_1+\frac{2}{9}\beta_2+\frac{4}{21}\beta_1\beta_2\right)\xi^0_2 -\frac{2}{9}\left(3\beta_1\beta_2\frac{\alpha_1}{\chi_1}\frac{\alpha_2}{\chi_2}+\gamma_1\beta_2+\beta_1 \gamma_2\right)\xi^2_2\; +\nonumber\\
&& + \frac{2}{3} \sin\varphi\sin \theta\beta_1\beta_2 \left(\frac{\alpha_1}{\chi_1}+\frac{\alpha_2}{\chi_2}\right) \xi_3^1 - \frac{8}{63}\beta_1 \beta_2 \xi^0_4\;, \nonumber \\
  C_{22} &=& - \left(\frac{8}{9}\beta_1+\frac{8}{9}\beta_2+\frac{16}{21}\beta_1\beta_2\right)\xi^0_2 +\frac{8}{9}\left(\gamma_1\beta_2+\beta_1 \gamma_2\right)\xi^2_2\;  + \frac{8}{63}\beta_1 \beta_2 \xi^0_4 \;, \nonumber \\  
 C_{13} &=& \frac{8}{25} \beta_1\beta_2 \left(\frac{\alpha_1}{\chi_1} -\frac{\alpha_2}{\chi_2}\right)  \xi^1_1  -  \frac{2}{25} \beta_1\beta_2 \left( \frac{\alpha_1}{\chi_1} -\frac{\alpha_2}{\chi_2}\right)  \xi^1_3\;, \nonumber \\
 C_{31} &=& - \frac{2}{5} \beta_1\beta_2 \left(\frac{\alpha_1}{\chi_1} -\frac{\alpha_2}{\chi_2}\right)  \xi^1_3\;, \nonumber \\
 C_{04} &=& \frac{64}{525}\beta_1 \beta_2 \xi^0_0 - \frac{64}{735}\beta_1 \beta_2 \xi^0_2 + \frac{24}{1225}\beta_1 \beta_2 \xi^0_4\;, \nonumber \\
 C_{40} &=& \frac{8}{35}\beta_1 \beta_2 \xi^0_4 \; , 
\eea
where a subscript $i$ denotes evaluation at $z_i$. 

At this stage, it is important to make the following comment; as it was pointed out in \cite{Bertacca:2012tp} for  $n=4$ and $L=0$, $\xi_L^{n}$ is divergence and  as correctly observed in \cite{Scaccabarozzi:2018vux}, it is not a real divergence\footnote{In order to  prove this (see also \cite{Bertacca:2012tp}), for simplicity, let us consider the angular correlation, with $z_1=z_2\equiv z$.
Rewriting $\xi_L^{n}$ as
\begin{equation}\label{xiLn2}
\xi_L^{n}(\chi_{12}; z) = \int_{k_{\rm min}}^{1/\chi_{12}}  \frac{dk}{2\pi^2} k^{2-n}
j_L(\chi_{12} k) \, P_\delta(k; z)+  \int_{1/\chi_{12}}^\infty  \frac{dk}{2\pi^2} k^{2-n}
j_L(\chi_{12} k) \, P_\delta(k; z),
\end{equation}
where we impose a large-scale cutoff $k_{\rm min} $, which we take as $k_{\rm min} < H_0/2$. 
Let us take $\chi_{12}  \simeq {2(1+z)/ H^(z)}$. 
 Starting from the second integral, for $L>0$, can be approximated  as
\begin{eqnarray}
&& \int_{1/\chi_{12}}^\infty  \frac{dk}{2\pi^2} k^{2-n} j_L(\chi_{12} k) \, P_\delta(k; z) \sim \frac{k_L^{2-n}}{2 \pi^2} \frac{P_\delta(k_L; z)}{\chi_{12}} I_L\;,
~~~ k_L= {(L+1/2)\over \chi_{12}},~~~
I_L = \int_0^\infty  j_L(y) dy = \frac{\sqrt{\pi}}{2}
\frac{\Gamma[(L+1)/2]}{\Gamma[(L+2)/2]}\;.
\end{eqnarray}
For $L=0$, the integral vanishes because ${k_L}|_{L=0}<1/\chi_{12}$. 
For the first integral in Eq. (\ref{xiLn2}), for $k_{\rm min} \le k \le 1/ \chi_{12}$, we have $P_\delta(k; z)\propto P_{\rm prim}(k) =A k^{n_s}$ since $k \ll k_{\rm eq}$, and let us simplify (at first approximation) $j_L(\chi_{12} k) \sim (\chi_{12} k)^L/ (2L+1)!!$. 
Then, for $k_{\rm min} \to 0$, the first integral  is divergent if
\begin{equation}\label{xiLn2}
n-L \ge 3 + n_s\;.
\end{equation}
It is clear that for $n=4$ and $L=0$, we should require an infrared (IR) cutoff $k_{\rm min} > 0$ since $\xi_0^4$ becomes power-law divergent for $n_s <1$. (If $n_s=1$, there is a logarithmic divergence.) The IR cutoff appears only in the terms of the correlation function
that contain $Y_{LM}$ with $M=L=0$. (In this case $Y_{00} \propto \mathcal{P}_0 \equiv 1$.) }.
However, this issue comes form the fact that we  have  neglected terms evaluated at the observer position, i.e. $\Delta_o$, in the above calculation. Instead, if we consider all terms in Eq.(\ref{delta_g}), the sum of all individually divergent contributions in the correlation function is instead finite in agreement with the equivalence principle \cite{Scaccabarozzi:2018vux}.  The above prescription is still correct if we safely impose an IR cut-off scale, as long as $k_{\rm min} \lesssim H_0$ and $k_{\rm min}  \le k$ when we compute the integrals in Eq. (\ref{xiLn}). (Let us point out that in \cite{Bertacca:2012tp} they  took as $k_{\rm min} \sim H_0/2$.)
Contrarily, as will we see later, the dipole which is a non divergent contribution in  $\Delta_o$  will play an important role within correlation function. 
Precisely, as we see in Section \ref{Analysis-dipole}, the dipole correction will add several new terms and effects on the galaxy two point correlation function.

\subsection{Non-local terms}

The remaining $\xi_{AB}$ all involve integrals along the lines
of sight. The spherical transforms of $\Delta_\kappa, \Delta_I$ are (for further details, see  \cite{Bertacca:2012tp})
\bea 
\Delta_\kappa ({\bf n}, z)&=& b(z)\int^{\chi}{d\tilde\chi} \; \sigma(z,\tilde{z})\left[\mathcal{A}^0_0(\tilde{\bf x},\tilde{z})- \mathcal{A}^0_2(\tilde{\bf x},\tilde{z})-\frac{3}{\tilde{\chi}}\mathcal{A}^1_1(\tilde{\bf x},\tilde{z})\right], \label{dkexp} \\
\Delta_{\rm I} ({\bf n}, z)&=& b(z)\int^{\chi}{d\tilde\chi} \; \mu(z,\tilde{z})\mathcal{A}^2_0(\tilde{\bf x},\tilde{z}), \label{diexp}
\eea
where 
\bea
\sigma( z,\tilde z) &\equiv & -2\frac{H^2(\tilde{z})}{(1+\tilde{z})^2} \frac{\left({\chi}-\tilde\chi\right)\tilde\chi}{ {\chi}}\frac{\big[1-\mathcal{Q}(z)\big]}{b( {z})}   \Omega_{\rm m} (\tilde z) , \label{sigma}\\
\mu( z,\tilde z) &\equiv & 3
\frac{H^2(\tilde{z})}{(1+\tilde{z})^2}\frac{\Omega_{\rm m}
(\tilde{z})}{b( {z})}\Bigg\{\frac{2}{ {\chi}} \big[1-\mathcal{Q}(z)
\big]   - \frac{H(\tilde{z})}{(1+\tilde{z})}
\big[f(\tilde{z})-1\big] \bigg[ b_e( {z}) - \big[1+2\mathcal{Q}(z)\big] +\frac{3}{2} \Omega_{\rm m} ( {z}) \nonumber\\
&&- \frac{2}{
{\chi}}\big[1-\mathcal{Q}( z )\big] \frac{(1+ {z})}{H(
{z})}\bigg]  \Bigg\},  \nonumber\\ \label{mu}
\eea
and $\chi=\chi(z), \tilde\chi=\chi(\tilde z)$. 
Let us point out that in Eq. (\ref{dkexp}) we used the definition (\ref{nabla^2_perp}).
Then the lensing-lensing correlation turns out
\be
{\xi}_{\kappa \kappa}( {\bf x}_1, {\bf x}_2)
=b(z_1)b(z_2) \int^{ {\chi}_1,\chi_2}
{d\tilde\chi_1}{d\tilde\chi_2} \sum_{\ell_1,\ell_2,L,n} \left(B_{\kappa
\kappa}\right)_n^{\ell_1\ell_2L}(
{\chi}_1,\tilde\chi_1; {\chi}_2, \tilde\chi_2)\,
S_{\ell_1\ell_2L}( { {{\bf n}}}_1, { {{\bf n}}}_2, {\tilde{\bf
n}}_{12})\, \xi_L^{n}(\tilde\chi_{12}; \tilde{z}_1, \tilde{z}_2),
\label{eq:xi_kk}
\end{equation}
and for the ${\rm II}$ correlation we find
\be
{\xi}_{\rm II}( {\bf x}_1, {\bf x}_2) =b(z_1)b(z_2)
\int^{ {\chi}_1,\chi_2} {d\tilde\chi_1}{d\tilde\chi_2}
\sum_{\ell_1,\ell_2,L,n} \left(B_{\rm II}\right)_n^{\ell_1\ell_2L}( {\chi}_1,\tilde\chi_1; {\chi}_2,
\tilde\chi_2)\, S_{\ell_1\ell_2L}( { {{\bf n}}}_1, { {{\bf n}}}_2,
{\tilde{\bf n}}_{12})\, \xi_L^{n}(\tilde\chi_{12}; \tilde{z}_1,
\tilde{z}_2). \label{eq:xi_II}
\end{equation}
The integration variables ${\tilde\chi_{12}},{\tilde{\bf n}}_{12}$
are given by
\be
{\tilde\chi_{12}}{\tilde{\bf n}}_{12}= {\chi}_{12} { {{\bf
n}}}_{12}+ \left(\tilde\chi_1 - {\chi}_1\right) { {{\bf
n}}}_1-\left(\tilde\chi_2 -
{\chi}_2\right) { {{\bf n}}}_2  ,  \quad\quad {\rm and} \quad\quad
\tilde\chi^2_{12} = { \tilde\chi_1^2+ \tilde\chi_2^2 + \frac{
\tilde\chi_1\tilde\chi_2}{ {\chi}_1 {\chi}_2}\left[
{\chi}_{12}^2-\left( {\tilde\chi}_1^2+ {\tilde\chi}_2^2\right)\right]}.
\ee

Similarly, we find:
\bea
{\xi}_{\rm loc\,I}( {\bf x}_1, {\bf x}_2) = b(z_1)b(z_2)
\int^{ {\chi}_2}{d\tilde\chi_2} \sum_{\ell_1,\ell_2,L,n} \left(B_{\rm loc\,I}\right)_n^{\ell_1\ell_2L}( {\chi}_1; {\chi}_2,
\tilde\chi_2) S_{\ell_1\ell_2L}( { {{\bf n}}}_1, { {{\bf n}}}_2,
{{\bf n}}_{1 \tilde{2}}) \, \xi_L^{n}(\chi_{1 \tilde{2}}; {z}_1,
\tilde{z}_2)  \;, \label{eq:xi_sI}\\
{\xi}_{\rm loc\,\kappa}( {\bf x}_1, {\bf x}_2) =
b(z_1)b(z_2) \int^{ {\chi}_2}{d\tilde\chi_2}
\sum_{\ell_1,\ell_2,L,n} \left(B_{\rm loc\,\kappa}\right)_n^{\ell_1\ell_2L}( {\chi}_1; {\chi}_2, \tilde\chi_2)
S_{\ell_1\ell_2L}( { {{\bf n}}}_1, { {{\bf n}}}_2, {{\bf n}}_{1\tilde{2}}) \, \xi_L^{n}(\chi_{1 \tilde{2}}; {z}_1, \tilde{z}_2) \;,
\label{eq:xi_sk}\\
{\xi}_{\rm \kappa\,I}( {\bf x}_1, {\bf x}_2) =
b(z_1)b(z_2) \int^{ {\chi}_1,\chi_2}
{d\tilde\chi_1}{d\tilde\chi_2} \sum_{\ell_1,\ell_2,L,n} \left(B_{\rm \kappa\, I}\right)_n^{\ell_1\ell_2L}(
{\chi}_1,\tilde\chi_1; {\chi}_2, \tilde\chi_2)\,
S_{\ell_1\ell_2L}( { {{\bf n}}}_1, { {{\bf n}}}_2, {\tilde{\bf
n}}_{12})\, \xi_L^{n}(\tilde\chi_{12}; \tilde{z}_1, \tilde{z}_2)\;,\nonumber\\
\label{eq:xi_kI}
\eea
where
\be
\chi_{1 \tilde{2}}{{\bf n}}_{ {1}\tilde 2}= {\left(
{\chi}_2-\tilde\chi_2 \right) { {{\bf n}}}_2 +  {\chi}_{12} {
{{\bf n}}}_{12}}\;, \quad\quad {\rm and} \quad\quad
\chi_{1 \tilde{2}}^2 ={  {\chi}_1^2+ \tilde\chi_2^2 + \frac{
\tilde\chi_2}{ {\chi}_2}\left[ {\chi}_{12}^2-\left( {\chi}_1^2+
{\chi}_2^2\right)\right]}.
\ee
We can obtain the remaining $\xi_{AB}$ by using the symmetry in \eqref{xiab}. In Ref.\ \cite{Bertacca:2012tp}, the authors have explicitly computed the above coefficients $B_{AB \; n}^{\phantom{AB \;}\ell_1\ell_2L}({\chi}_1,\tilde\chi_1; {\chi}_2, \tilde\chi_2)$, where $A,B={\rm loc},\, \kappa,\, {\rm I}$. (As we have already pointed out above, we have replaced the subscript $\rm s$ used in \cite{Bertacca:2012tp} with $\rm loc$.)

\section{Analysis  of dipole term} \label{Analysis-dipole}

In this section we discuss the main part of this work, where we will discuss the effect of the local group through the dipole at the observer on two point correlation function.
From Eq. (\ref{EqE}) we know
\be
E_o'=(E')_o({\bf n})=-H_0f_0 \left[\nabla^{-2}\delta({\bf x},0)\right]_{{\bf x}\to {\bf 0}}\;,
\ee
where we note that ${\bf x}\to {\bf 0}$ is equivalently to $\chi \to 0$.
Then
\bea
(\p_\| E')_o&=&-H_0f_0 \left[n^i\p_i\nabla^{-2}\delta({\bf x},0)\right]_{{\bf x}\to {\bf 0}}=-H_0f_0 
\left[\int  \frac{d^3k}{(2\pi)^3} \frac{\mathcal{P}_{1}( {\hat{\bf k}} \cdot   {\bf n}) }{ik} \;
\delta({\bf k},0)~ {\rm e}^{i{\bf k} \cdot {\bf x}} \right]_{{\bf x}\to {\bf 0}}\nonumber\\
&=& - H_0f_0 
\int  \frac{d^3k}{(2\pi)^3} \frac{\mathcal{P}_{1}( {\hat{\bf k}} \cdot   {\bf n}) }{ik}\delta({\bf k},0) =-H_0f_0 \left[ {\mathcal A}_1^1 ({\bf x},0)\right]_{{\bf x}\to {\bf 0}}
\eea
and Eq. (\ref{Dv_||o}) turns out
\be
 {\Delta}_{v_\| o}= b(z)\omega_o(z) \left[ {\mathcal A}_1^1 ({\bf x},0)\right]_{{\bf x}\to {\bf 0}} \;.
 \ee
 
In GR, the {rocket effect} contains new terms that depends also on the magnification bias and the expansion rate:
\be
 {\Delta}_{v_\| o}= b(z)\omega_o(z) \int  \frac{d^3k}{(2\pi)^3} \frac{\mathcal{P}_{1}( {\hat{\bf k}} \cdot   {\bf n}) }{ik}\delta({\bf k},0) 
 \ee
where
\be \label{omega0}
 \omega_o(z)=-{H_0f_0\over b(z)}  \left[3-b_e(z) - {3 \over 2} \Omega_{\rm m}(z)
+ \frac{2(1+ {z})}{ \chi(z) H(z)}\left(1-\mathcal{Q}(z)\right)\right]\;.
\ee

At this point if we want to correlate $\Delta$ with $\Delta_o$, we should compute the tensor $\mathcal S_{\ell 1 \ell_2}^{n_1 n_2} ({\bf x}_1,{\bf x}_2)$ in three different regimes\footnote{We have slightly changed the argument of this tensor in order to simplify  the analysis that we are making here below.}:
\bea
{\rm 1)}\; {\mathcal S}_{\ell 1 \ell_2}^{n_1 n_2} ({\bf x}_1,{\bf x}_2)\Big|_{\small \begin{array}{c} \chi_1 \to 0\\ \chi_2 \to 0
\end{array}}\;, \quad \quad {\rm 2)}\;{\mathcal S}_{\ell 1 \ell_2}^{n_1 n_2} ({\bf x}_1,{\bf x}_2)\Big|_{ \chi_2 \to 0}\;, \quad \quad {\rm 3)}\; {\mathcal S}_{\ell 1 \ell_2}^{n_1 n_2} ({\bf x}_1,{\bf x}_2)\Big|_{ \chi_1 \to 0}\;.
\eea

\subsection{${\mathcal S}_{\ell 1 \ell_2}^{n_1 n_2} ({\bf x}_1,{\bf x}_2)$ for $ \chi_1 \to 0$ and $\chi_2 \to 0$}
Using Eq.\ (\ref{calS}) we have
\bea \label{calS(chi0chi0)}
&&{\mathcal S}_{\ell_1\ell_2}^{n_1+n_2}(0,0;{\bf n}_1,{\bf n}_2)=
(-1)^{\ell_2} \int  \frac{d^3k}{(2\pi)^3} (ik)^{-(n_1+n_2)}
\mathcal{P}_{\ell_1}( {\hat{\bf k}} \cdot   {\bf n}_1)
\mathcal{P}_{\ell_2}( {\hat{\bf k}} \cdot   {\bf n}_2)
 \; P_\delta(k;
{z}_1, {z}_2)\nonumber\\
&&=
(-1)^{\ell_2} \frac{(4\pi)^2}{(2\ell_1+1)(2\ell_2+1)} \sum_{m_1\,m_2}\left\{ Y_{\ell_1 m_1} ( {\bf n}_1) Y_{\ell_2 m_2} ( {\bf n}_2)\left[\int  \frac{d^3k}{(2\pi)^3} (ik)^{-(n_1+n_2)}
Y^*_{\ell_1 m_1} (\hat {\bf k}) \;Y^*_{\ell_2 m_2} (\hat {\bf k})
 \; P_\delta(k;{z}_1, {z}_2)\right]\right\}\nonumber\\
  &&=
(-1)^{\ell_1} \frac{(4\pi)^2}{(2\ell_1+1)^2} \delta^K_{\ell_1 \ell_2}\left[  \sum_{m_1} (-1)^{m_1}Y_{\ell_1 m_1} ( {\bf n}_1) Y_{\ell_1 -m_1} ( {\bf n}_2)\right]\left[\int  \frac{k^2 d k}{(2\pi)^3} (ik)^{-(n_1+n_2)}
 \; P_\delta(k;0, 0)\right]\nonumber\\
   &&=
\frac{(-1)^{\ell_1} i^{-(n_1+n_2)}}{(2\ell_1+1)^2} \delta^K_{\ell_1 \ell_2} \left[\sum_{m_1\,m_2}  Y_{\ell_1 m_1} ( {\bf n}_1) Y^*_{\ell_1 m_1} ( {\bf n}_2)\right]\left[\frac{2}{\pi}\int  k^{2-(n_1+n_2)} dk 
 \; P_\delta(k;0, 0)\right]\nonumber\\
    &&=
\frac{(-1)^{\ell_1} i^{-(n_1+n_2)}}{(2\ell_1+1)} \delta^K_{\ell_1 \ell_2} {\mathcal P}_{\ell_1}({\bf n}_1 \cdot {\bf n}_2)\left[\frac{1}{2\pi^2}\int  k^{2-(n_1+n_2)} dk 
 \; P_\delta(k;0, 0)\right] 
\eea
where in the second  and the last line we used Eq. (\ref{LegandrePol}), in the third  and forth line $Y^*_{\ell m}({\bf n})=(-1)^m Y_{\ell - m}({\bf n})$ and
\be
\int {\rm d}^2 \hat {\bf k} ~ Y_{\ell_1 m_1} (\hat {\bf k})  Y_{\ell_2 m_2} (\hat {\bf k}) =  (-1)^{m_2}\delta^K_{\ell_1 \ell_2} \delta^K_{m_1 -m_2} \;.
\ee
In particular,  for $\langle  {\Delta}_{v_\| o} ({\bf n}_1, z_1) {\Delta}_{v_\| o}({\bf n}_2, z_2) \rangle$, the above relation will be 
\be 
{\mathcal S}_{11}^{2}(0,0;{\bf n}_1,{\bf n}_2)= \frac{1}{3}{\mathcal P}_{1}({\bf n}_1 \cdot {\bf n}_2)\left[\frac{1}{2\pi^2}\int   dk 
 \; P_\delta(k;0, 0)\right]=\frac{1}{3}\xi_0^{\, 2}(0; 0, 0){\mathcal P}_{1}({\bf n}_1 \cdot {\bf n}_2) \;.
\ee
This result is well known and, for the relativistic analysis, it has recently been studied in details in Ref. \cite{Scaccabarozzi:2018vux}.
\subsection{${\mathcal S}_{\ell 1 \ell_2}^{n_1 n_2} ({\bf x}_1,{\bf x}_2)$ for $\chi_2 \to 0$}

In similar way, we have
\bea \label{calS(chi1chi0)}
&&{\mathcal S}_{\ell_1\ell_2}^{n_1+n_2}(z_1,0;{\bf n}_1,{\bf n}_2)=
(-1)^{\ell_2} \int  \frac{d^3k}{(2\pi)^3} (ik)^{-(n_1+n_2)}
\mathcal{P}_{\ell_1}( {\hat{\bf k}} \cdot   {\bf n}_1)
\mathcal{P}_{\ell_2}( {\hat{\bf k}} \cdot   {\bf n}_2)
 \; P_\delta(k; {z}_1, 0)\, {\rm e}^{i {\bf k}\cdot {\bf x}_1}\nonumber\\
 &&=(-1)^{\ell_2} \int  \frac{d^3k}{(2\pi)^3} (ik)^{-(n_1+n_2)}P_\delta(k; {z}_1, 0)
 \Bigg[\frac{4\pi}{2\ell_1+1}\sum_{m_1=-\ell_1}^{\ell_1}  Y^*_{\ell_1 m_1} (\hat {\bf k}) Y_{\ell_1 m_1} ( {\bf n}_1)\Bigg] \Bigg[\frac{4\pi}{2\ell_2+1}\sum_{m_2=-\ell_2}^{\ell_2}  Y^*_{\ell_2 m_2} (\hat {\bf k}) Y_{\ell_2 m_2} ( {\bf n}_2) \Bigg]\nonumber\\
 && \quad\quad\quad\quad \quad \times\quad  ~ \sum_{\ell=0}^{\infty} \sum_{m=-\ell}^\ell ( 4\pi )  i^\ell j_\ell (\chi_1 k)Y^*_{\ell m} (\hat {\bf k}) Y_{\ell m} ( {\bf n}_1)\nonumber\\
 &&={(4\pi)^{3/2} (-1)^{\ell_2}\over \sqrt{(2\ell_1+1)(2\ell_2+1)}} \sum_\ell i^{\ell-(n_1 + n_2)}\sqrt{2\ell+1} \left( \begin{array} {ccc} \ell_1 & \ell_2 & \ell \\ 0 & 0 & 0 \end{array} \right)  \xi_\ell^{\, n_1+n_2}(\chi_{1}; z_1, 0) \sum_{m_2} Y_{\ell_2 m_2} ( {\bf n}_2)  \nonumber\\
  &&\quad\quad\quad \quad\quad \times \quad \sum_{m_1 m} \left( \begin{array}{ccc} \ell_1 &\ell_2& \ell \\ m_1 & m_2 &m \end{array} \right) Y_{\ell_1 m_1} ( {\bf n}_1)Y_{\ell m} ( {\bf n}_1) \nonumber\\
 && ={(4\pi) (-1)^{\ell_2}\over (2\ell_2+1)} \sum_\ell i^{\ell-(n_1 + n_2)}(2\ell+1) {\left( \begin{array} {ccc} \ell_1 & \ell_2 & \ell \\ 0 & 0 & 0 \end{array} \right)}^2  \xi_\ell^{\, n_1+n_2}(\chi_{1}; z_1, 0) \sum_{m_2} (-1)^{m_2} Y_{\ell_2 m_2} ( {\bf n}_2)  Y_{\ell_2 - m_2} ( {\bf n}_1)\nonumber\\
  && ={(-1)^{\ell_2}} ~ {\mathcal P}_{\ell_2}({\bf n}_1 \cdot {\bf n}_2) ~ \sum_\ell i^{\ell-(n_1 + n_2)}(2\ell+1) {\left( \begin{array} {ccc} \ell_1 & \ell_2 & \ell \\ 0 & 0 & 0 \end{array} \right)}^2  \xi_\ell^{\, n_1+n_2}(\chi_{1}; z_1, 0)\,,
\eea
where at the second and line we used Eqs. (\ref{LegandrePol}) and (\ref{expkx}), at fourth line the Gaunt integral, at the sixth line the following identity
\bea 
 \sum_{m_1 m_2} \left( \begin{array}{ccc} \ell_1 &\ell_2& \ell \\ m_1 & m_2 &m \end{array} \right) Y_{\ell_1 m_1} ( {\bf n})Y_{\ell_2 m_2} ( {\bf n}) = (-1)^m 
 \sqrt{
 \frac{(2\ell_1+1)(2\ell_2+1)}{4 \pi (2\ell+1)}
 }Y_{\ell -m} ( {\bf n}) 
\eea
and in the last line we applied again Eq. (\ref{LegandrePol}).
Finally for the dipole term in $\{{\bf n}_2, z_2\}$, we have $\ell_2=n_2=1$ and we find
\be \label{calS(chi1chi0)-2}
{\mathcal S}_{\ell_1 1}^{n_1+1}(z_1,0;{\bf n}_1,{\bf n}_2)=- {\mathcal P}_{1}({\bf n}_1 \cdot {\bf n}_2) ~ \sum_\ell i^{\ell-(n_1 + 1)}(2\ell+1) {\left( \begin{array} {ccc} \ell_1 & 1 & \ell \\ 0 & 0 & 0 \end{array} \right)}^2  \xi_\ell^{\, n_1+1}(\chi_{1}; z_1, 0)\;.
\ee
To check, let us make the following comment;
it is useful to see that for $\chi_1 \to 0$ (and $z_1 \to 0$) we recover Eq. (\ref{calS(chi0chi0)}). Indeed defining  $ \epsilon =  \chi_1 k \ll 1$, i.e. $ \chi_1  \ll 1/k $, we have 
$$ \xi_\ell^{\, n_1+n_2} \sim \frac{ (\chi_1 k)^\ell}{(2\ell+1)!!}\;.$$
 Therefore  it is not zero if $\ell=0$. Then using 
\bea
 \left( \begin{array}{ccc} \ell_1 &\ell_2& 0 \\ 0 & 0 &0 \end{array} \right)={(-1)^{\ell_1}\over \sqrt{2\ell_1+1}} \delta_{\ell_1 \ell_2}
\eea
we recover the previous result.

Now, let us come back to the result in Eq. (\ref{calS(chi1chi0)-2}). 
We note that, from $\Delta$, we have $\ell_1=0,1,2$  and  it  immediately turns out
\bea
{\rm for } ~ ~ \ell_1=0 \quad\quad &\to& \quad\quad {\mathcal S}_{0 1}^{n_1+1}(z_1,0;{\bf n}_1,{\bf n}_2)=- i^{-n_1} \xi_1^{\, n_1+1}(\chi_{1}; z_1, 0) {\mathcal P}_{1}({\bf n}_1 \cdot {\bf n}_2) \nonumber\\
{\rm for } ~ ~ \ell_1=1 \quad\quad &\to& \quad\quad {\mathcal S}_{1 1}^{n_1+1}(z_1,0;{\bf n}_1,{\bf n}_2)=- i^{-(n_1+1)}\left[\frac{1}{3}\xi_0^{\, n_1+1}(\chi_{1}; z_1, 0) -\frac{2}{3}\xi_2^{\, n_1+1}(\chi_{1}; z_1, 0) \right]  {\mathcal P}_{1}({\bf n}_1 \cdot {\bf n}_2)\nonumber\\
{\rm for } ~ ~ \ell_1=2 \quad\quad &\to& \quad\quad {\mathcal S}_{2 1}^{n_1+1}(z_1,0;{\bf n}_1,{\bf n}_2)=- i^{-n_1}\left[\frac{2}{5}\xi_1^{\, n_1+1}(\chi_{1}; z_1, 0) -\frac{3}{5}\xi_3^{\, n_1+1}(\chi_{1}; z_1, 0) \right]  {\mathcal P}_{1}({\bf n}_1 \cdot {\bf n}_2)\;.\nonumber\\
\eea

\subsection{${\mathcal S}_{\ell 1 \ell_2}^{n_1 n_2} ({\bf x}_1,{\bf x}_2)$ for $ \chi_1 \to 0$ }

Here using the results  obtained in the previous subsections, the expression for ${\mathcal S}_{\ell 1 \ell_2}^{n_1 n_2} ({\bf x}_1,{\bf x}_2)$ for $ \chi_1 \to 0$ is straightforward. In this case the trivial calculation will be done  if  we ``repalce'' $\chi_1 \to \chi_2$, $\ell_1 \to \ell_2$ (and {\it viceversa}) and $z_1 \to z_2$. Then we find
\bea \label{calS(chi0chi2)}
{\mathcal S}_{\ell_1\ell_2}^{n_1+n_2}(0,z_2;{\bf n}_1,{\bf n}_2) ={(-1)^{\ell_1}} ~ {\mathcal P}_{\ell_2}({\bf n}_1 \cdot {\bf n}_2) ~ \sum_\ell i^{\ell-(n_1 + n_2)}(2\ell+1) {\left( \begin{array} {ccc} \ell_1 & \ell_2 & \ell \\ 0 & 0 & 0 \end{array} \right)}^2  \xi_\ell^{\, n_1+n_2}(\chi_{2}; 0, z_2)\,.
\eea
For the dipole term in $\{{\bf n}_1, z_1\}$, we have $\ell_1=n_1=1$ and Eq. (\ref{calS(chi0chi2)}) becomes
\bea \label{calS(chi0chi2)-1}
{\mathcal S}_{1 \ell_2}^{1+n_2}(0,z_2;{\bf n}_1,{\bf n}_2) =- ~ {\mathcal P}_{1}({\bf n}_1 \cdot {\bf n}_2) ~ \sum_\ell i^{\ell-(1 + n_2)}(2\ell+1)
\left( \begin{array} {ccc} 1 & \ell_2 & \ell \\ 0 & 0 & 0\end{array}  \right)^2
  \xi_\ell^{\, n_1+n_2}(\chi_{2}; 0, z_2)\;.
\eea
Then, for $\ell_2=0,1,2$, we have
\bea
{\rm for } ~ ~ \ell_2=0 \quad\quad &\to& \quad\quad {\mathcal S}_{1 0}^{n_2+1}(0,z_2;{\bf n}_1,{\bf n}_2)=- i^{-n_2} \xi_1^{\, n_2+1}(\chi_{2}; 0, z_2) {\mathcal P}_{1}({\bf n}_1 \cdot {\bf n}_2) \nonumber\\
{\rm for } ~ ~ \ell_2=1 \quad\quad &\to& \quad\quad {\mathcal S}_{1 1}^{n_2+1}(0,z_2;{\bf n}_1,{\bf n}_2)=- i^{-(n_2+1)}\left[\frac{1}{3}\xi_0^{\, n_2+1}(\chi_{2}; 0, z_2) -\frac{2}{3}\xi_2^{\, n_2+1}(\chi_{2}; 0, z_2) \right]  {\mathcal P}_{1}({\bf n}_1 \cdot {\bf n}_2)\nonumber\\
{\rm for } ~ ~ \ell_2=2 \quad\quad &\to& \quad\quad {\mathcal S}_{1 2}^{n_2+1}(0,z_2;{\bf n}_1,{\bf n}_2)=- i^{-n_2}\left[\frac{2}{5}\xi_1^{\, n_2+1}(\chi_{2}; 0, z_2) -\frac{3}{5}\xi_3^{\, n_2+1}(\chi_{2}; 0, z_2) \right]  {\mathcal P}_{1}({\bf n}_1 \cdot {\bf n}_2)\;.\nonumber\\
\eea
Now we have all ingredients to compute the wide-angle  two-point correlation function in GR with dipole effect. Let us add another comment. From the above results immediately we note that we cannot obtain the tripolar spherical harmonic basis for the dipole terms.

\section{Dipole effect on two-point correlation function}\label{dipole-corr}

In this section we are going to compute all  possible terms that contain the dipole term ${\Delta}_{v_\| o}$, i.e.
\bea \label{xi-v||0}
\xi_{v_{\| o}v_{\| o}}( {\bf n}_1, {\bf n}_2, z_1,z_2) &=& \langle{\Delta}_{v_\| o}( {\bf n}_1,z_1) \Delta_{v_\| o}( {\bf n}_2,z_2) \rangle \,,  \\
\xi_{v_{\| o}\Delta}( {\bf n}_1, {\bf n}_2, z_1,z_2)  &=&  \langle{\Delta}_{v_\| o}( {\bf n}_1,z_1) \Delta( {\bf n}_2,z_2)  \rangle=\sum_A \langle \Delta_A( {\bf n}_1,z_1) {\Delta}_{v_{\| o}}( {\bf n}_2,z_2)  \rangle =\sum_A \xi_{v_{\| o} A}( {\bf n}_1, {\bf n}_2, z_1,z_2)\;, \nonumber \\\\
& {\rm and}&\nonumber\\\nonumber\\
\xi_{\Delta\, v_{\| o}}( {\bf n}_1, {\bf n}_2, z_1,z_2) &=&  \langle \Delta( {\bf n}_1,z_1) \Delta_{v_\| o}( {\bf n}_2,z_2)  \rangle=\sum_B \langle {\Delta}_{v_\| o}( {\bf n}_1,z_1) \Delta_B( {\bf n}_2,z_2)  \rangle=\sum_B \xi_{B v_{\| o}}( {\bf n}_1, {\bf n}_2, z_1,z_2) \,, \nonumber \\
\eea
where $A,B={\rm loc},\, \kappa,\, {\rm I}$. Here below we analyse in details all these relations.

\subsection{$\xi_{v_{\| o}v_{\| o}}( {\bf n}_1, {\bf n}_2, z_1,z_2) $}
Using the prescription in Sec. \ref{Analysis-dipole} we find immediately
\be\label{eq_dipolo}
\xi_{v_{\| o}v_{\| o}}( {\bf n}_1, {\bf n}_2, z_1,z_2) = \frac{1}{3}\omega_{o1} \omega_{o2}\;\xi_0^{\, 2}(0; 0, 0)\;{\mathcal P}_{1}({\bf n}_1 \cdot {\bf n}_2) 
\ee
where $\omega_{on}=\omega_{o}(z_n)$ for $n=1,2$.
\subsection{$\xi_{\Delta \, v_{\| o}}( {\bf n}_1, {\bf n}_2, z_1,z_2) $}
 It easy to see that
\bea
\xi_{{\rm loc} v_{\| o}}( {\bf n}_1, {\bf n}_2, z_1,z_2) &=& b_1 b_2 \omega_{o2} \Bigg\{-\left(1+\frac{3}{5}\beta_1\right)  \xi_1^{\, 1}(\chi_{1}; z_1, 0) +{1\over3}\frac{\beta_1\alpha_1}{\chi_1}\left[ \xi_0^{\, 2}(\chi_{1}; z_1, 0) -2  \xi_2^{\, 2}(\chi_{1}; z_1, 0)\right] \nonumber\\
&&+\gamma_1 \xi_1^{\, 3}(\chi_{1}; z_1, 0) + \frac{2}{5}\beta_1 \xi_3^{\, 1}(\chi_{1}; z_1, 0) \Bigg\} {\mathcal P}_{1}({\bf n}_1 \cdot {\bf n}_2) \;,
\eea
\bea
\xi_{\kappa v_{\| o}}( {\bf n}_1, {\bf n}_2, z_1,z_2) &=& b_1 b_2 \omega_{o2} \Bigg\{ \int_0^{\chi_1} d\tilde \chi_1\; \sigma_{1\tilde 1}\bigg[-\frac{3}{5} \xi_1^{\, 1}(\tilde \chi_{1}; \tilde z_1, 0) -\frac{3}{5}  \xi_3^{\, 1}(\tilde \chi_{1}; \tilde z_1, 0)  -{1\over \tilde \chi_1} \xi_0^{\, 2}(\tilde\chi_{1}; \tilde z_1, 0)  \nonumber\\
&&\quad\quad + {2\over \tilde \chi_1}  \xi_2^{\, 2}(\chi_{1}; z_1, 0) \bigg] \Bigg\} {\mathcal P}_{1}({\bf n}_1 \cdot {\bf n}_2) 
\eea
and
\bea
\xi_{{\rm I} v_{\| o}}( {\bf n}_1, {\bf n}_2, z_1,z_2) &=& b_1 b_2 \omega_{o2} \Bigg[ \int_0^{\chi_1} d\tilde \chi_1 \; \mu_{1\tilde 1}\; \xi_1^{\, 3}(\tilde \chi_{1}; \tilde z_1, 0) \Bigg] {\mathcal P}_{1}({\bf n}_1 \cdot {\bf n}_2) \;.
\eea

\subsection{$\xi_{v_{\| o}\Delta}( {\bf n}_1, {\bf n}_2, z_1,z_2) $}
Using the symmetries of the two point correlation function we obtain
\be
\xi_{v_{\| o}{\rm loc}}( {\bf n}_1, {\bf n}_2, z_1,z_2) = \langle{\Delta}_{v_\| o}( {\bf n}_1,z_1) \Delta_{\rm loc}( {\bf n}_2,z_2)  \rangle
\ee
\bea
\xi_{v_{\| o}{\rm loc} }( {\bf n}_1, {\bf n}_2, z_1,z_2) &=& b_1 b_2 \omega_{o1} \Bigg\{-\left(1+\frac{3}{5}\beta_2\right)  \xi_1^{\, 1}(\chi_{2}; 0, z_2) +{1\over3}\frac{\beta_2\alpha_2}{\chi_2}\left[ \xi_0^{\, 2}(\chi_{2}; 0, z_2)  -2  \xi_2^{\, 2}(\chi_{2}; 0, z_2)  \right] \nonumber\\
&&+\gamma_2 \xi_1^{\, 3}(\chi_{2}; 0, z_2)  + \frac{2}{5}\beta_2 \xi_3^{\, 1}(\chi_{2}; 0, z_2) \Bigg\} {\mathcal P}_{1}({\bf n}_1 \cdot {\bf n}_2) \;,
\eea
\bea
\xi_{v_{\| o} \kappa }( {\bf n}_1, {\bf n}_2, z_1,z_2) &=& b_1 b_2 \omega_{o1} \Bigg\{ \int_0^{\chi_2} d\tilde \chi_2\; \sigma_{2\tilde 2}\bigg[-\frac{3}{5} \xi_1^{\, 1}(\tilde \chi_{2}; 0, \tilde z_2) -\frac{3}{5}  \xi_3^{\, 1}(\tilde \chi_{2}; 0, \tilde z_2) -{1\over \tilde \chi_2} \xi_0^{\, 2}(\tilde \chi_{2}; 0, \tilde z_2)  \nonumber\\
&&\quad\quad + {2\over \tilde \chi_2}  \xi_2^{\, 2}(\tilde \chi_{2}; 0, \tilde z_2)   \bigg] \Bigg\} {\mathcal P}_{1}({\bf n}_1 \cdot {\bf n}_2) 
\eea
and
\bea
\xi_{v_{\| o} {\rm I} }( {\bf n}_1, {\bf n}_2, z_1,z_2) &=& b_1 b_2 \omega_{o1} \Bigg[ \int_0^{\chi_2} d\tilde \chi_2  \; \mu_{2\tilde 2} \;\xi_1^{\, 3}(\tilde \chi_{2}; \tilde z_2, 0) \Bigg] {\mathcal P}_{1}({\bf n}_1 \cdot {\bf n}_2) \;.
\eea

\section{Analysis using only local terms}\label{localtermswithdipole}

First of all, taking into account that trivially $${\mathcal P}_{1}({\bf n}_1 \cdot {\bf n}_2) = \cos2\theta={4 \over 3}{\mathcal P}_{2}(\cos \theta) -{1 \over 3}{\mathcal P}_{0}(\cos \theta)\;,$$
we note that we have to modify $C_{00}$ and $C_{02}$ in Eq. (\ref{eq:psis}). Then, we can rewrite the decomposition in Eq. (\ref{xisum}) in the following way
\bea\label{xi_totloc}
\xi( {\bf n}_1, {\bf n}_2, z_1,z_2)= \langle \Delta_{\rm g}( {\bf n}_1,z_1) \Delta_{\rm g}( {\bf n}_2,z_2)  \rangle &=& \xi_{\rm loc} +\xi_{v_{\| o}{\rm loc}}+ \xi_{{\rm loc} v_{\| o}}+\xi_{v_{\| o}v_{\| o}} \nonumber \\ 
&=& b(z_1) b(z_2) \sum_{\tilde{L} , \tilde{\ell}} \; D_{\tilde{L} \tilde{\ell}_3} (z_2, \theta, \varphi) \; \mathcal{P}_{\tilde{L} }(\cos\varphi) \; \mathcal{P}_{ \tilde{\ell_3}}(\cos\theta) \; ,
\eea 
where the new coefficients are
\bea
D_{00} &=& C_{00}-\frac{1}{9}\omega_{o1} \omega_{o2}
-\frac{1}{3}\omega_{o2} \Bigg\{-\left(1+\frac{3}{5}\beta_1\right)  \xi_1^{\, 1}(\chi_{1}; z_1, 0) +{1\over3}\frac{\beta_1\alpha_1}{\chi_1}\left[ \xi_0^{\, 2}(\chi_{1}; z_1, 0) -2  \xi_2^{\, 2}(\chi_{1}; z_1, 0)\right] \nonumber\\
&&+\gamma_1 \xi_1^{\, 3}(\chi_{1}; z_1, 0) + \frac{2}{5}\beta_1 \xi_3^{\, 1}(\chi_{1}; z_1, 0) \Bigg\} -{1 \over 3} \omega_{o1} \Bigg\{-\left(1+\frac{3}{5}\beta_2\right)  \xi_1^{\, 1}(\chi_{2}; 0, z_2)  \nonumber\\
&&+{1\over3}\frac{\beta_2\alpha_2}{\chi_2}\left[ \xi_0^{\, 2}(\chi_{2}; 0, z_2)  -2  \xi_2^{\, 2}(\chi_{2}; 0, z_2)  \right] +\gamma_2 \xi_1^{\, 3}(\chi_{2}; 0, z_2)  + \frac{2}{5}\beta_2 \xi_3^{\, 1}(\chi_{2}; 0, z_2) \Bigg\}\;,
\eea
\bea
D_{02} &=& C_{02}+\frac{4}{9}\omega_{o1} \omega_{o2}
+\frac{4}{3}\omega_{o2} \Bigg\{-\left(1+\frac{3}{5}\beta_1\right)  \xi_1^{\, 1}(\chi_{1}; z_1, 0) +{1\over3}\frac{\beta_1\alpha_1}{\chi_1}\left[ \xi_0^{\, 2}(\chi_{1}; z_1, 0) -2  \xi_2^{\, 2}(\chi_{1}; z_1, 0)\right] \nonumber\\
&&+\gamma_1 \xi_1^{\, 3}(\chi_{1}; z_1, 0) + \frac{2}{5}\beta_1 \xi_3^{\, 1}(\chi_{1}; z_1, 0) \Bigg\}  + {4 \over 3} \omega_{o1} \Bigg\{-\left(1+\frac{3}{5}\beta_2\right)  \xi_1^{\, 1}(\chi_{2}; 0, z_2)  \nonumber\\
&&+{1\over3}\frac{\beta_2\alpha_2}{\chi_2}\left[ \xi_0^{\, 2}(\chi_{2}; 0, z_2)  -2  \xi_2^{\, 2}(\chi_{2}; 0, z_2)  \right] +\gamma_2 \xi_1^{\, 3}(\chi_{2}; 0, z_2)  + \frac{2}{5}\beta_2 \xi_3^{\, 1}(\chi_{2}; 0, z_2) \Bigg\}\;,
\eea
\bea
D_{11}=C_{11}\;, \quad  D_{20}=C_{20} \;, \quad D_{22}=C_{22} \;,  \quad D_{13}=C_{13} \;,  \quad D_{31}=C_{31} \;,  \quad D_{04}=C_{04} \;,  \quad D_{40}=C_{40}  \;.
\eea
In the following section we are going to discuss and quantify the effects of these new terms in the two-point correlation function. First of all, we will study the  start plane-parallel limit, then we will  study the wide angle effect due to the Kaiser Rocket effect.

\subsection{Flat-sky (plane-parallel) limit}

At this point, it might be interesting to consider the dipole terms  at  very small angle, i.e. for small galaxy separation, we have  ${\bf n}_1$ and ${\bf n}_2$ are (almost) parallel. In other words, $\theta \to 0$ and, in the configuration space, we can also generalise the result obtained in \cite{Bertacca:2012tp} in plane-parallel limit:
\bea
{\xi}_{\rm loc}(z,\chi_{12}) &=& b^2  \Bigg\{  \Bigg[ \left(1+\frac{2}{3}\beta+ \frac{1}{5}\beta^2 \right) \xi_0^{0}(  {\chi}_{12}; {z},{z})
   -   \bigg[ 2 \left(1+\frac{1}{3}\beta \right) \gamma  -\frac{\beta^2 \alpha^2}{3\chi^2} \bigg] \xi_0^{2}(  {\chi}_{12}; {z},{z}) \nonumber \\
   &&~+   \gamma^2  \xi_0^{4}( {\chi}_{12}; {z},{z}) +  \frac{1}{3}\omega_{o}^2 \;\xi_0^{\, 2}(0; 0) + 2\omega_{o}\bigg[-\left(1+\frac{3}{5}\beta\right)  \xi_1^{\, 1}(\chi; z) +{1\over3}\frac{\beta\alpha}{\chi}\left[ \xi_0^{\, 2}(\chi; z)  -2  \xi_2^{\, 2}(\chi; z)  \right] \nonumber \\
   &&+\gamma \xi_1^{\, 3}(\chi;  z)  + \frac{2}{5}\beta \xi_3^{\, 1}(\chi;  z) \bigg]\Bigg] \mathcal{P}_{0}( { {{\bf n}}}_1 \cdot  { {{\bf n}}}_{12})   +  \bigg(- 4 \beta \left( \frac{1}{3}+ \frac{1}{7}\beta\right)\xi_2^{0}( {\chi}_{12}; {z},{z}) \nonumber \\
   &&~+ \frac{2}{3} \beta \left( 2 \gamma - \frac{\beta\alpha^2}{\chi^2}\right)\xi_2^{2}( {\chi}_{12}; {z},{z})\bigg) \mathcal{P}_{2}( { {{\bf n}}} \cdot  { {{\bf n}}}_{12})
   + \frac{8}{35} \beta^2 \xi_4^{0}( {\chi}_{12}; {z},{z}) \mathcal{P}_{4}( {\bf n} \cdot  { \bf n}_{12}) \Bigg\}, \label{ppxiss}
 \eea
where $\xi_\ell^{\, n}(\chi; z)\equiv \xi_\ell^{\, n}(\chi; 0, z)=\xi_\ell^{\, n}(\chi; z, 0)$. It is trivial to see that, at fixed redshift, it is only a constant term of the monopole.

\section{Numerical Results}\label{sec:numresults}

\begin{figure*}[htb!]
\includegraphics[width=0.5\columnwidth]{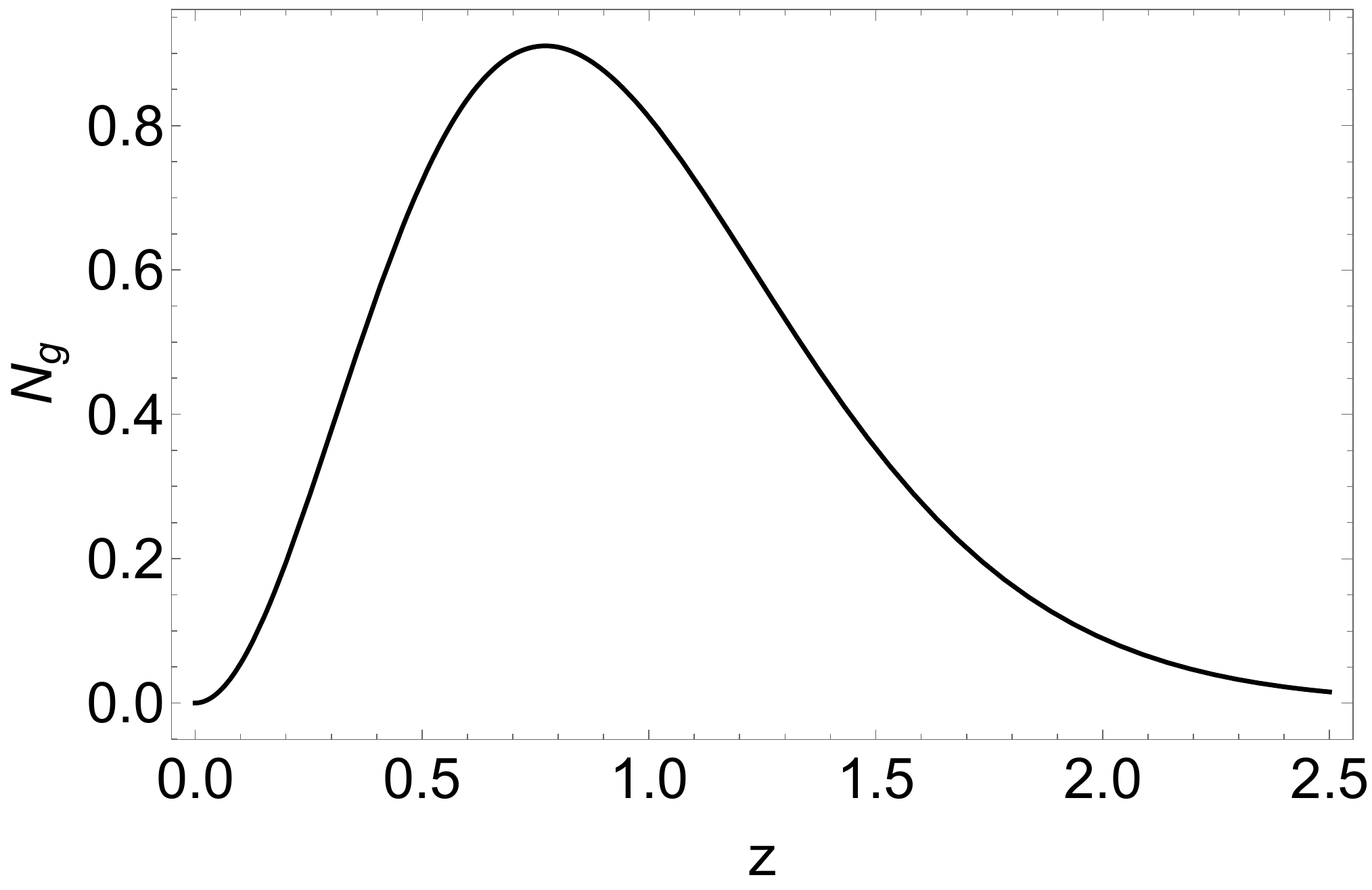}
\caption{Redshift distribution, normalized to unity for the generic survey considered in this paper.}\label{Nz}
\end{figure*}

\begin{center}
\begin{figure*}[htb!]
\includegraphics[width=0.5\columnwidth]{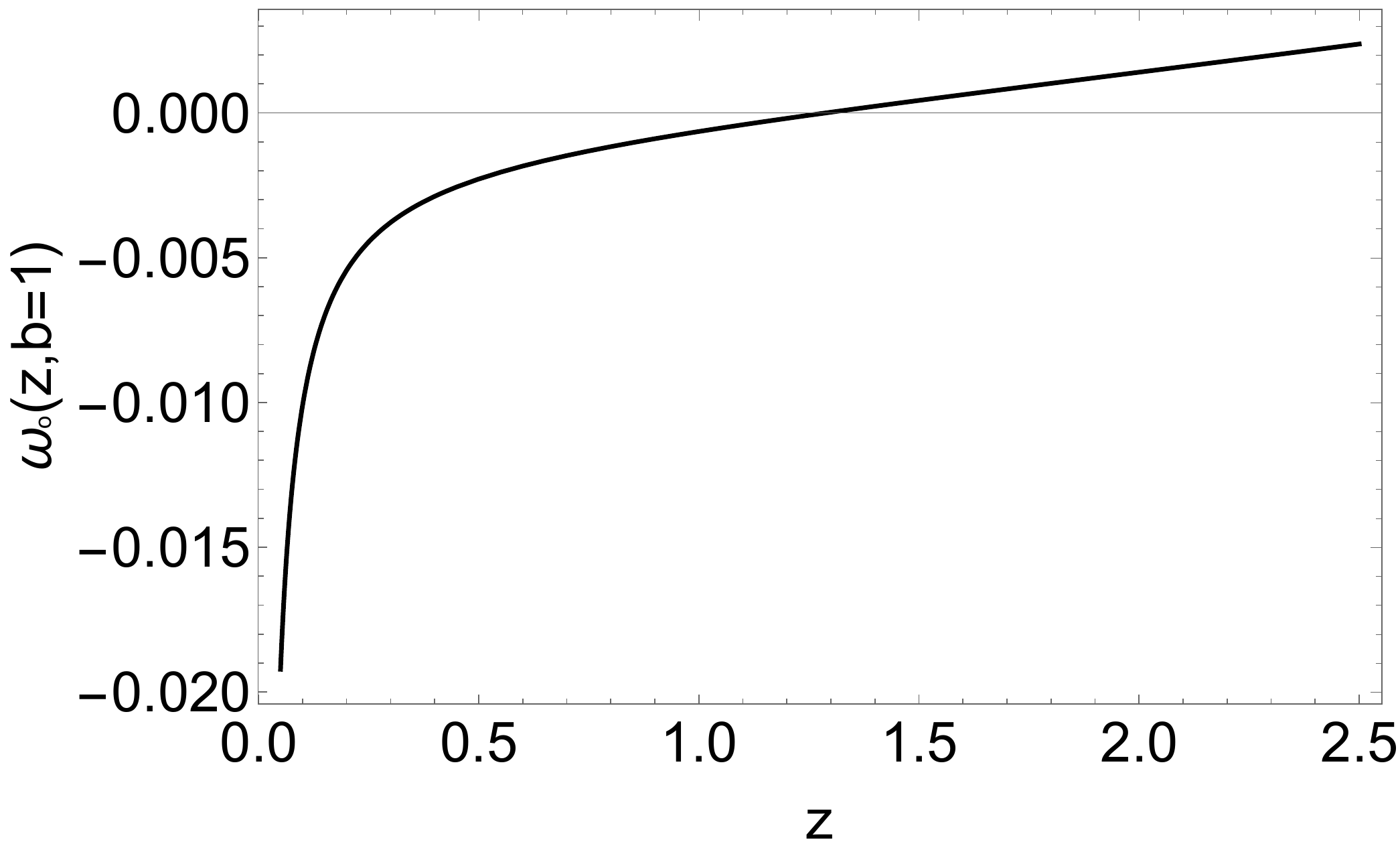}
\caption{The function $\omega_o$ in \eqref{omega0}, assuming a concordance model, and with ${\cal Q}=0$,  $b= 1$ .}\label{omega-z}
\end{figure*}
\end{center}

\begin{center}
\begin{figure*}[htb!]
\includegraphics[width=0.65\columnwidth]{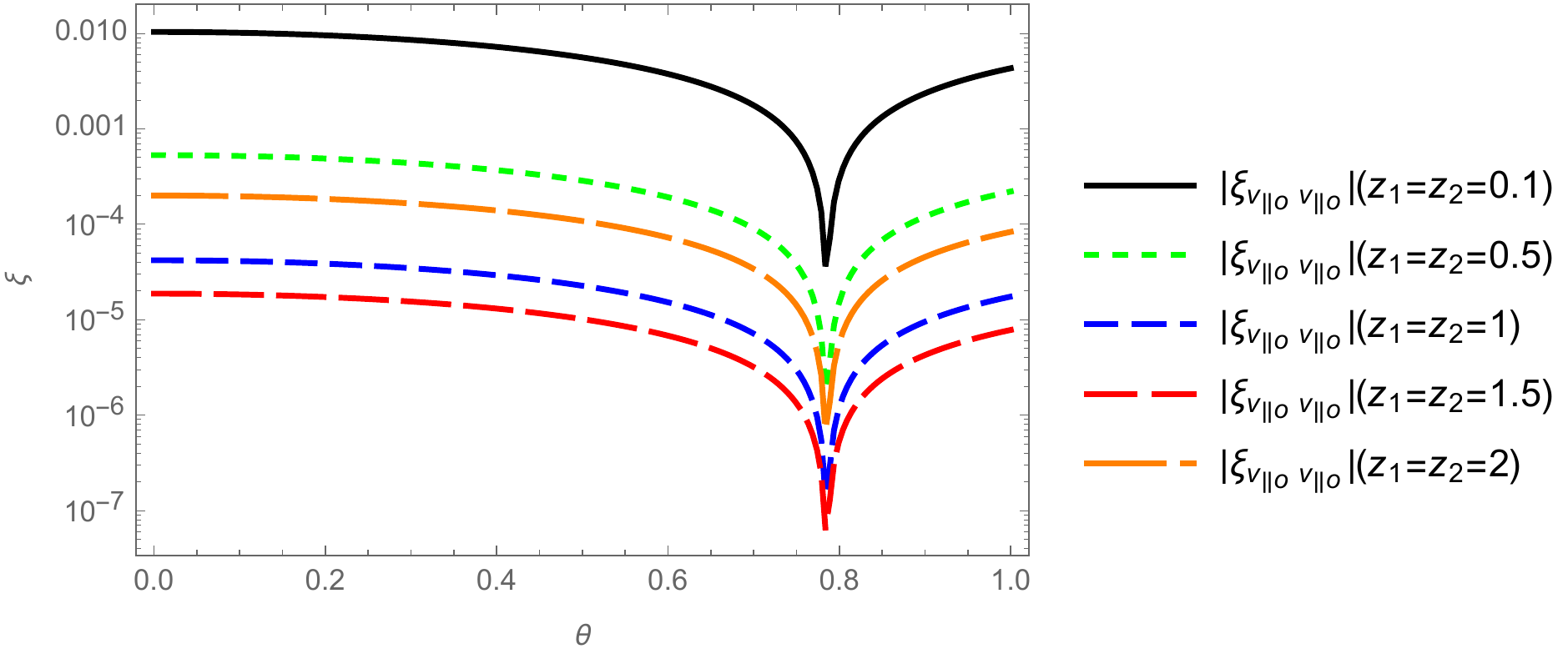}
\caption{Absolute value of the correlation  $\xi_{v_{\| o}v_{\| o}}$  as a function of the separation angle $\theta$, for different values of $z_1=z_2$.}\label{Fig:xi_vo}
\end{figure*}
\end{center}

For a reference survey, we take a generic  survey  which aims to measure galaxy spectra up to $z\sim 2.5$. Fig.\ \ref{Nz} shows a generic  redshift normalised distribution that we are assuming.
In the following sections, we  set the spatial curvature $K=0$ and, for the magnification bias Eq.\ \eqref{Q}, we assume $\mathcal{Q} = 0$. Finally, we choose the fiducial values $ w_0=-1$, $ w_a=0$
[where $\{w_0,w_a\}$ parameterise the dark energy equation of state, as $w= w_0 +w_a(1-a)$], $ h= 0.6766$ (where $h$ parameterises the present Hubble parameter, $H_0 = h 100$km/s/Mpc), $\Omega_{\rm cdm}= 0.3111$ and $\Omega_{\rm b}= 0.0490$ (see \cite{Aghanim:2018eyx}). 
In Fig.\ \ref{omega-z} we show the evolution of $\omega_o$ at different $z$. As we pointed out in Section \ref{Analysis-dipole}, $\omega_o$ encodes the effects of Hubble expansion and  the galaxy redshift distribution.
In Fig.\ \ref{Fig:xi_vo} we plot the rocket Kaiser contribution at wide-angle scales of $\xi_{v_{\| o}v_{\| o}}$  at different $z$ and how the contribution in \eqref{eq_dipolo} depends on the separation angle $2\theta$. 
Indeed, it easy to see that for $\theta \to 0$ is constant (as we pointed out in the flat-sky regime) and is zero when $\theta=\pi/4$ because $\mathcal{P}_{1}(\cos(\pi/2))=0$. 

In this section, we ignore the integrated part of $\xi$ and focus on the local part of Eq.\ (\ref{xi})\footnote{We think that for this topic  a deep and further investigation, including also the integrated part of  Eq.\ (\ref{xi}), should be done soon; for example using LIGER method \cite{Borzyszkowski:2017ayl}. We will postpone this in a future work.}.
Now, in order to understand which local term is more important and in which configuration,  we separate the correlation in Eq.\ (\ref{xi_totloc})  in several parts. Precisely, let us divide $\xi$ as follows:
\begin{equation}
\label{locp}
\xi =  \xi_{{\rm loc}-K}+  \xi_{\rm loc-wide}
+\xi_{v_{\| o}{\rm loc}}+ \xi_{{\rm loc}\; v_{\| o}}+\xi_{v_{\| o}v_{\| o}},
\end{equation}
where we have split $\xi_{\rm loc}= \xi_{{\rm loc}-K}+\xi_{\rm loc-wide}$. In particular, 
\begin{quote}
$ \xi_{{\rm loc}-K}$ encodes the effect of the matter overdensity $\delta$ and peculiar velocity $\beta$ due to {\it the Kaiser effect}. (Here,  the Kaiser effect represents in {\it Kaiser boost}, see \cite{Kaiser:1987qv}). In general, this is the  correlation function that it is considered in most of the literature (in \cite{Raccanelli:2013multipoli, Raccanelli:2013gja} is also called $\xi_\beta$).

$\xi_{\rm loc-wide}$  includes all terms that receive contributions from all of $\alpha$ and $\gamma$. Therefore it gives both the wide-angle and mode-coupling contributions, and the relativistic corrections due to potential terms to wide-angle effects (see also \cite{Bertacca:2012tp, Raccanelli:2013multipoli, Raccanelli:2013gja,Raccanelli:2016avd}).

$\xi_{v_{\| o}v_{\| o}}$ is the main contribution of the Kaiser Rocket effect.

$\xi_{v_{\| o}{\rm loc}} ~~ \& ~~ \xi_{{\rm loc}\; v_{\| o}}$  describe the correlation of the local terms (i.e. that depend on $\delta$, $\alpha$, $\beta$ and $\gamma$ ) with the dipole.

\end{quote}

\begin{figure*}[!htbp]
\includegraphics[width=0.49 \linewidth]{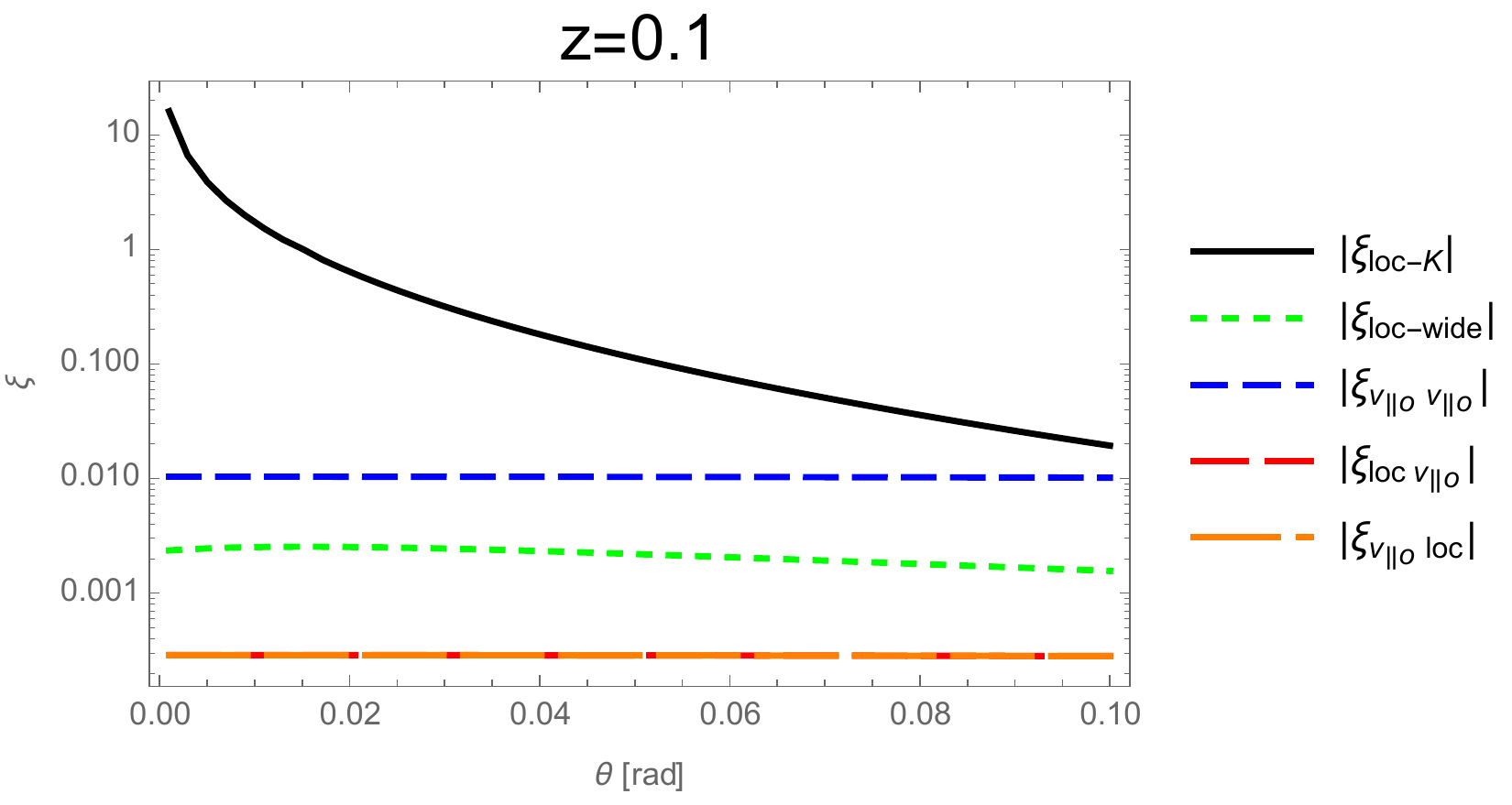}
\includegraphics[width=0.49 \linewidth]{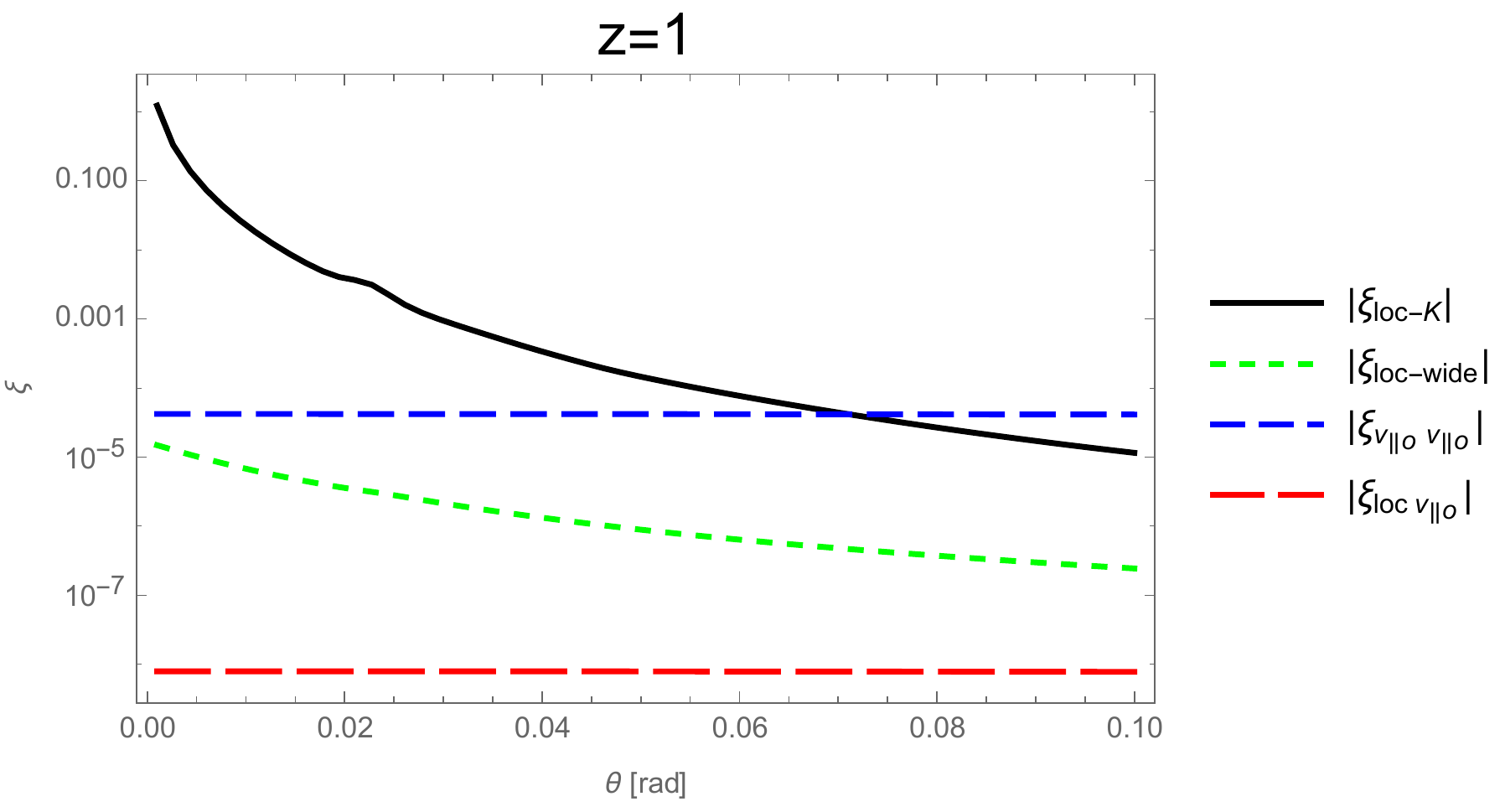}
\caption{Absolute value of all contributions. Left side with  $z_1=z_2=0.1$, right side with  $z_1=z_2=1$. \label{fig:z_1=z_2}}
\end{figure*}
The relative importance of the dipole/Rocket effect depends on the particular configuration, i.e. on $\{z_1, z_2, \theta\}$. Here below we will study the dependence of these terms for different separation angles, scales, and redshifts of the two galaxies. First of all, let us consider  pairs of galaxies transverse to the line of sight, i.e. for $z_1=z_2$. Fig.\ \ref{fig:z_1=z_2}  shows how the dipole contributions depend on the separation angle $2\theta$.
We note that at low redshift, e.g.  $z_1=z_2=0.1$, the  contribution $ \xi_{{\rm loc}-K}$ is dominant for $\theta<0.1$. Instead, at $z_1=z_2=1$, $\xi_{v_{\| o}v_{\| o}}> \xi_{{\rm loc}-K}$ for $\theta>0.07$. 

\begin{figure*}[!htbp]
\includegraphics[width=0.49 \linewidth]{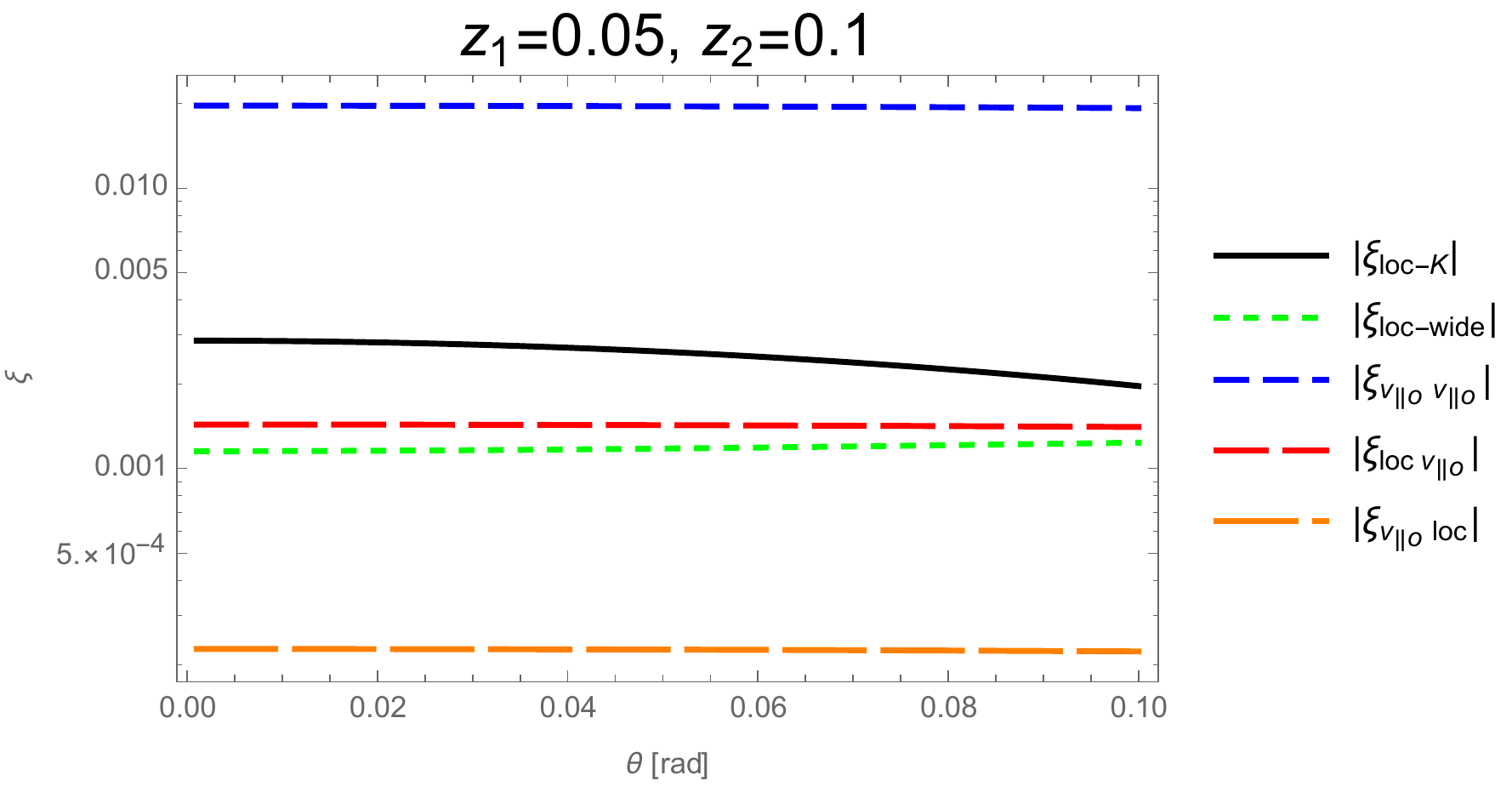}
\includegraphics[width=0.49 \linewidth]{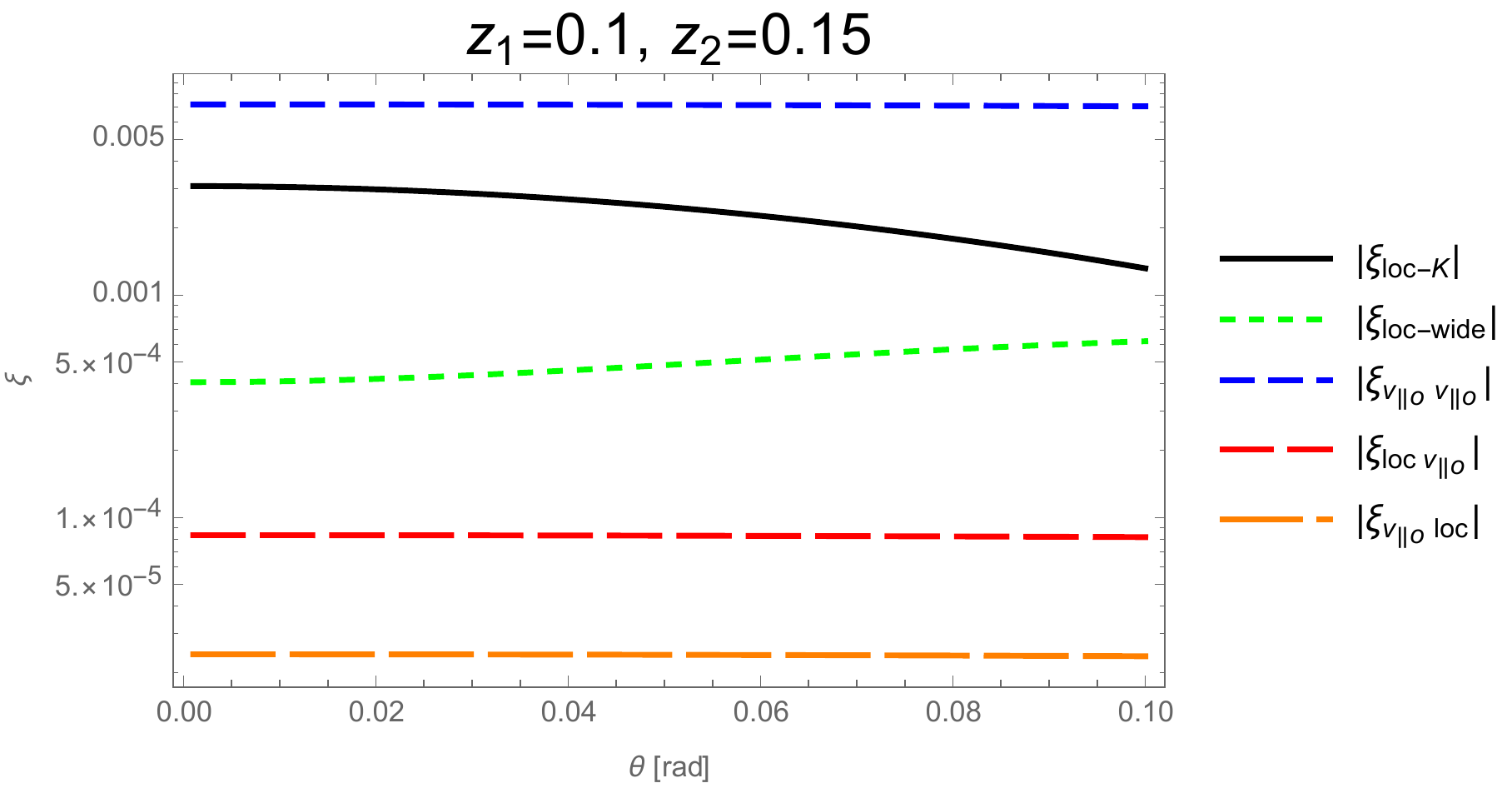}
\includegraphics[width=0.49 \linewidth]{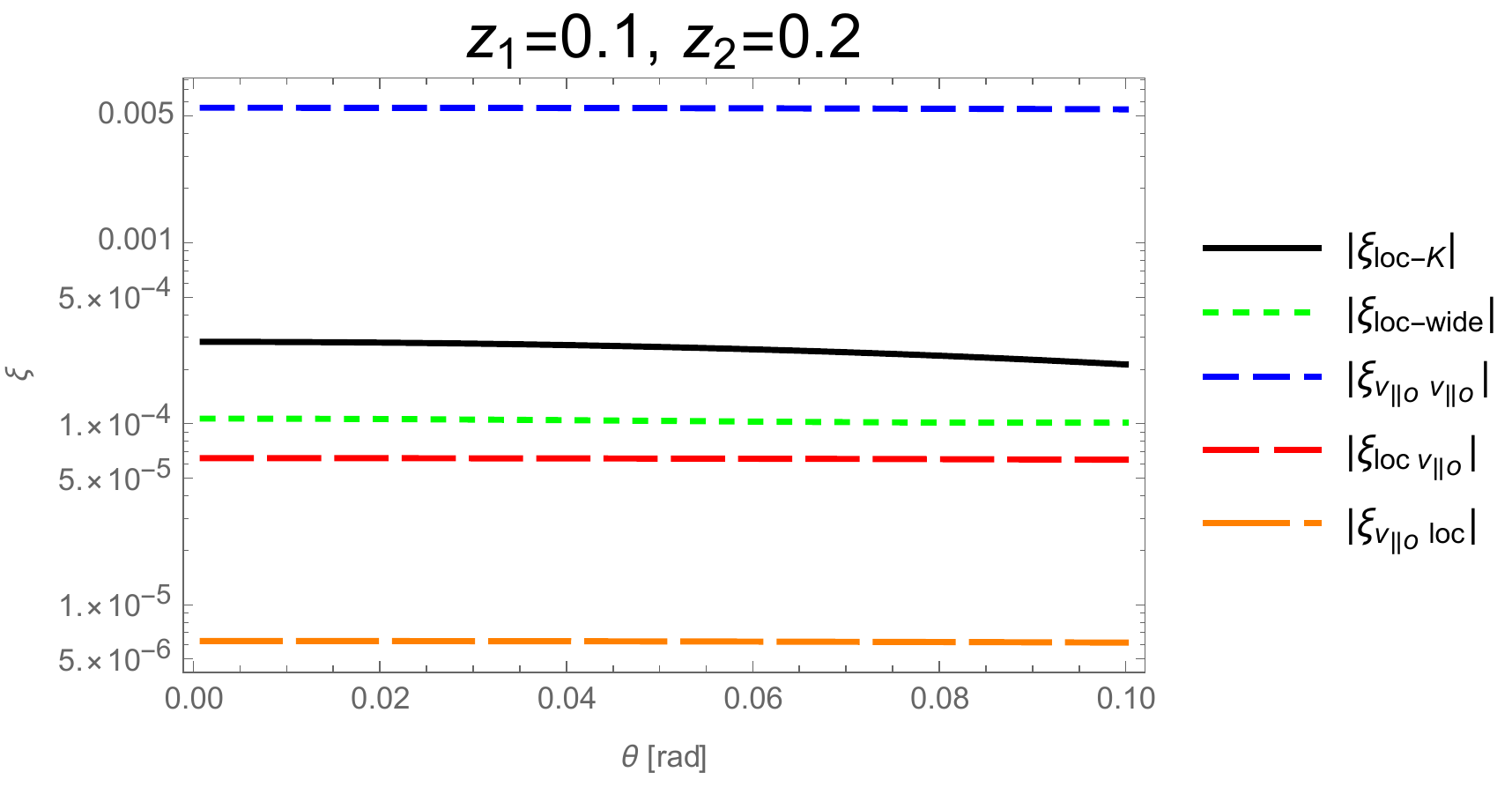}
\includegraphics[width=0.49 \linewidth]{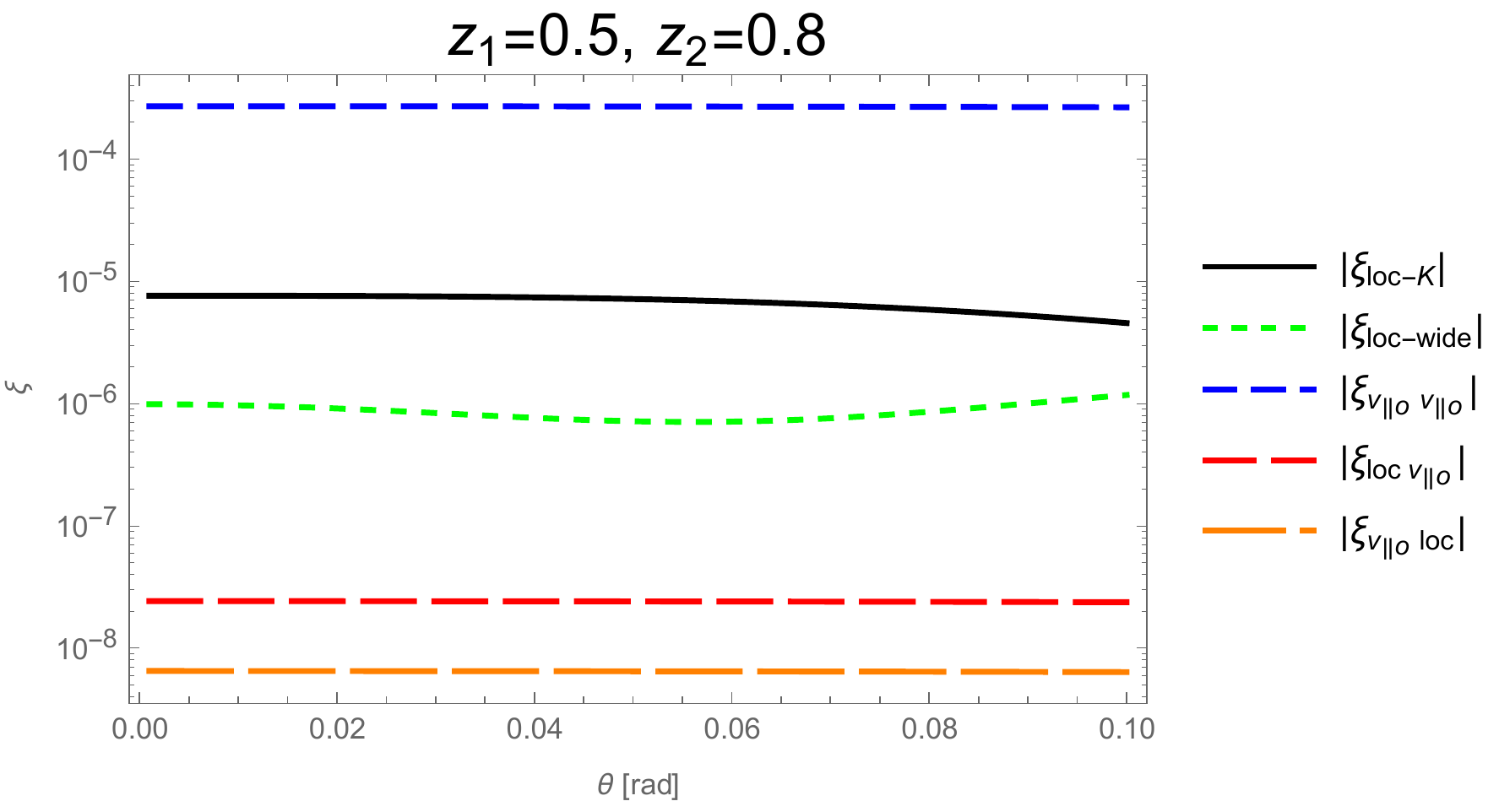}
\includegraphics[width=0.49 \linewidth]{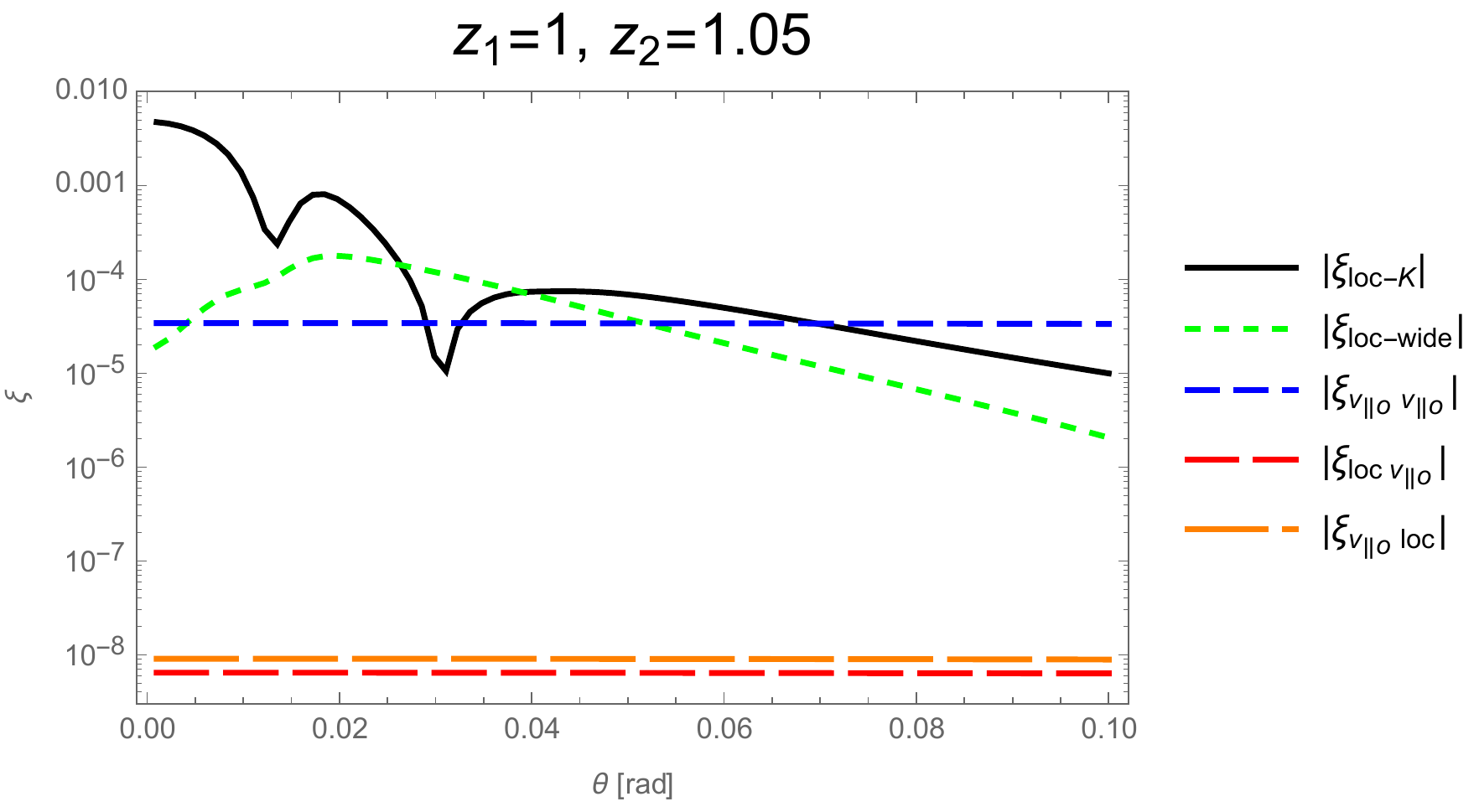}
\includegraphics[width=0.49 \linewidth]{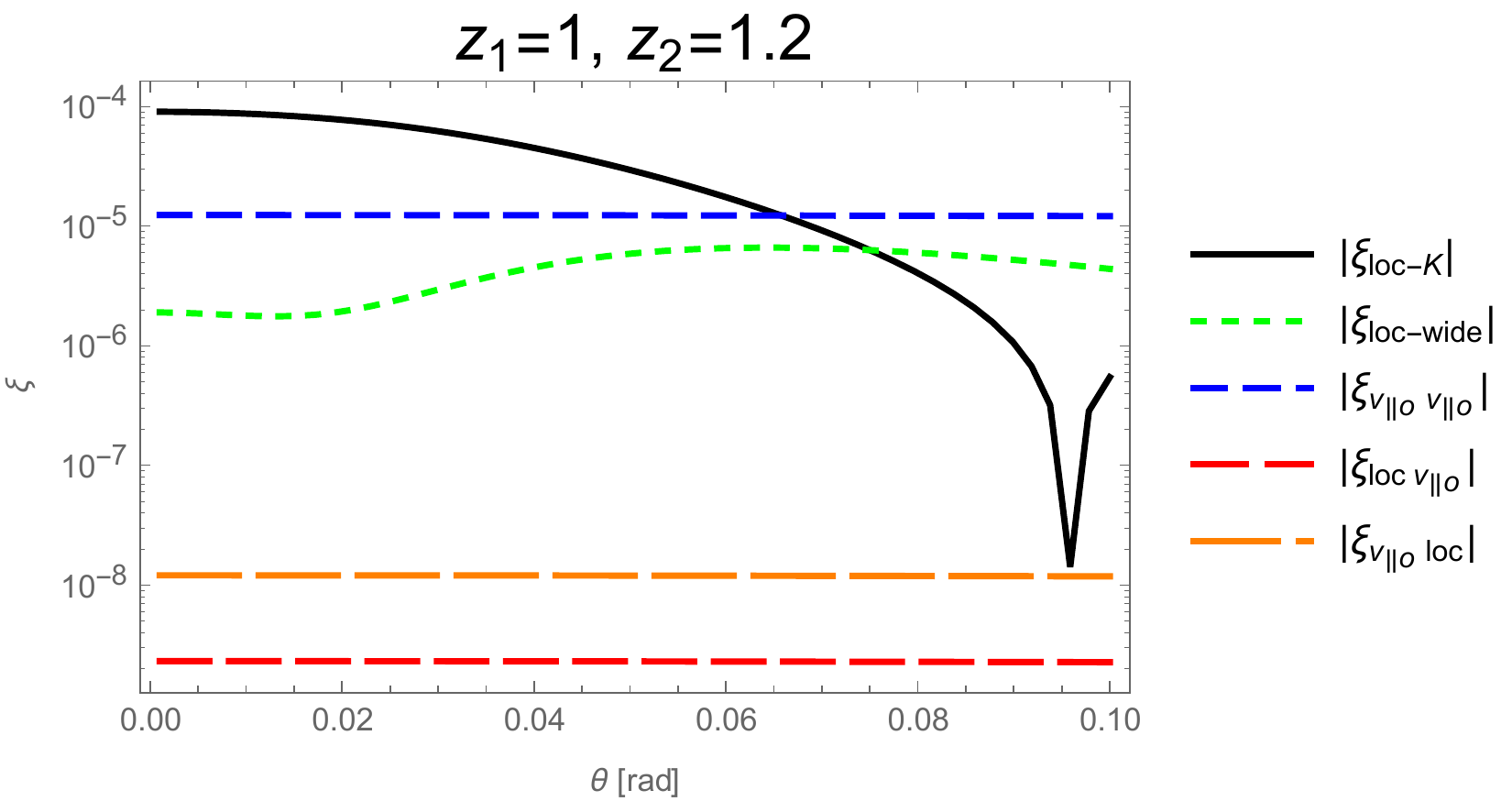}
\caption{Absolute value of all contributions where we have fixed $z_1\neq z_2$ and with $\theta$ varying. \label{fig:z_1neqz_2}}
\end{figure*}
In Fig.\ \ref{fig:z_1neqz_2} we still fix $z_1$ and $z_2$, but with two different values of redshift,  galaxies with non-transverse separation.
It is clear from the first four panels ( i.e. for $z_1=0.05~z_2=0.1$, $z_1=0.1~z_2=0.15$, $z_1=0.1~z_2=0.2$ and $z_1=0.5~z_2=0.8$)  that $\xi_{v_{\| o}v_{\| o}}$ is the dominant contribution of the correlation function on large-scales. In the bottom-left panel, i.e. for $z_1=1~z_2=1.05$,  we have a non negligible effect of $\xi_{\rm loc-wide}$ at BAO scales. In general,  for most of above panels,  $\xi_{\rm loc-wide}$ is subdominant.

\begin{figure*}[!htbp]
\includegraphics[width=0.49 \linewidth]{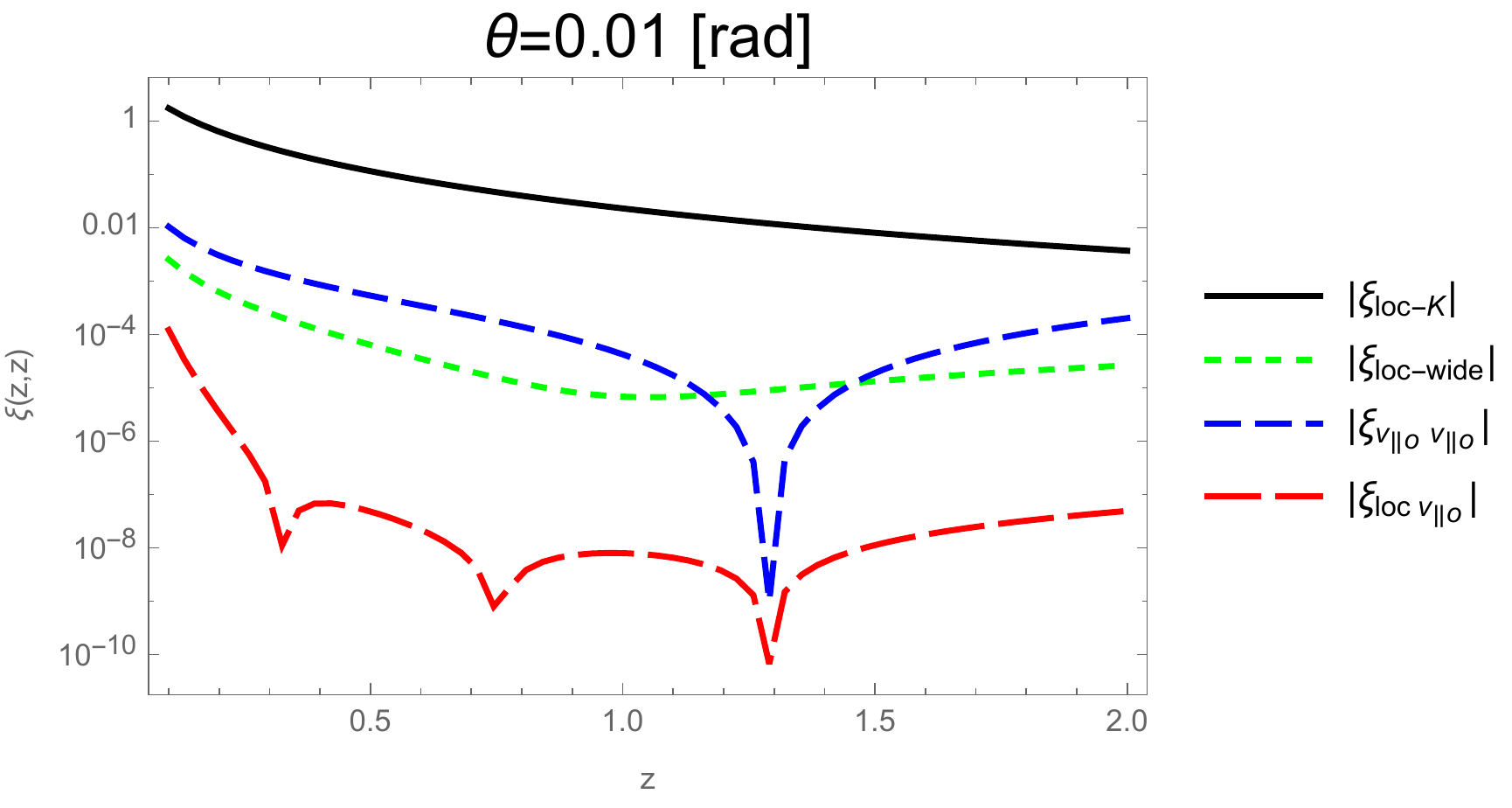}
\includegraphics[width=0.49 \linewidth]{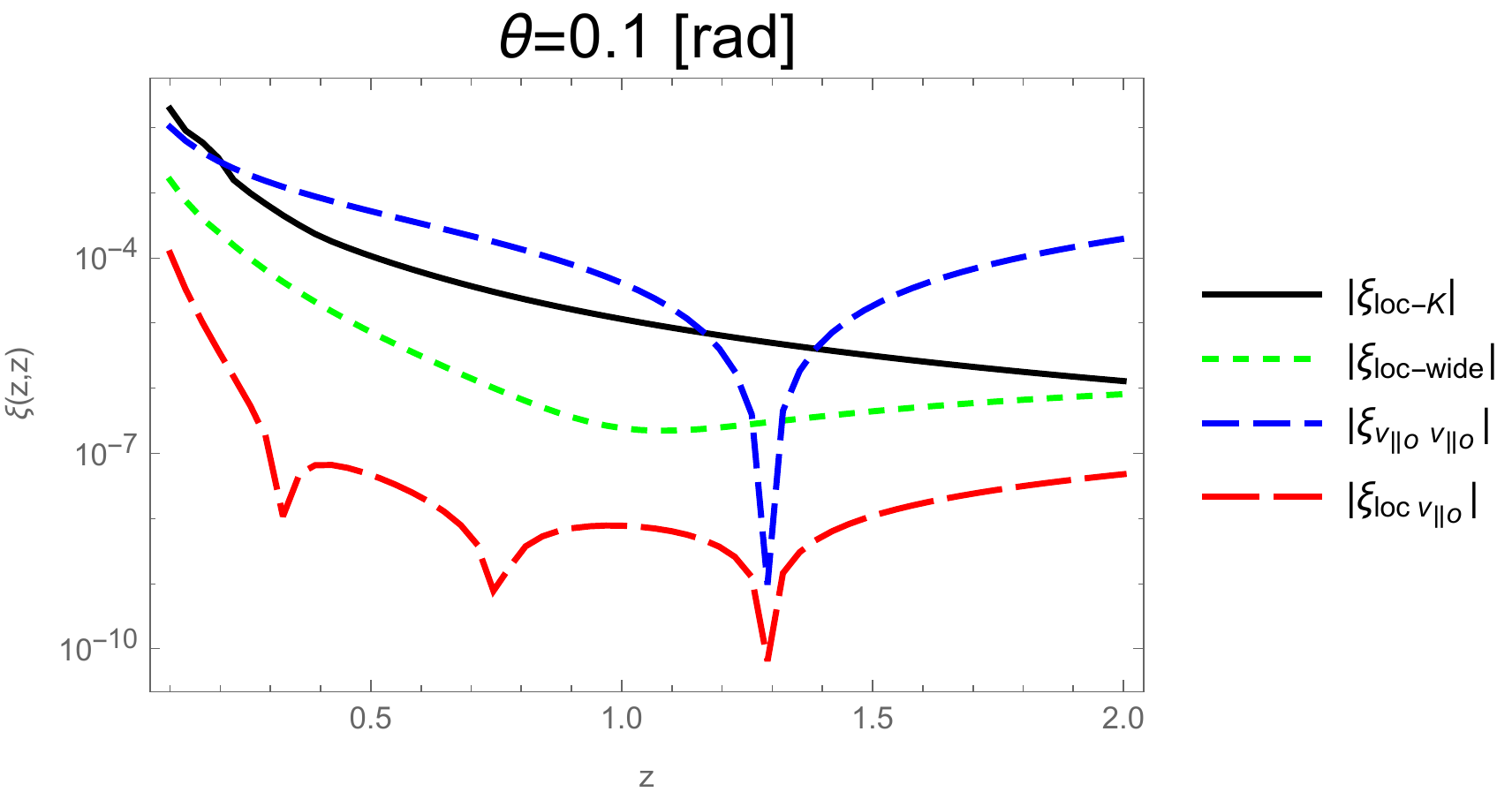}
\caption{Absolute value of all contributions where we have fixed $\theta$ and with $z= z_1= z_2$ varying. \label{fig:-theta_xi_z1=z2}}
\end{figure*}

Another interesting configuration is to set $\theta$ constant and with $z_1=z_2$ varying. 
In Fig.\ \ref{fig:-theta_xi_z1=z2}  we put $\theta=0.01$rad on the left panel and $\theta=0.1$rad on right panel.
 As expected for small $\theta$ the dominant contribution here is the standard Kaiser component. Conversely, for large separation angle (e.g. $\theta=0.1$rad),  unless around $z=1.3$ [because $\omega_o(z=1.3)=0$, e.g. see Fig.\ \ref{omega-z}] the local correlation is weak so  the Rocket effect is the only relevant component.

\begin{figure*}[!htbp]
 \includegraphics[width=0.49 \linewidth]{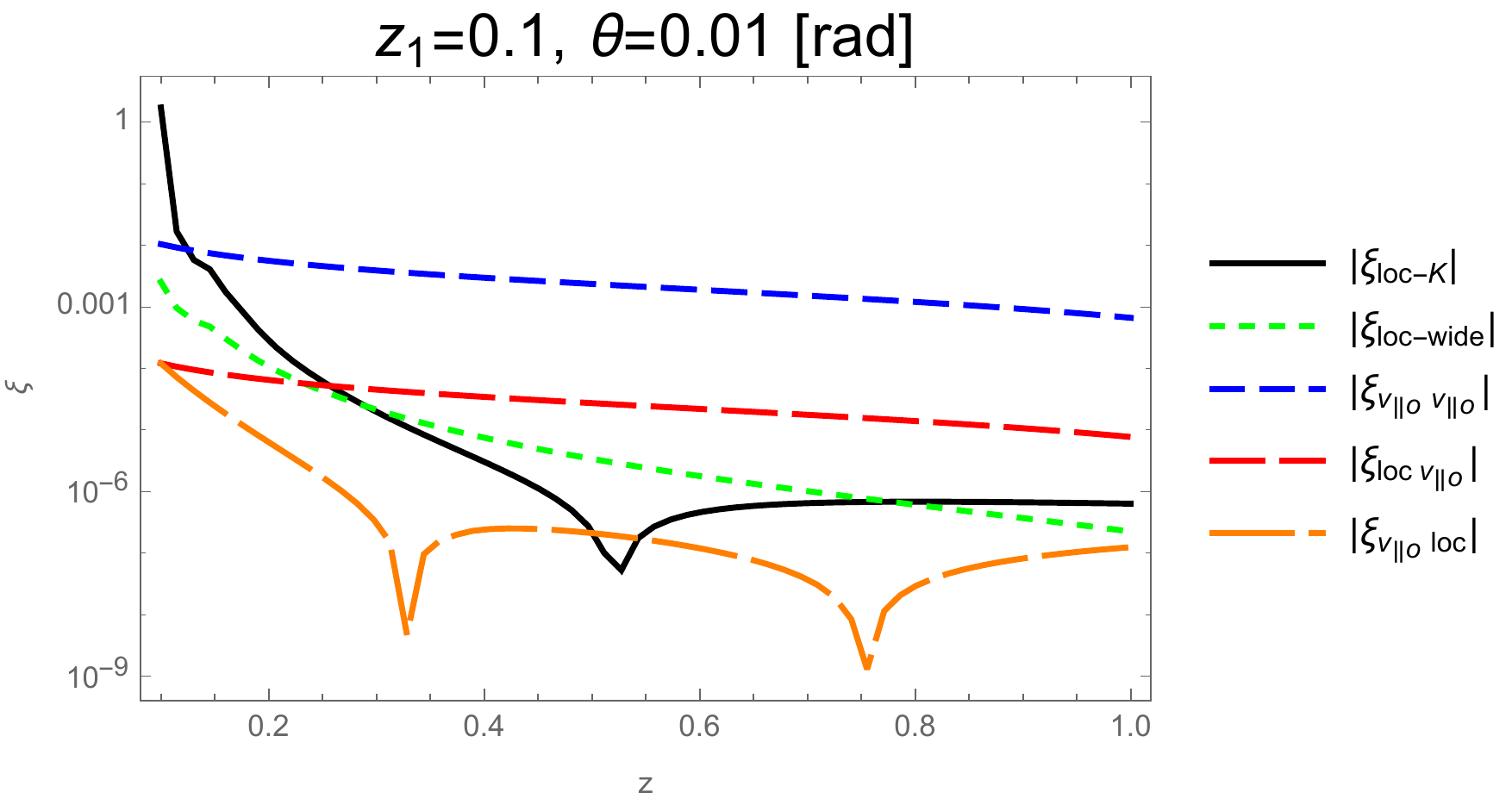}
\includegraphics[width=0.49 \linewidth]{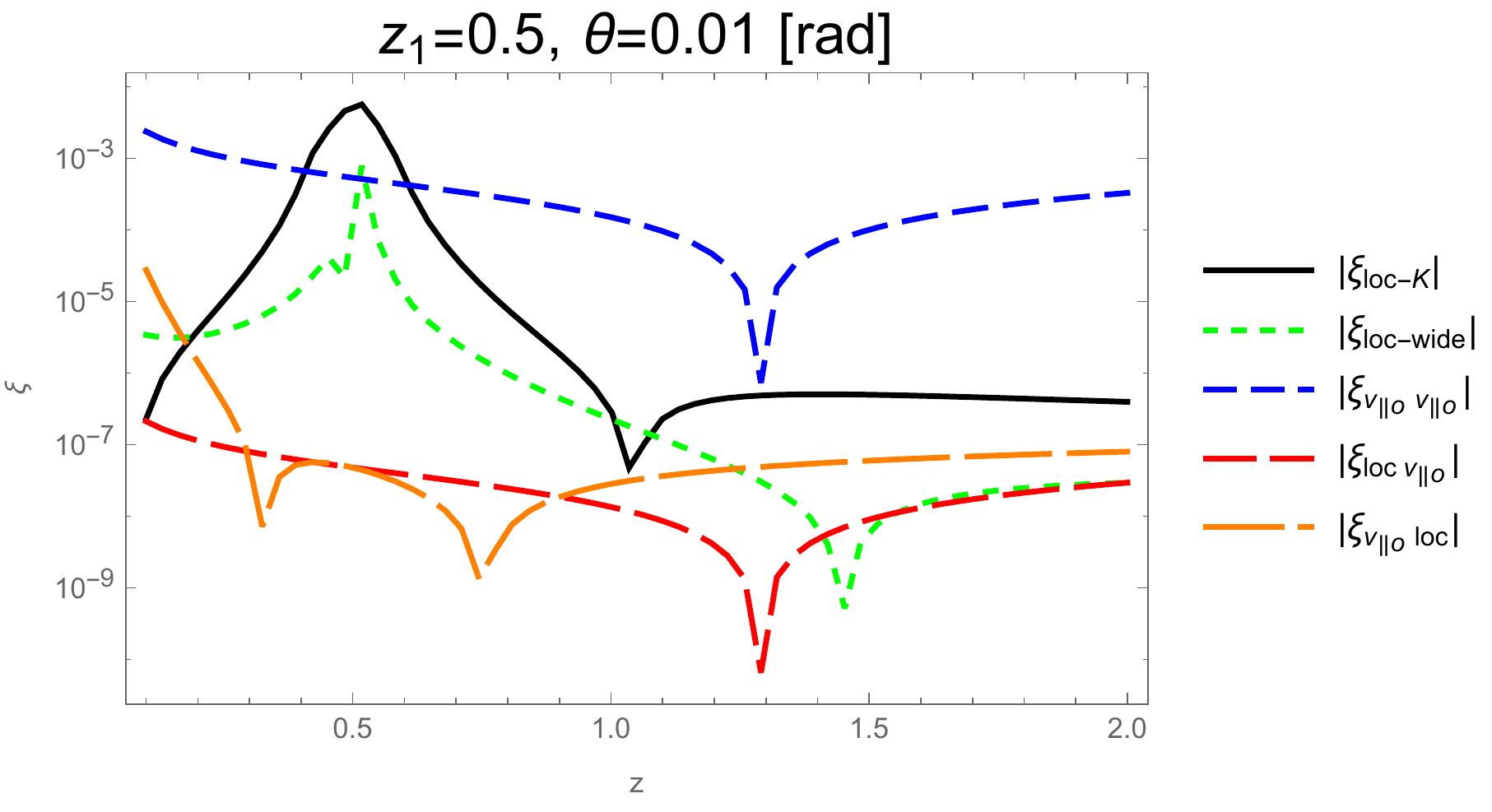}
\includegraphics[width=0.49 \linewidth]{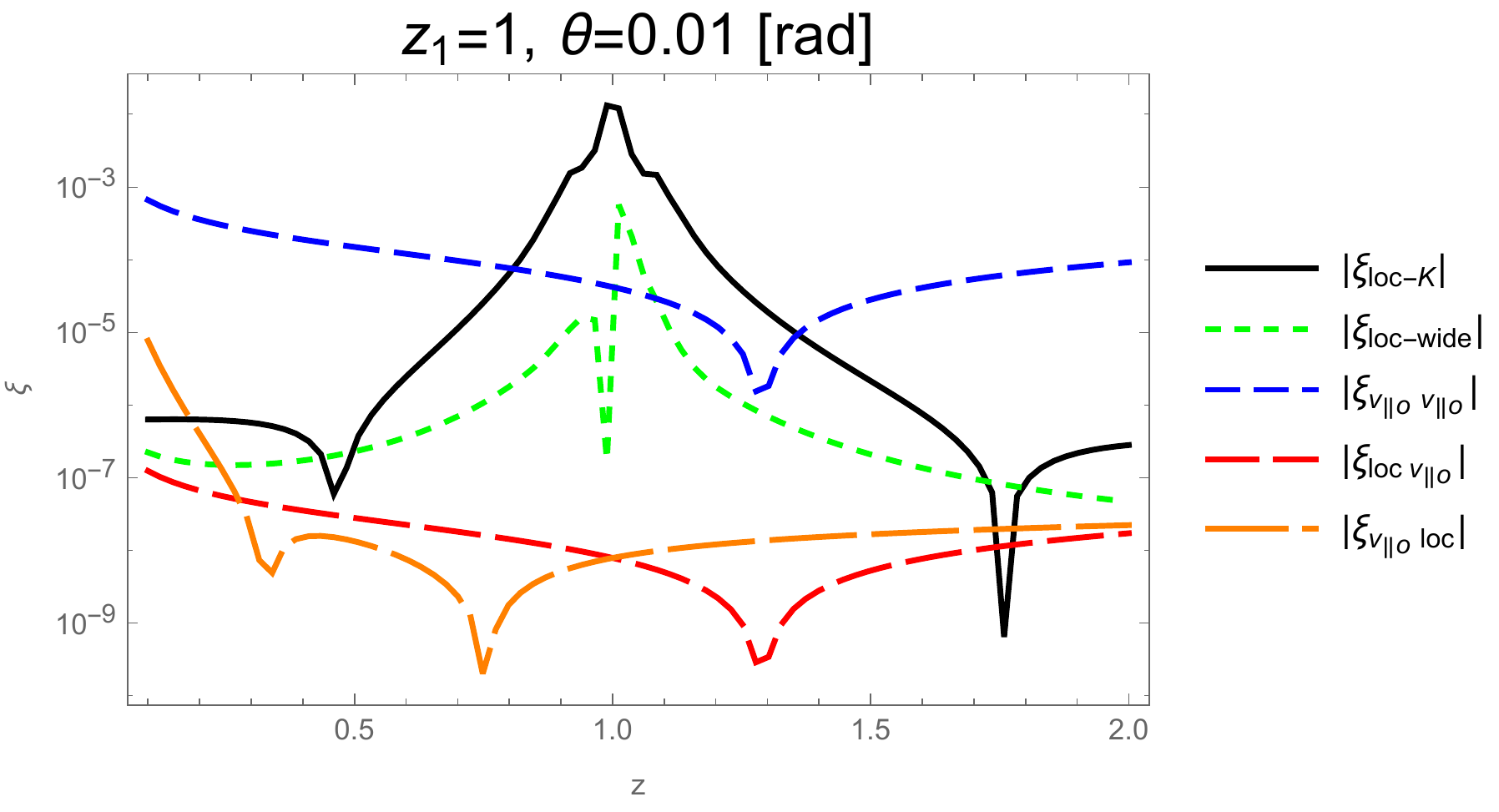}
\includegraphics[width=0.49 \linewidth]{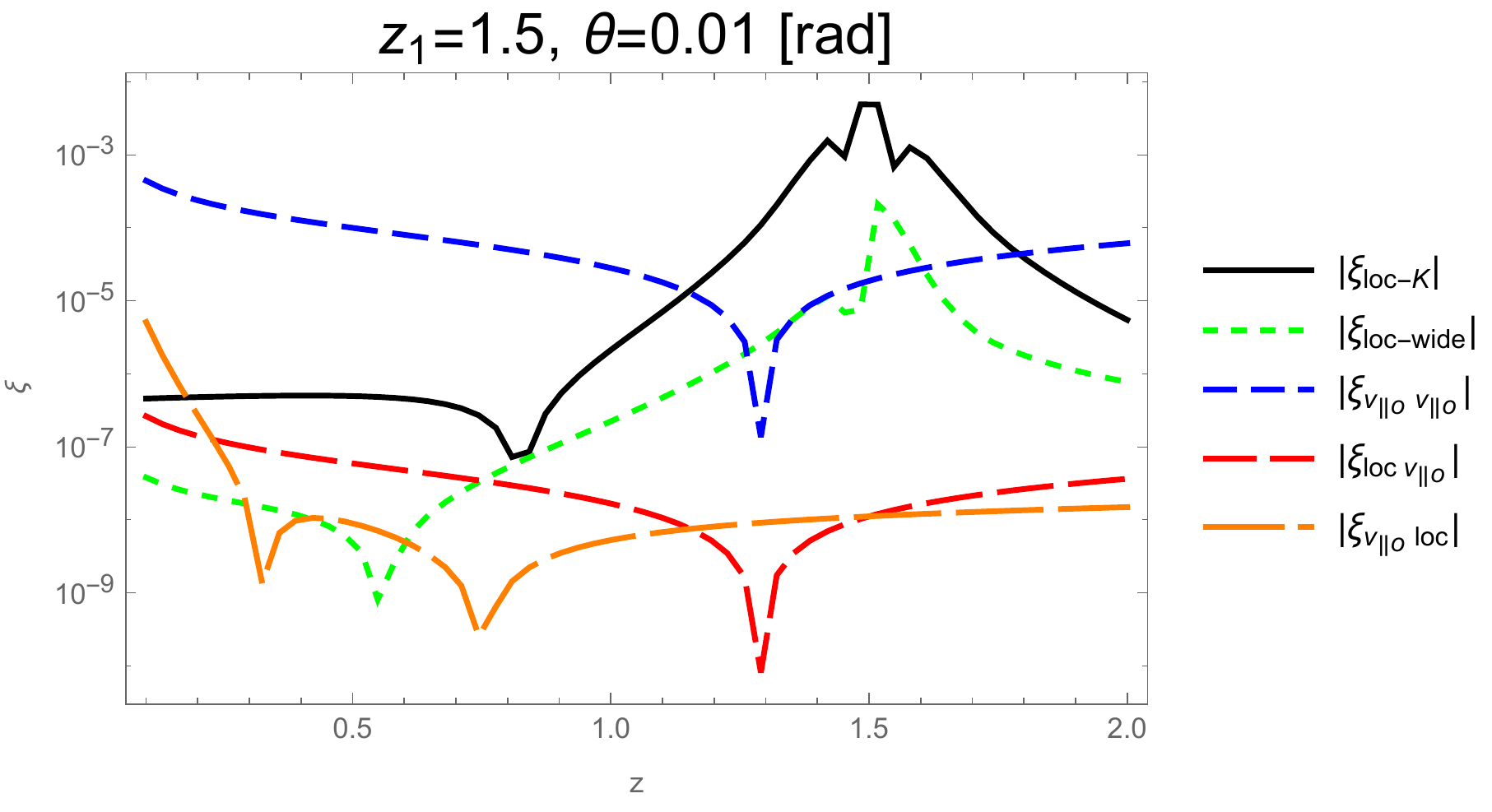}
\caption{Absolute value of all contributions as a function of $z=z_2$, where we have fixed $z_1$ and the separation angle $\theta$.} \label{fig:non-transverse-smalltheta}
\end{figure*}

\begin{center}
\begin{figure*}[htb!]
\includegraphics[width=0.49\columnwidth]{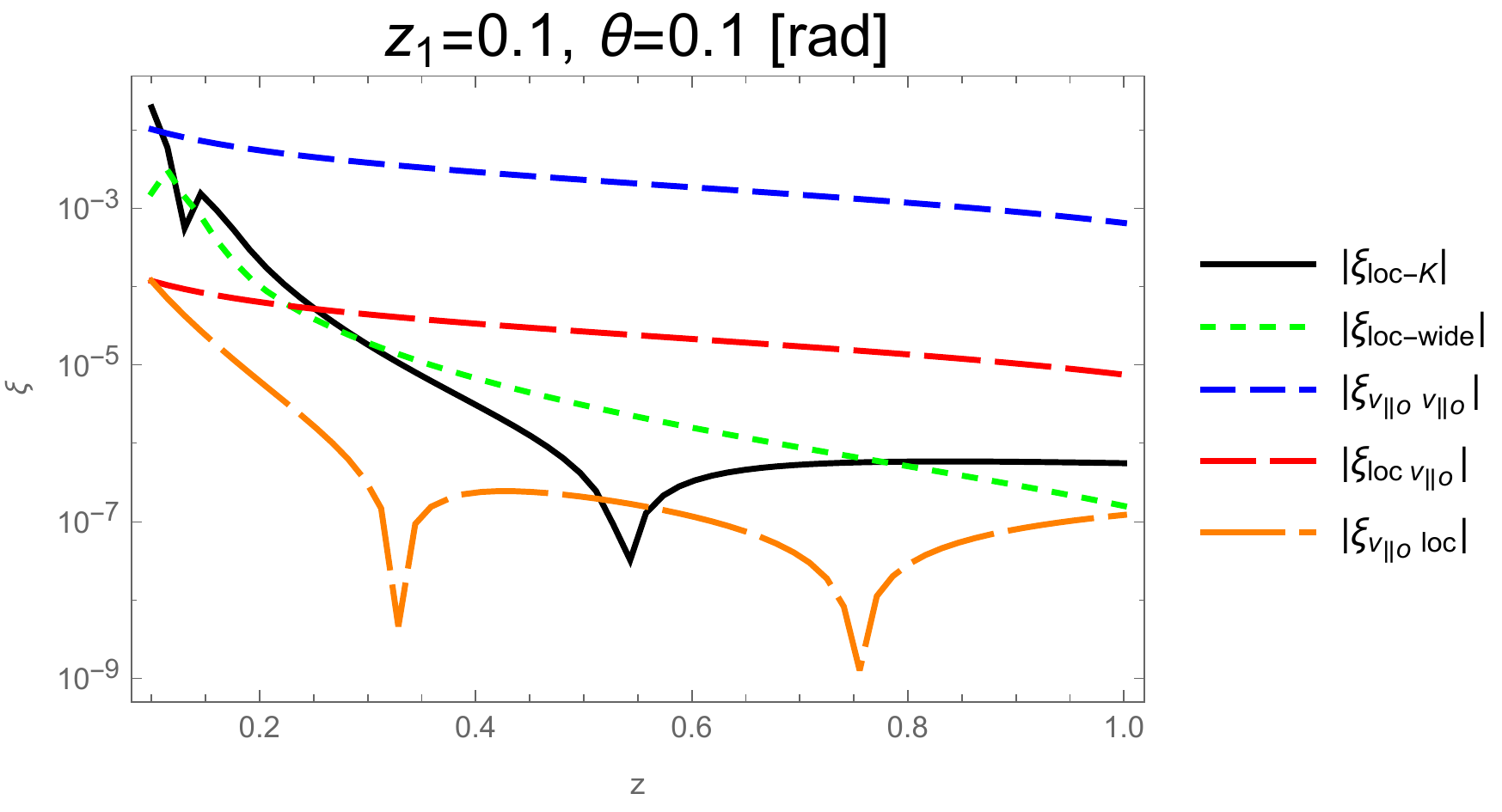}
\includegraphics[width=0.49\columnwidth]{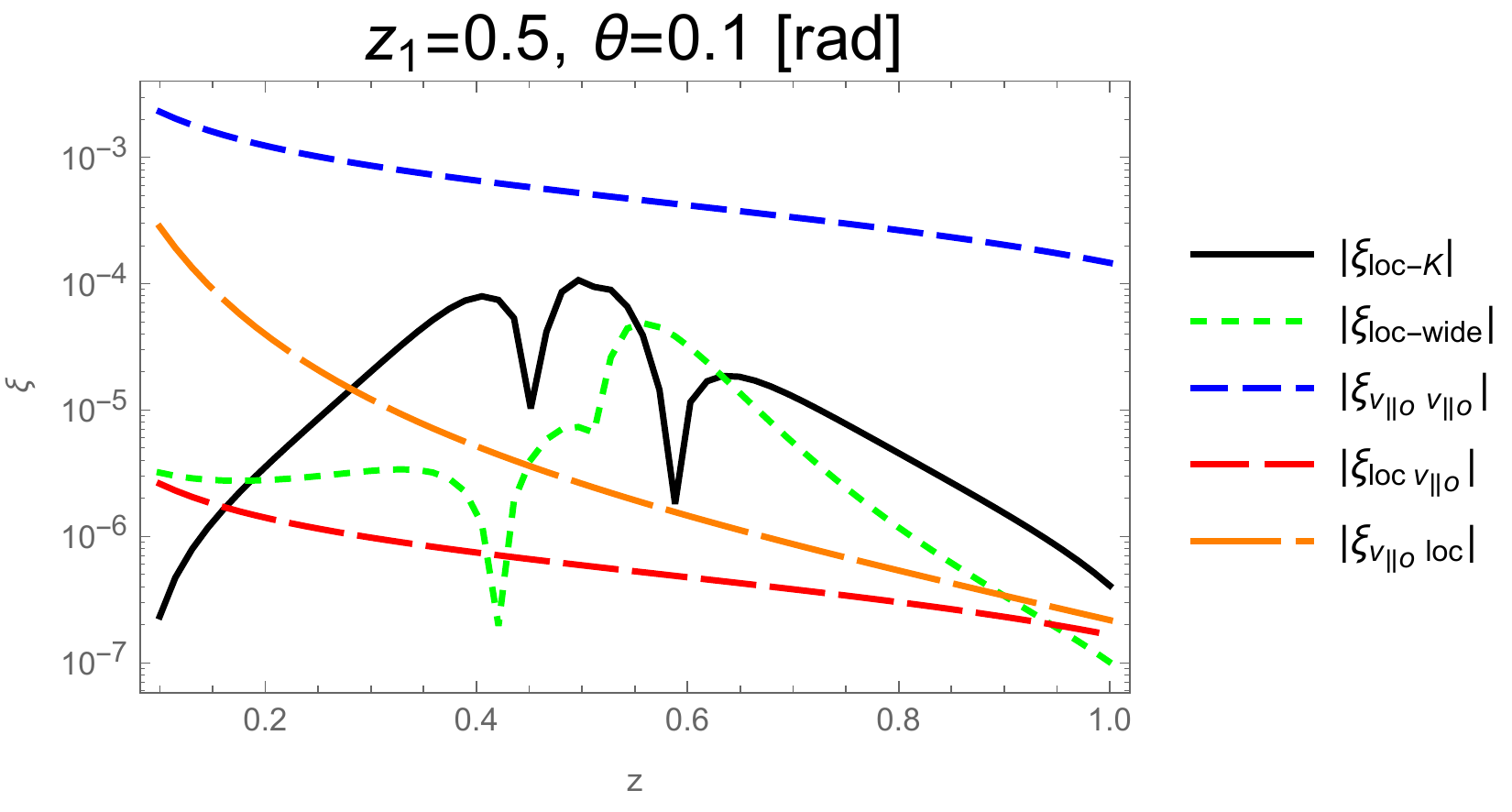}
\includegraphics[width=0.49\columnwidth]{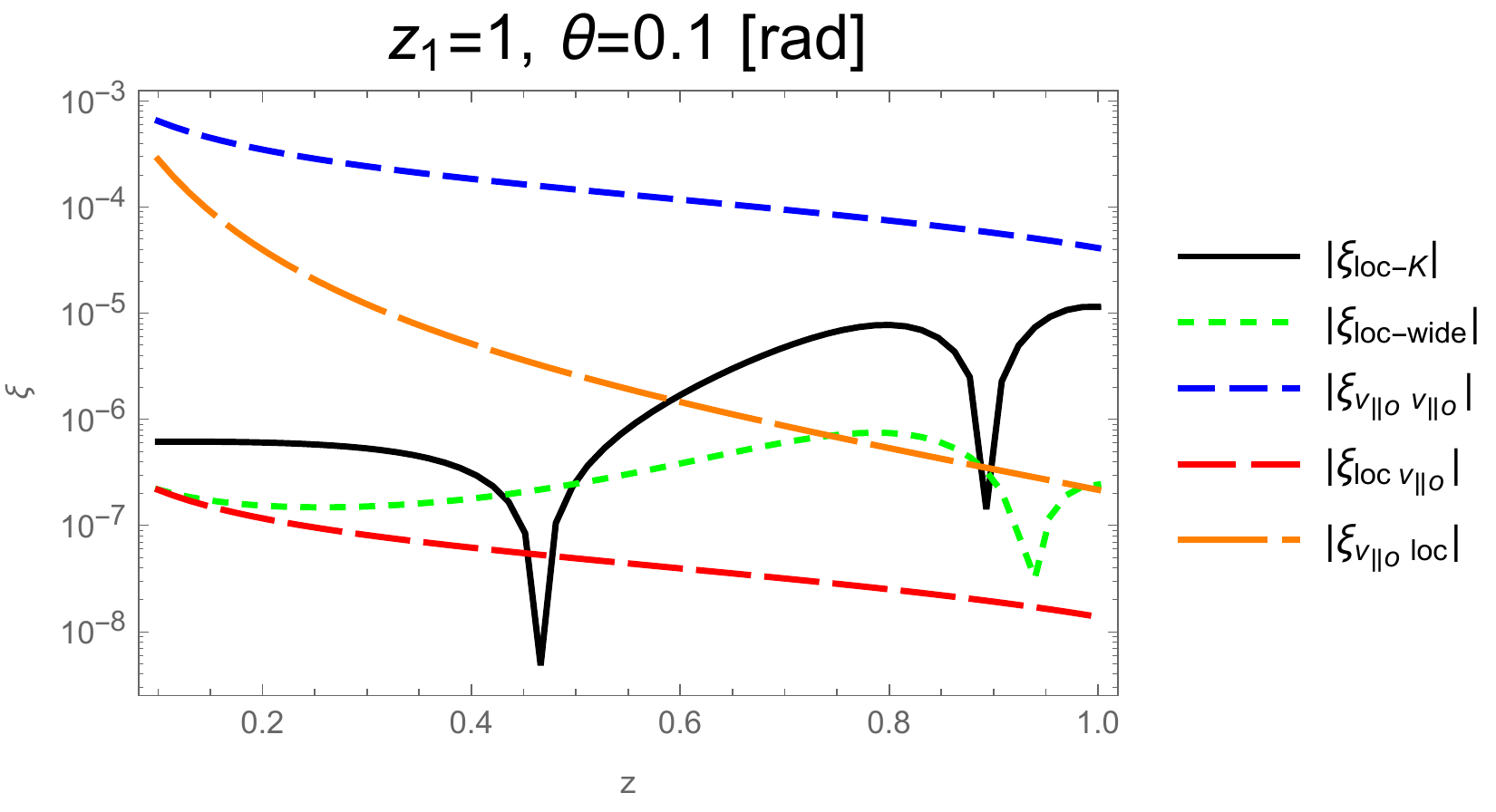}
\includegraphics[width=0.49\columnwidth]{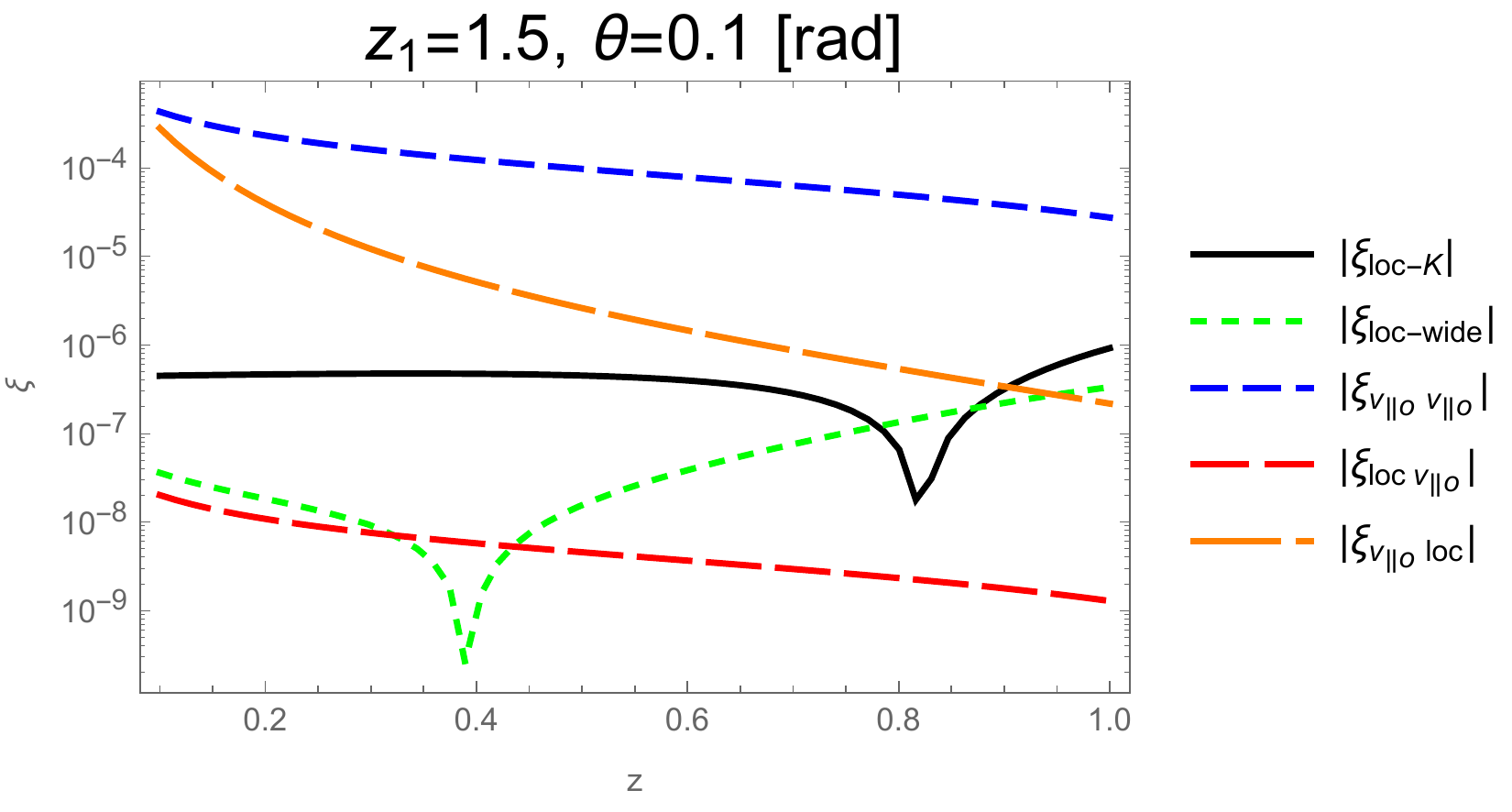}
\caption{Absolute value of all contributions as a function of $z=z_2$, where we have fixed $z_1$ and $\theta$ .}\label{fig:non-transverse-largetheta}
\end{figure*}
\end{center}

 Now let us focus on configurations where we fix $z_1$ and varying $z_2$, both  for a small separation angle (see Fig.\ \ref{fig:non-transverse-smalltheta}) and   for a large separation angle (see Fig.\ \ref{fig:non-transverse-largetheta}). Also in these cases,  for distances larger than the Baryon Acoustic Oscillations (BAO) scales the dipole  contribution on $\xi$ dominates, as expected. 

As we observe in Eq.\ (\ref{xi_totloc}), the contribution of the Kaiser Rocket effect is mainly in the monopole over the pair orientation angle $\varphi$, i.e. for $\tilde L=0$.
Therefore, it is useful to focus  in detail the corrections to the local correlation function, due to the Rocket effect. Due to the fact that this contribution  might be important in  wide and deep surveys, we have to consider carefully the geometry of the system.
Precisely, we follow the approach suggested in \cite{Raccanelli:2013multipoli} where the authors introduced a suitable modification in the 
  argument of the Legendre polynomials, i.e. in the angular dependence of the multipole expansion. For the monopole we have to use the following relations
\bea
\label{eq:mono-loc}
\xi_{{\rm loc}0}(z_2, \theta) &=& \frac{1}{2} \frac{\pi}{\pi-2\theta} \int_{\theta}^{\pi-\theta} ~ d\varphi ~ \xi_{\rm loc}(z_2, \theta, \varphi) ~ \mathcal{P}_0 
 \left\{\cos\left[\frac{\pi \left( \varphi - \theta \right)}{\pi - 2\theta}\right]\right\} \, \rm sin \left[\frac{\pi \left( \varphi - \theta \right)}{\pi - 2\theta}\right]\;,\\
 \Delta\xi_{0}(z_2, \theta) &=& \frac{1}{2} \frac{\pi}{\pi-2\theta} \int_{\theta}^{\pi-\theta} ~ d\varphi ~ \left[\xi(z_2, \theta, \varphi)- \xi_{\rm loc}(z_2, \theta, \varphi)\right] ~ \mathcal{P}_0 
 \left\{\cos\left[\frac{\pi \left( \varphi - \theta \right)}{\pi - 2\theta}\right]\right\} \, \rm sin \left[\frac{\pi \left( \varphi - \theta \right)}{\pi - 2\theta}\right]\;,\
\eea
where we have defined $\varphi$ in the following way
\be
\varphi(z_1,z_2,\theta) = \cos^{-1}\left\{ \sqrt{\frac{1+\cos 2 \theta}{2}}\left[\frac{\chi(z_1)-\chi(z_2)}{\chi_{12}(z_1,z_2,\theta)}\right]\right\} \;.
\ee
\begin{figure*}[!htbp]
\includegraphics[width=0.40 \linewidth]{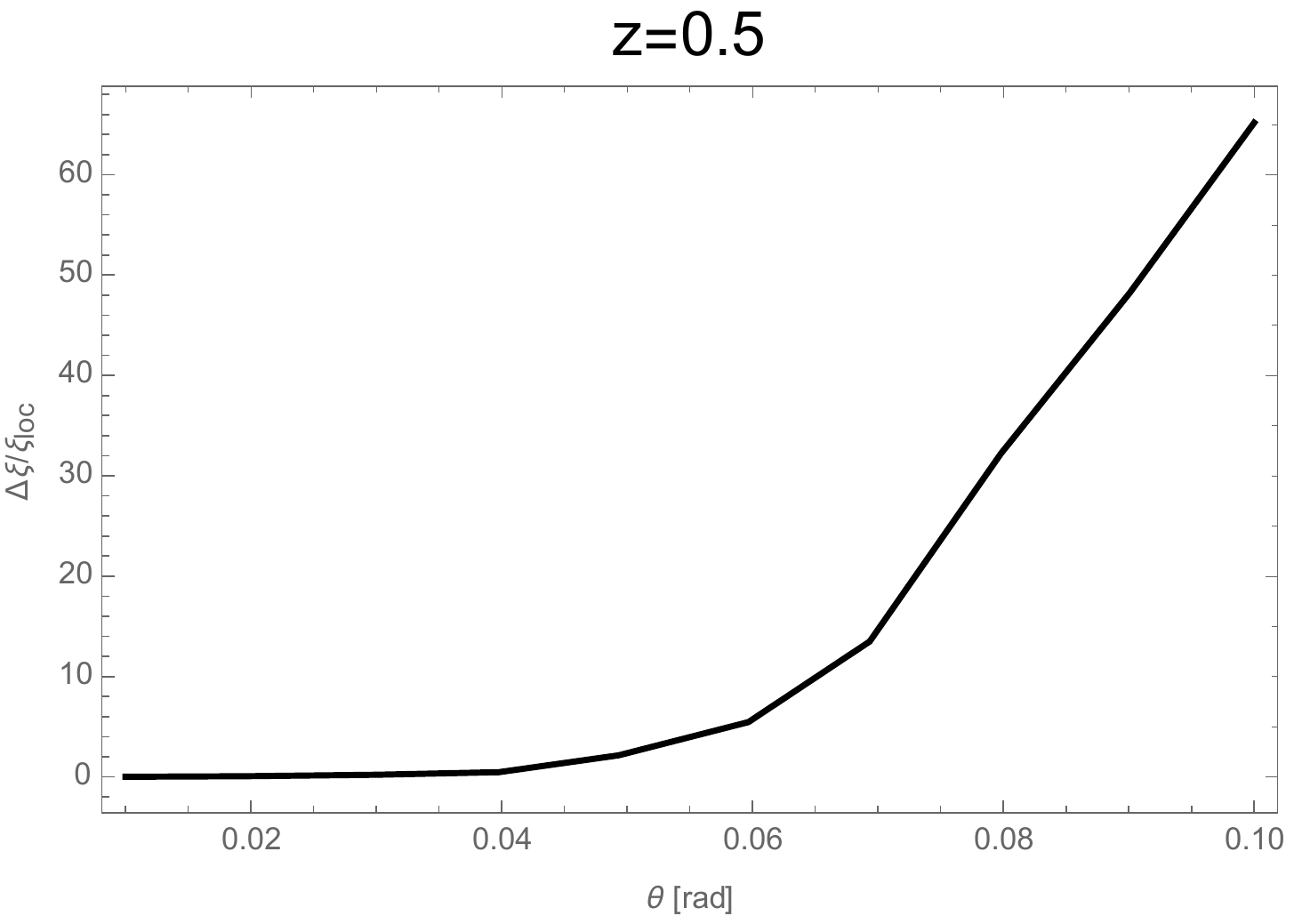}
\includegraphics[width=0.40 \linewidth]{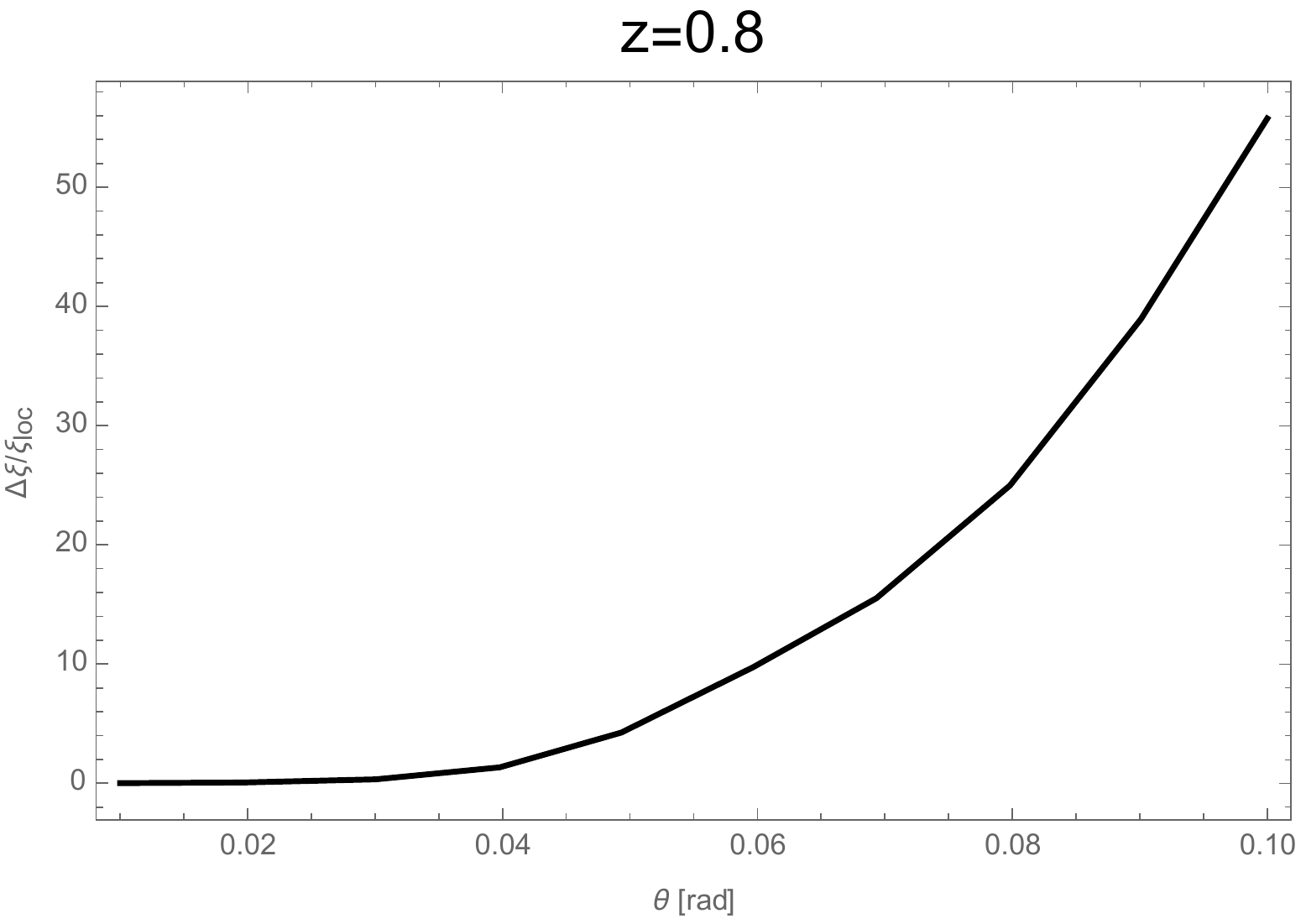}
\includegraphics[width=0.40 \linewidth]{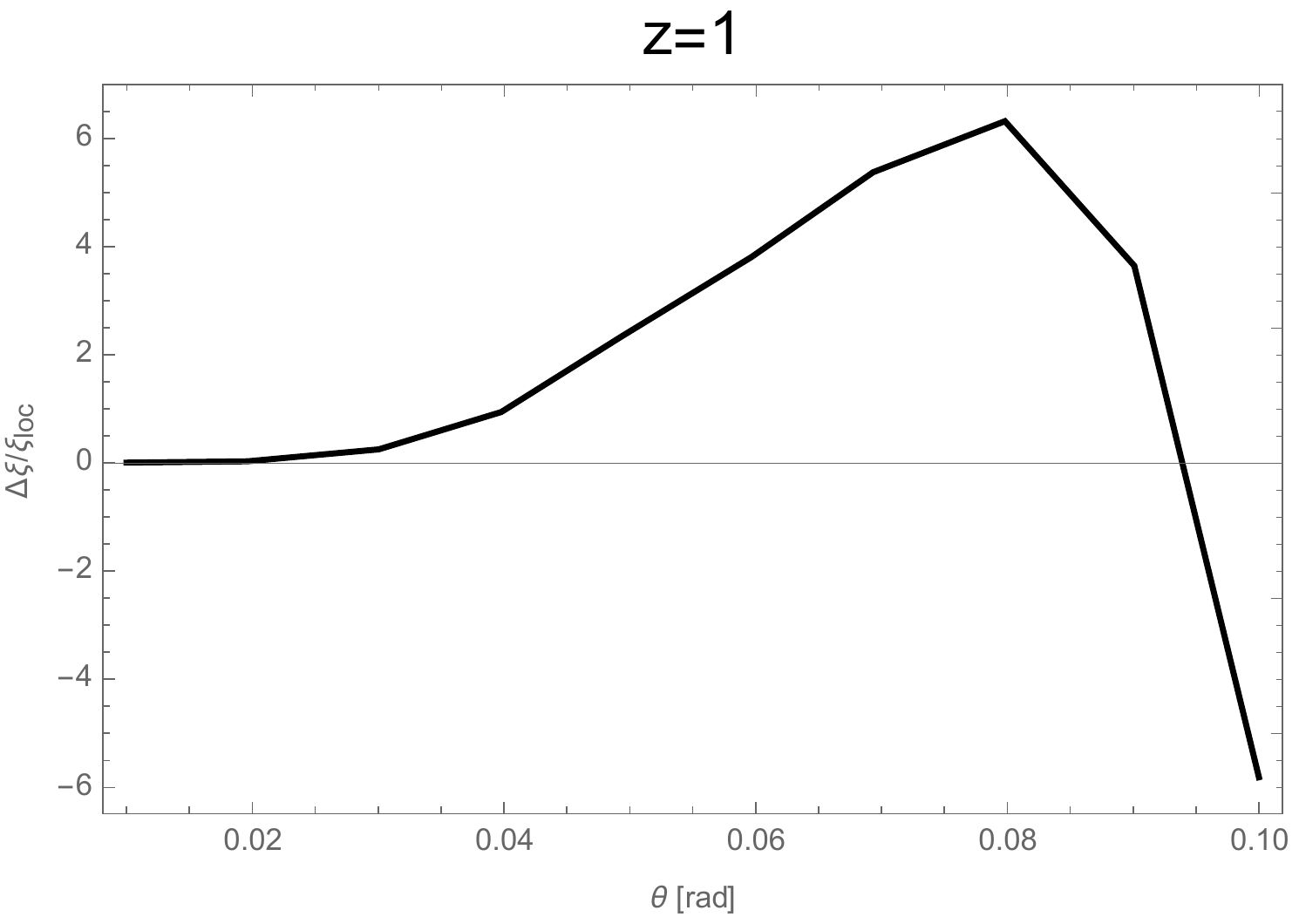}
\includegraphics[width=0.40 \linewidth]{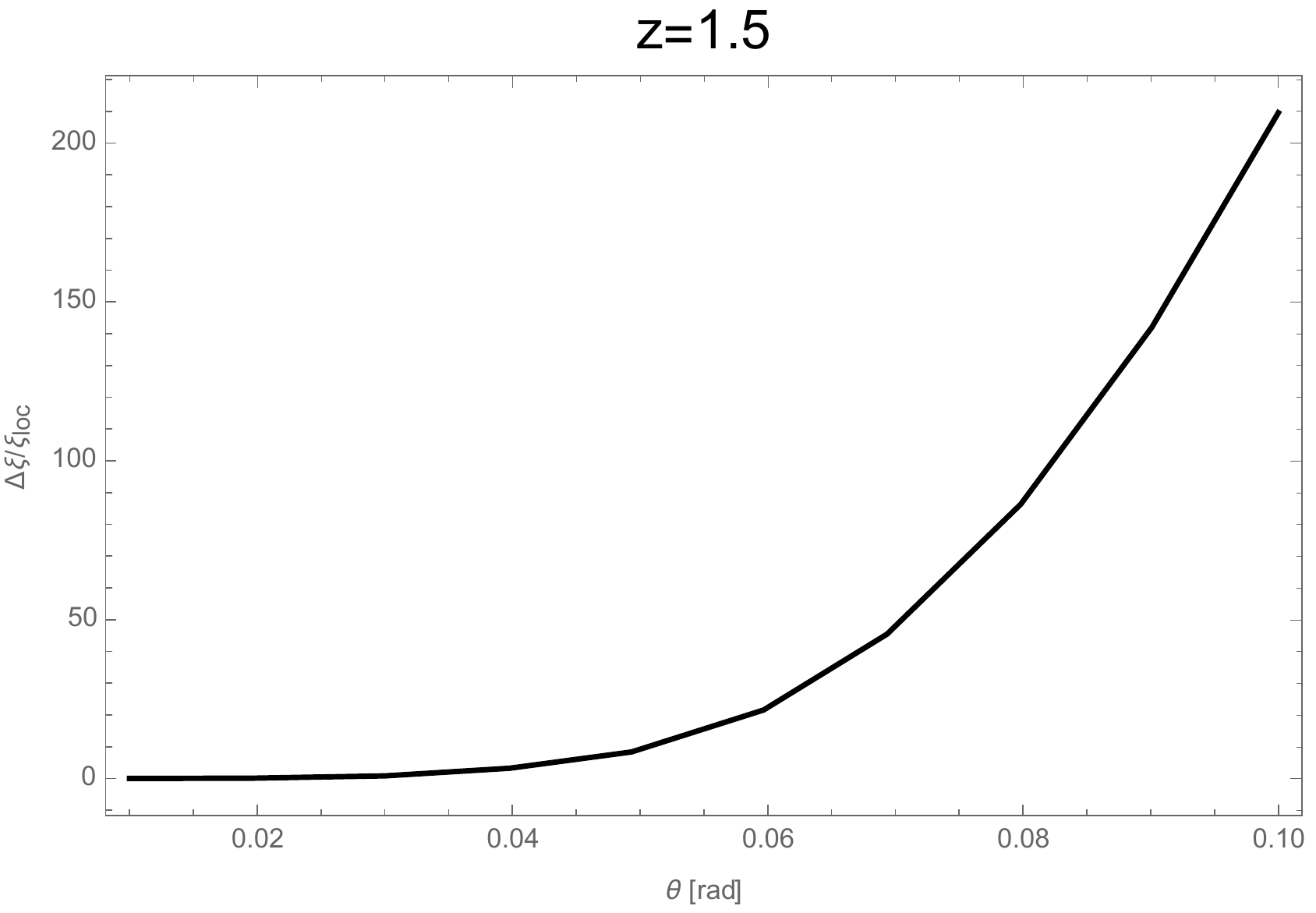}
\caption{$ \Delta\xi_{0}/\xi_{{\rm loc}0} $ as a function of the angular separation $\theta$, for different values of $z_2$. \label{fig:monopole-theta}}
\end{figure*}
In Fig.\ \ref{fig:monopole-theta} we plot $ \Delta\xi_{0}/\xi_{{\rm loc}0} $ as a function of the angular separation $\theta$, for different
values of $z_2$.  Also in this case the amplitude of the Rocket effect  quickly dominates the local contribution at large angular separations.
\begin{figure*}[!htbp]
\includegraphics[width=0.40 \linewidth]{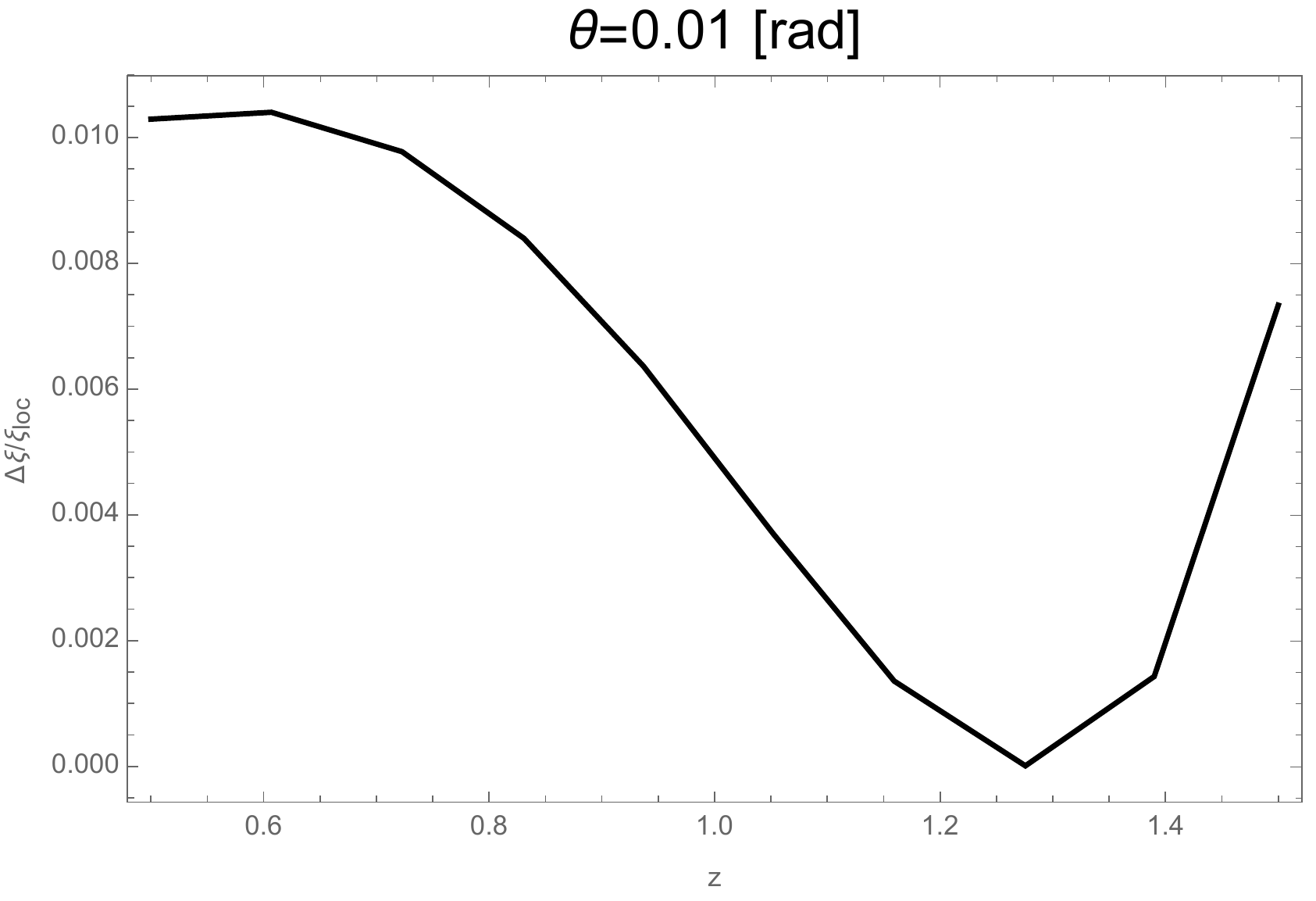}
\includegraphics[width=0.40 \linewidth]{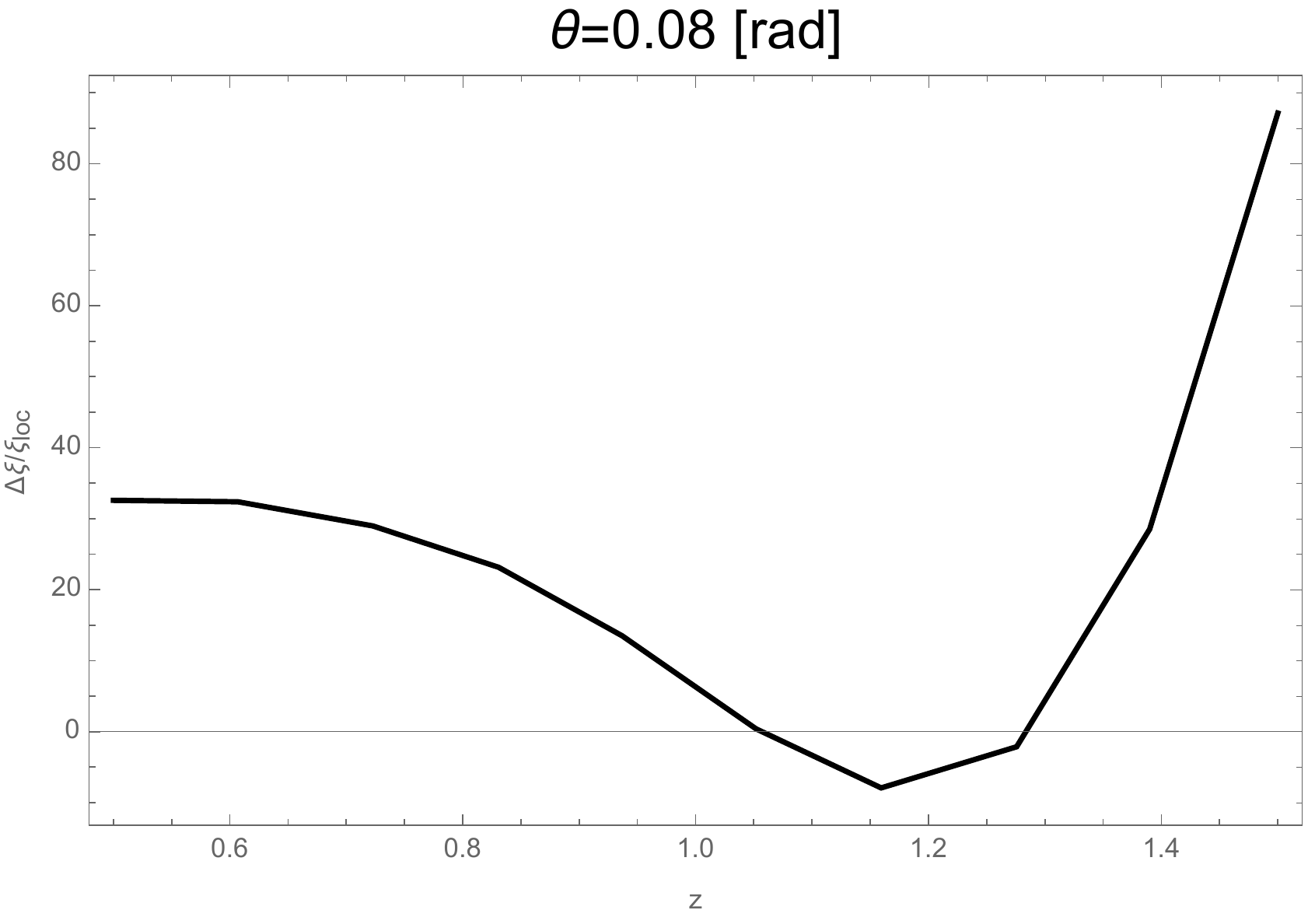}
\includegraphics[width=0.40 \linewidth]{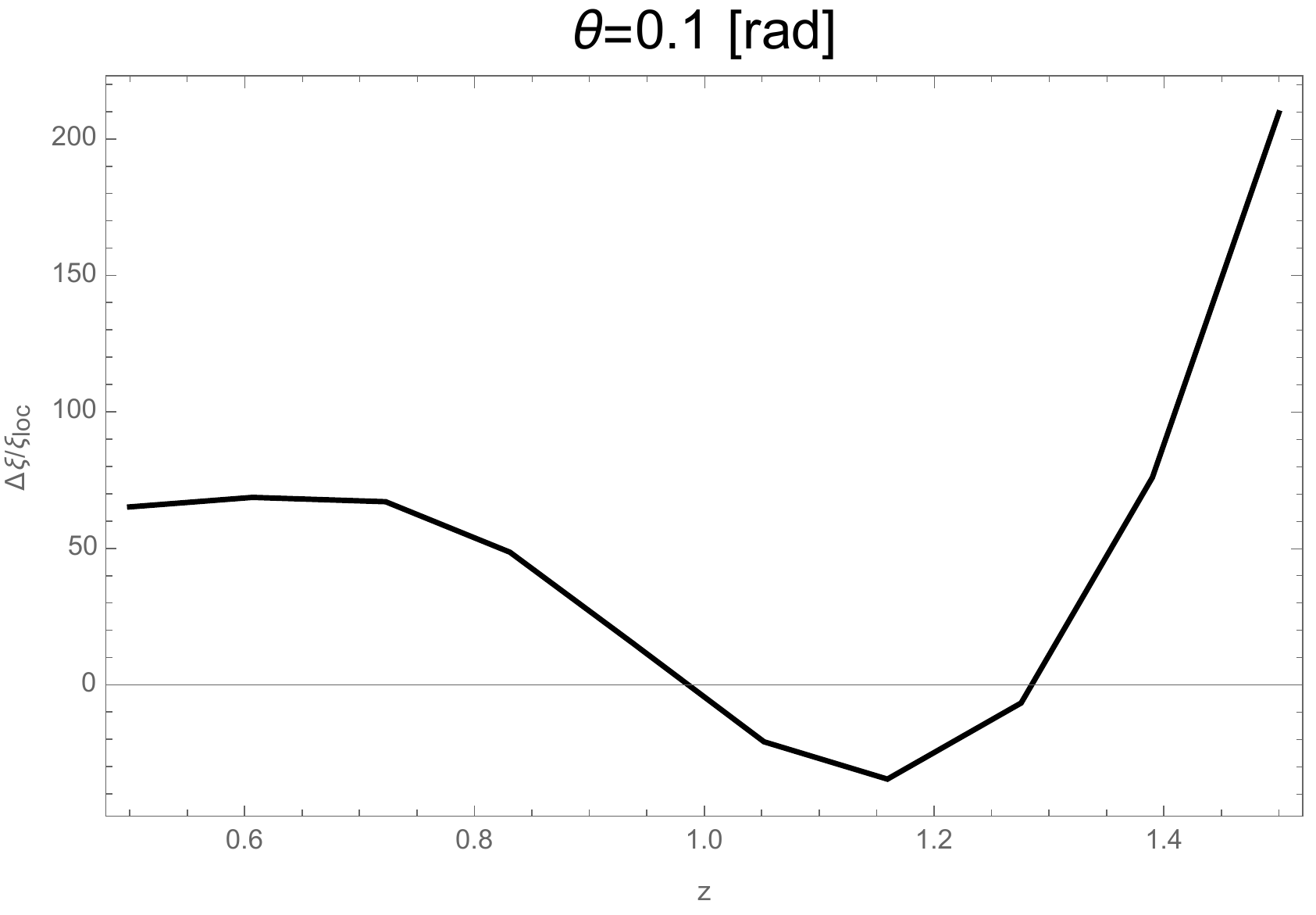}
\includegraphics[width=0.40 \linewidth]{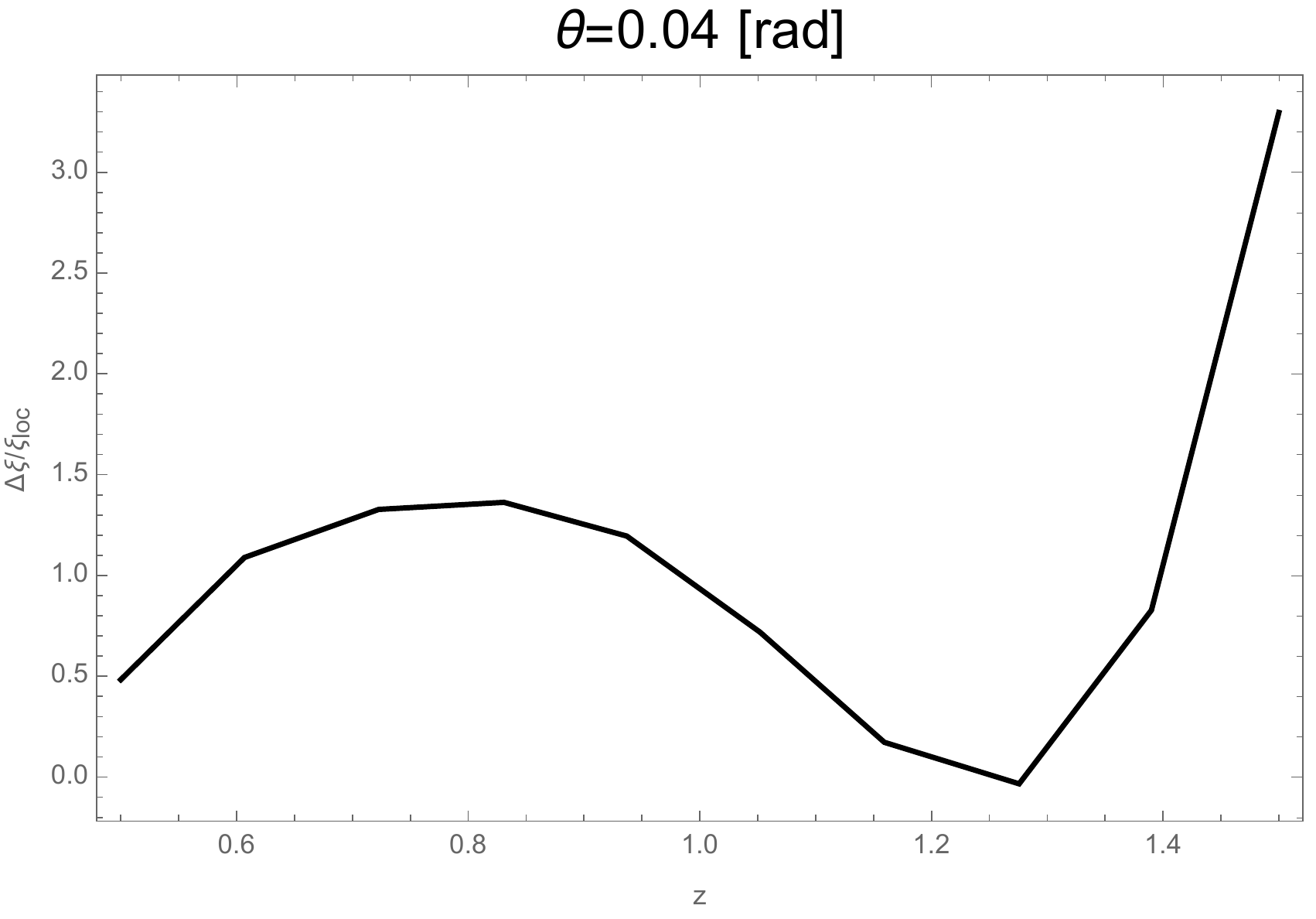}
\caption{$ \Delta\xi_{0}/\xi_{{\rm loc}0} $ as a function of $z_2$, for different values of the angular separation $\theta$. \label{fig:monopole}}
\end{figure*}
The positive or negative values at large $\theta$ could be motivated  by the plots in Fig.\ \ref{fig:monopole} where we are showing $ \Delta\xi_{0}/\xi_{{\rm loc}0} $ as a function of $z_2$, for different values of $\theta$. Precisely, in the bottom-left panel of Fig. \ref{fig:monopole-theta}, we note that $\Delta\xi_{0}/\xi_{{\rm loc}0}$ increases until $\theta \simeq 0.08$ rad and then rapidly decreases into negative values. This curve trend can be explained by the plots in Fig. Ref \ref{Fig:xi_vo} where $\xi_{v_{\| o}v_{\| o}}$, for $ \theta \ge 0.08$ rad, becomes negative. This should produce a maximum for $ \Delta \xi_{0} / \xi_{{\rm loc} 0} $.

  The evaluation of how this impacts on  galaxy clustering measurements  is beyond the scope of this paper  and is left to a future work.

\section{Conclusions}\label{sec:Conlusions}

In this paper we investigate the wide-angle correlations in the galaxy power spectrum in redshift space, including both all general relativistic effects and the Kaiser Rocket effect (which depends on the dipole of the observer).

 We showed via illustrative examples that the Rocket effect on large scales  could  in principle dominate the local signal  of the 2-point correlation function of galaxies at very large scales (see also \cite{Scaccabarozzi:2018vux}). As we can see from Figs.\ \ref{fig:z_1=z_2}, \ref{fig:z_1neqz_2}, \ref{fig:-theta_xi_z1=z2}, \ref{fig:non-transverse-smalltheta}, \ref{fig:non-transverse-largetheta}, \ref{fig:monopole-theta}, \ref{fig:monopole}, the Rocket effect depends on the redshift of the galaxies, the magnification bias $\Q$, the evolution bias $b_e$ and the angular separation  $2\theta$.

 From this work we understood that it is important to evaluate the Kaiser rocket effect well. In particular,  it is important to understand if it is only  a possible source of systematic effects or, if isolated and measured, it allows us to estimate cosmological parameters and break degenerations.
  Future wide and deep surveys will need to utilise a more precise modelling, including all geometry, relativistic and the dipole corrections.
 The next step will be to implement this effect in LIGER \cite{Borzyszkowski:2017ayl} where, building mock galaxy catalogues including all general relativistic corrections at linear order in the cosmological perturbations, we can quantify the impact and  investigate the detectability of the Kaiser Rocket effect in the angular clustering of galaxies  from forthcoming survey data.

In addition, for future surveys, it might be important to quantify if the Rocket terms could contaminate $f_{\rm NL}$ constraints  and, at the same time, how to disentangle these two effects. Note that the Rocket effects will depend on $\omega_{o}$ (which are proportional to the evolution and the magnification bias). Therefore, it is also useful to study in details the relation between window function/selection function ({\it via} $\omega_{o}$) of the surveys and the Rocket effect at different redshift. Finally it is also important to understand if these results are sensitive to the fiducial model. We leave these efforts to a future work.

\acknowledgments
We thank Enzo Branchini, Alan Heavens,  Benedict Kalus, Sabino Matarrese, Cristiano Porciani and Licia Verde for useful discussions. 
DB acknowledges partial financial support by ASI Grant No. 2016-24-H.0.


\begin{thebibliography}{99}

\bibitem{Kogut:1993ag}
  A.~Kogut {\it et al.},
  Astrophys.\ J.\  {\bf 419} (1993) 1
  doi:10.1086/173453
  [astro-ph/9312056].
  
\bibitem{Yahil:1977zz}
  A.~Yahil, G.~A.~Tammann and A.~Sandage,
  Astrophys.\ J.\  {\bf 217} (1977) 903.
  doi:10.1086/155636
  
  
\bibitem{Fixsen:1996nj}
  D.~J.~Fixsen, E.~S.~Cheng, J.~M.~Gales, J.~C.~Mather, R.~A.~Shafer and E.~L.~Wright,
  Astrophys.\ J.\  {\bf 473} (1996) 576
  doi:10.1086/178173
  [astro-ph/9605054].
  
  
    
\bibitem{Hinshaw:2008kr}
  G.~Hinshaw {\it et al.} [WMAP Collaboration],
  Astrophys.\ J.\ Suppl.\  {\bf 180} (2009) 225
  doi:10.1088/0067-0049/180/2/225
  [arXiv:0803.0732 [astro-ph]].
  
\bibitem{Gibelyou:2012ri}
  C.~Gibelyou and D.~Huterer,
  Mon.\ Not.\ Roy.\ Astron.\ Soc.\  {\bf 427} (2012) 1994
  doi:10.1111/j.1365-2966.2012.22032.x
  [arXiv:1205.6476 [astro-ph.CO]].

\bibitem{Nusser:2014sha}
  A.~Nusser, M.~Davis and E.~Branchini,
  Astrophys.\ J.\  {\bf 788} (2014) 157
  doi:10.1088/0004-637X/788/2/157
  [arXiv:1402.6566 [astro-ph.CO]].
  
\bibitem{Maartens:2017qoa}
  R.~Maartens, C.~Clarkson and S.~Chen,
  JCAP {\bf 1801} (2018) no.01,  013
  doi:10.1088/1475-7516/2018/01/013
  [arXiv:1709.04165 [astro-ph.CO]].
  
\bibitem{Pant:2018smd}
  N.~Pant, A.~Rotti, C.~A.~P.~Bengaly and R.~Maartens,
  JCAP {\bf 1903} (2019) 023
  doi:10.1088/1475-7516/2019/03/023
  [arXiv:1808.09743 [astro-ph.CO]].
  
  
\bibitem{Baleisis:1997wx}
  A.~Baleisis, O.~Lahav, A.~J.~Loan and J.~V.~Wall,
  Mon.\ Not.\ Roy.\ Astron.\ Soc.\  {\bf 297} (1998) 545
  doi:10.1046/j.1365-8711.1998.01536.x
  [astro-ph/9709205].
  
\bibitem{Blake:2002gx}
  C.~Blake and J.~Wall,
  Nature {\bf 416} (2002) 150
  doi:10.1038/416150a
  [astro-ph/0203385].


\bibitem{Singal:2011dy}
  A.~K.~Singal,
  Astrophys.\ J.\  {\bf 742} (2011) L23
  doi:10.1088/2041-8205/742/2/L23
  [arXiv:1110.6260 [astro-ph.CO]].
  
  
\bibitem{Rubart:2013tx}
  M.~Rubart and D.~J.~Schwarz,
  Astron.\ Astrophys.\  {\bf 555} (2013) A117
  doi:10.1051/0004-6361/201321215
  [arXiv:1301.5559 [astro-ph.CO]].

\bibitem{Kothari:2013gya}
  P.~Tiwari, R.~Kothari, A.~Naskar, S.~Nadkarni-Ghosh and P.~Jain,
  Astropart.\ Phys.\  {\bf 61} (2014) 1
  doi:10.1016/j.astropartphys.2014.06.004
  [arXiv:1307.1947 [astro-ph.CO]].
  
\bibitem{Schwarz:2015pqa}
  D.~J.~Schwarz {\it et al.},
  PoS AASKA {\bf 14} (2015) 032
  doi:10.22323/1.215.0032
  [arXiv:1501.03820 [astro-ph.CO]].
  
\bibitem{Tiwari:2015tba}
  P.~Tiwari and A.~Nusser,
  JCAP {\bf 1603} (2016) 062
  doi:10.1088/1475-7516/2016/03/062
  [arXiv:1509.02532 [astro-ph.CO]].
  
\bibitem{Colin:2017juj}
  J.~Colin, R.~Mohayaee, M.~Rameez and S.~Sarkar,
  Mon.\ Not.\ Roy.\ Astron.\ Soc.\  {\bf 471} (2017) no.1,  1045
  doi:10.1093/mnras/stx1631
  [arXiv:1703.09376 [astro-ph.CO]].
  
\bibitem{Bengaly:2017slg}
  C.~A.~P.~Bengaly, R.~Maartens and M.~G.~Santos,
  JCAP {\bf 1804} (2018) no.04,  031
  doi:10.1088/1475-7516/2018/04/031
  [arXiv:1710.08804 [astro-ph.CO]].
  
\bibitem{Peebles1980}
 P.~J.~E., Peebles,
`` The large-scale structure of the universe",
Princeton, N.J., Princeton University Press, 1980.~435 p.

\bibitem{Davis:2010sw}
  M.~Davis, A.~Nusser, K.~Masters, C.~Springob, J.~P.~Huchra and G.~Lemson,
  Mon.\ Not.\ Roy.\ Astron.\ Soc.\  {\bf 413} (2011) 2906
  doi:10.1111/j.1365-2966.2011.18362.x
  [arXiv:1011.3114 [astro-ph.CO]].
  
\bibitem{Strauss:1995fz}
  M.~A.~Strauss and J.~A.~Willick,
  Phys.\ Rept.\  {\bf 261} (1995) 271
  doi:10.1016/0370-1573(95)00013-7
  [astro-ph/9502079].
  
\bibitem{Juszkiewicz1990}
R. Juszkiewicz, N. Vittorio, R.~F.~G. Wyse
Astrophys.\ J.\  {\bf 349} (1990) 408-414


 \bibitem{Lahav-Kaiser-Hoffman1990}
 O. Lahav, N. Kaiser and Y. Hoffman,
Astrophys.\ J.\  {\bf 352} (1990) 448-456
doi:10.1086/168550 
 
\bibitem{Peacock1992}
J. A. Peacock,
  Mon.\ Not.\ Roy.\ Astron.\ Soc.\  {\bf 258} (1992) 581-586.
  
\bibitem{Kaiser:1987qv}
  N.~Kaiser,
  Mon.\ Not.\ Roy.\ Astron.\ Soc.\  {\bf 227} (1987) 1.

\bibitem{Hamilton:1997zq}
  A.~J.~S.~Hamilton,
  astro-ph/9708102.
  
\bibitem{Yoo:2008tj}
  J.~Yoo,
  Phys.\ Rev.\ D {\bf 79} (2009) 023517
  [arXiv:0808.3138 [astro-ph]].
  
\bibitem{Yoo:2009au}
  J.~Yoo, A.~L.~Fitzpatrick and M.~Zaldarriaga,
  Phys.\ Rev.\ D {\bf 80}, 083514 (2009)
  [arXiv:0907.0707].
  
    
\bibitem{Yoo:2010ni}
  J.~Yoo,
  Phys.\ Rev.\ D {\bf 82}, 083508 (2010) 
  [arXiv:1009.3021].
  


\bibitem{Bonvin:2011bg}
  C.~Bonvin and R.~Durrer,
  Phys.\ Rev.\ D {\bf 84}, 063505 (2011)
  [arXiv:1105.5280].
  
   
\bibitem{Challinor:2011bk}
  A.~Challinor and A.~Lewis,
  Phys.\ Rev.\ D {\bf 84}, 043516 (2011)
  [arXiv:1105.5292].
  
\bibitem{Bruni:2011ta}
  M.~Bruni, R.~Crittenden, K.~Koyama, R.~Maartens, C.~Pitrou and D.~Wands,
  Phys. Rev. D{\bf 85}  041301 (2012)
  [arXiv:1106.3999].

\bibitem{Baldauf:2011bh}
  T.~Baldauf, U.~Seljak, L.~Senatore and M.~Zaldarriaga,
  arXiv:1106.5507.


\bibitem{Jeong:2011as}
  D.~Jeong, F.~Schmidt and C.~M.~Hirata,
  Phys.\ Rev.\ D {\bf 85}, 023504 (2012)
  [arXiv:1107.5427].
  
  
\bibitem{Yoo:2011zc}
  J.~Yoo, N.~Hamaus, U.~Seljak and M.~Zaldarriaga,
  arXiv:1109.0998.


\bibitem{Bertacca:2012tp}
  D.~Bertacca, R.~Maartens, A.~Raccanelli and C.~Clarkson,
  JCAP {\bf 1210} (2012) 025
  [arXiv:1205.5221 [astro-ph.CO]].
  
\bibitem{Yoo:2012se}
  J.~Yoo, N.~Hamaus, U.~Seljak and M.~Zaldarriaga,
  Phys.\ Rev.\ D {\bf 86} (2012) 063514
  [arXiv:1206.5809 [astro-ph.CO]].
  
\bibitem{Yoo:2013tc}
  J.~Yoo and V.~Desjacques,
  Phys.\ Rev.\ D {\bf 88} (2013) no.2,  023502
  [arXiv:1301.4501 [astro-ph.CO]].


\bibitem{Raccanelli:2013multipoli}
  A.~Raccanelli, D.~Bertacca,  O.~Dor\'{e}  and R.~Maartens,
  JCAP {\bf 1408} (2014) 022
  [arXiv:1306.6646 [astro-ph.CO]].
  
\bibitem{DiDio:2013bqa}
  E.~Di Dio, F.~Montanari, J.~Lesgourgues and R.~Durrer,
  JCAP {\bf 1311} (2013) 044
  [arXiv:1307.1459 [astro-ph.CO]].
  
\bibitem{Yoo:2013zga}
  J.~Yoo and U.~Seljak,
  Mon.\ Not.\ Roy.\ Astron.\ Soc.\  {\bf 447} (2015) no.2,  1789
  [arXiv:1308.1093 [astro-ph.CO]].
  
\bibitem{DiDio:2013sea}
  E.~Di Dio, F.~Montanari, R.~Durrer and J.~Lesgourgues,
  JCAP {\bf 1401} (2014) 042
  [arXiv:1308.6186 [astro-ph.CO]].
  
  
\bibitem{Bonvin:2013ogt}
  C.~Bonvin, L.~Hui and E.~Gaztanaga,
  Phys.\ Rev.\ D {\bf 89} (2014) no.8,  083535
  [arXiv:1309.1321 [astro-ph.CO]].
  
\bibitem{Raccanelli:2013gja}
  A.~Raccanelli, D.~Bertacca, R.~Maartens, C.~Clarkson and O.~Dor\'{e},
  Gen.\ Rel.\ Grav.\  {\bf 48} (2016) no.7,  84
  [arXiv:1311.6813 [astro-ph.CO]].
  
\bibitem{Bacon:2014uja}
  D.~J.~Bacon, S.~Andrianomena, C.~Clarkson, K.~Bolejko and R.~Maartens,
  Mon.\ Not.\ Roy.\ Astron.\ Soc.\  {\bf 443} (2014) no.3,  1900
  [arXiv:1401.3694 [astro-ph.CO]].
  
\bibitem{Chen:2014bba}
  S.~Chen and D.~J.~Schwarz,
  Phys.\ Rev.\ D {\bf 91} (2015) no.4,  043507
  doi:10.1103/PhysRevD.91.043507
  [arXiv:1407.4682 [astro-ph.GA]].

  
\bibitem{Raccanelli:2015vla}
  A.~Raccanelli, F.~Montanari, D.~Bertacca, O.~Dor\'{e} and R.~Durrer,
  JCAP {\bf 1605} (2016) no.05,  009
  [arXiv:1505.06179 [astro-ph.CO]].
  
\bibitem{Alonso:2015uua}
  D.~Alonso, P.~Bull, P.~G.~Ferreira, R.~Maartens and M.~Santos,
  Astrophys.\ J.\  {\bf 814} (2015) no.2,  145
  [arXiv:1505.07596 [astro-ph.CO]].
  
\bibitem{Montanari:2015rga}
  F.~Montanari and R.~Durrer,
  JCAP {\bf 1510} (2015) no.10,  070
  [arXiv:1506.01369 [astro-ph.CO]].
  
\bibitem{Chen:2015wga}
  S.~Chen and D.~J.~Schwarz,
  Astron.\ Astrophys.\  {\bf 591} (2016) A135
  doi:10.1051/0004-6361/201526956
  [arXiv:1507.02160 [astro-ph.CO]].
  
\bibitem{Alonso:2015sfa}
  D.~Alonso and P.~G.~Ferreira,
  Phys.\ Rev.\ D {\bf 92} (2015) no.6,  063525
  doi:10.1103/PhysRevD.92.063525
  [arXiv:1507.03550 [astro-ph.CO]].
  
  
\bibitem{Fonseca:2015laa}
  J.~Fonseca, S.~Camera, M.~Santos and R.~Maartens,
  Astrophys.\ J.\  {\bf 812} (2015) no.2,  L22
  [arXiv:1507.04605 [astro-ph.CO]].
  
\bibitem{Bonvin:2015kuc}
  C.~Bonvin, L.~Hui and E.~Gaztanaga,
  JCAP {\bf 1608} (2016) no.08,  021
  [arXiv:1512.03566 [astro-ph.CO]].
  
\bibitem{Gaztanaga:2015jrs}
  E.~Gaztanaga, C.~Bonvin and L.~Hui,
  JCAP {\bf 1701} (2017) no.01,  032
  [arXiv:1512.03918 [astro-ph.CO]].
  
\bibitem{Cardona:2016qxn}
  W.~Cardona, R.~Durrer, M.~Kunz and F.~Montanari,
  Phys.\ Rev.\ D {\bf 94} (2016) no.4,  043007
  [arXiv:1603.06481 [astro-ph.CO]].
  
\bibitem{Raccanelli:2016avd}
  A.~Raccanelli, D.~Bertacca, D.~Jeong, M.~C.~Neyrinck and A.~S.~Szalay,
  Phys.\ Dark Univ.\  {\bf 19} (2018) 109
  doi:10.1016/j.dark.2017.12.003
  [arXiv:1602.03186 [astro-ph.CO]].
  
\bibitem{DiDio:2016ykq}
  E.~Di Dio, F.~Montanari, A.~Raccanelli, R.~Durrer, M.~Kamionkowski and J.~Lesgourgues,
  JCAP {\bf 1606} (2016) no.06,  013
  [arXiv:1603.09073 [astro-ph.CO]].
  
\bibitem{Borzyszkowski:2017ayl}
  M.~Borzyszkowski, D.~Bertacca and C.~Porciani,
  Mon.\ Not.\ Roy.\ Astron.\ Soc.\  {\bf 471} (2017) no.4,  3899
  doi:10.1093/mnras/stx1423
  [arXiv:1703.03407 [astro-ph.CO]].
  
  
\bibitem{Scaccabarozzi:2017ncm}
  F.~Scaccabarozzi and J.~Yoo,
  JCAP {\bf 1706} (2017) no.06,  007
  doi:10.1088/1475-7516/2017/06/007
  [arXiv:1703.08552 [gr-qc]].
  
  

\bibitem{Abramo:2017xnp}
  L.~R.~Abramo and D.~Bertacca,
  Phys.\ Rev.\ D {\bf 96} (2017) no.12,  123535
  doi:10.1103/PhysRevD.96.123535
  [arXiv:1706.01834 [astro-ph.CO]].



  
\bibitem{Scaccabarozzi:2018vux}
  F.~Scaccabarozzi, J.~Yoo and S.~G.~Biern,
  JCAP {\bf 1810} (2018) no.10,  024
  doi:10.1088/1475-7516/2018/10/024
  [arXiv:1807.09796 [astro-ph.CO]].



  

\bibitem{Tansella:2017rpi}
  V.~Tansella, C.~Bonvin, R.~Durrer, B.~Ghosh and E.~Sellentin,
  JCAP {\bf 1803} (2018) no.03,  019
  [arXiv:1708.00492 [astro-ph.CO]].
  
  
\bibitem{Lepori:2017twd}
  F.~Lepori, E.~Di Dio, E.~Villa and M.~Viel,
  JCAP {\bf 1805} (2018) no.05,  043
  doi:10.1088/1475-7516/2018/05/043
  [arXiv:1709.03523 [astro-ph.CO]].
  
  
\bibitem{Villa:2017yfg}
  E.~Villa, E.~Di Dio and F.~Lepori,
  JCAP {\bf 1804} (2018) no.04,  033
  doi:10.1088/1475-7516/2018/04/033
  [arXiv:1711.07466 [astro-ph.CO]].


  
%
%


\bibitem{Tansella:2018hdm}
  V.~Tansella, C.~Bonvin, G.~Cusin, R.~Durrer, M.~Kunz and I.~Sawicki,
  Phys.\ Rev.\ D {\bf 98} (2018) no.10,  103515
  [arXiv:1807.00731 [astro-ph.CO]].

  
\bibitem{Schoneberg:2018fis}
  N.~Schöneberg, M.~Simonović, J.~Lesgourgues and M.~Zaldarriaga,
  JCAP {\bf 1810} (2018) 047
  [arXiv:1807.09540 [astro-ph.CO]].

  
\bibitem{Tansella:2018sld}
  V.~Tansella, G.~Jelic-Cizmek, C.~Bonvin and R.~Durrer,
  JCAP {\bf 1810} (2018) 032
  [arXiv:1806.11090 [astro-ph.CO]].
  

\bibitem{Breton:2018wzk}
  M.~A.~Breton, Y.~Rasera, A.~Taruya, O.~Lacombe and S.~Saga,
  Mon.\ Not.\ Roy.\ Astron.\ Soc.\  {\bf 483} (2019) no.2,  2671
  [arXiv:1803.04294 [astro-ph.CO]].
  
  \bibitem{Yahil-Strauss-Davis-Huchra1991}
A. Yahil,  M.~A. Strauss, M. Davis and  J.~P. Huchra,
Astrophys.\ J.\  {\bf 372} (1991) 380-393,
      doi:10.1086/169985
      
  \bibitem{Davis-Strauss-Yahil1991}
M. Davis, M.~A. Strauss, and A. Yahil,
  Astrophys.\ J.\  {\bf 372} (1991) 394-409,
      doi:10.1086/169986

\bibitem{Fisher:1994cm}
  K.~B.~Fisher, O.~Lahav, Y.~Hoffman, D.~Lynden-Bell and S.~Zaroubi,
  Mon.\ Not.\ Roy.\ Astron.\ Soc.\  {\bf 272} (1995) 885.







%
%

\bibitem{Nusser:1993sx}
  A.~Nusser and M.~Davis,
  Astrophys.\ J.\  {\bf 421} (1994) L1
  doi:10.1086/187172
  [astro-ph/9309009].
  
\bibitem{Nusser:2011vz}
  A.~Nusser, E.~Branchini and M.~Davis,
  Astrophys.\ J.\  {\bf 744} (2012) 193
  doi:10.1088/0004-637X/744/2/193
  [arXiv:1106.6145 [astro-ph.CO]].
      
\bibitem{Ma:1995ey}
  C.~P.~Ma and E.~Bertschinger,
  Astrophys.\ J.\  {\bf 455}  7 (1995)
  [arXiv:astro-ph/9506072].

\bibitem{Matarrese:1997ay}
  S.~Matarrese, S.~Mollerach and M.~Bruni,
  Phys.\ Rev.\  D {\bf 58}  043504 (1998)
  [arXiv:astro-ph/9707278].

\bibitem{Schmidt:2009rh} 
  F.~Schmidt, E.~Rozo, S.~Dodelson, L.~Hui and E.~Sheldon,
  Phys.\ Rev.\ Lett.\  {\bf 103}, 051301 (2009)
  doi:10.1103/PhysRevLett.103.051301
  [arXiv:0904.4702 [astro-ph.CO]].

\bibitem{Abdalla:2004ah} 
  F.~B.~Abdalla and S.~Rawlings,
  Mon.\ Not.\ Roy.\ Astron.\ Soc.\  {\bf 360}, 27 (2005)
  doi:10.1111/j.1365-2966.2005.08650.x
  [astro-ph/0411342].


\bibitem{Hall:2012wd} 
  A.~Hall, C.~Bonvin and A.~Challinor,
  Phys.\ Rev.\ D {\bf 87}, no. 6, 064026 (2013)
  doi:10.1103/PhysRevD.87.064026
  [arXiv:1212.0728 [astro-ph.CO]].
  
\bibitem{Szalay:1997cc}
  A.~S.~Szalay, T.~Matsubara and S.~D.~Landy,
  astro-ph/9712007.

\bibitem{Bharadwaj:1998bq}
  S.~Bharadwaj,
  Astrophys.\ J.\  {\bf 516} (1999) 507
  [astro-ph/9812274].
  
\bibitem{Matsubara:1999du}
  T.~Matsubara,
  astro-ph/9908056.

\bibitem{Szapudi:2004gh}
  I.~Szapudi,
  Astrophys.\ J.\  {\bf 614} 51 (2004)
  [astro-ph/0404477].

\bibitem{Papai:2008bd}
  P.~Papai and I.~Szapudi,
  arXiv:0802.2940.

\bibitem{Raccanelli:2010hk}
  A.~Raccanelli, L.~Samushia and W.~J.~Percival,
  arXiv:1006.1652.

\bibitem{Samushia:2011cs}
  L.~Samushia, W.~J.~Percival and A.~Raccanelli,
  arXiv:1102.1014.

  
\bibitem{Aghanim:2018eyx}
  N.~Aghanim {\it et al.} [Planck Collaboration],
  arXiv:1807.06209 [astro-ph.CO].
  
  
  

%
%
%
%
%
%
%
%
%
%
%
%
%
%
%
%
%
%
%
%
%
%
%
%
%
%
%
%
%
%
%
%
%
%
%
%
%
%
%
%
%
%
%
%
%
%
%
%
%
%
%
%
%
%
%
%
%
%
%
%
%
%
%
%
%
%
%
%
%
%
%
%
%
%
%
%
%
%
%
%
%
%

  
  
%
%
%


  
\end{thebibliography}
\end{document}